\documentclass[aps,prd,twocolumn,groupedaddress,nofootinbib,floatfix]{revtex4-2}

\usepackage{amsmath,amssymb,amsthm}
\usepackage{graphicx}
\usepackage{epstopdf}
\usepackage{esint}
\usepackage{tikz}

\usetikzlibrary{arrows.meta,decorations.pathreplacing}
\usetikzlibrary{decorations.pathmorphing}

\newcommand{\txt}{\textstyle}
\newcommand{\dsp}{\displaystyle}

\newcommand{\half} {{\textstyle \frac{1}{2}}}
\newcommand{\third}{{\txt \frac{1}{3}}}
\newcommand{\quarter}{{\txt\frac{1}{4}}}

\newcommand{\e}{{\rm e}}   
\newcommand{\ri}{{\rm i}}
\newcommand{\rd}{{\rm d}}
\renewcommand{\Re}{{\rm Re}\,}

\newcommand{\tr}{{\rm tr}\,}

\begin{document}
	
	\title{Scalar Casimir Effect on a Two-Dimensional Sphere with a Wu--Yang Magnetic Monopole}
	
	\author{Jun-Cheng Zhang}
	\author{Xiao-Yin Pan}
	\email{panxiaoyin@nbu.edu.cn}
	\affiliation{Department of Physics, Ningbo University, Ningbo 315211, Zhejiang Province, China}
	
	\author{Juhao Wu}
	\email{jhwusu@outlook.com}
	\affiliation{Stanford University, Stanford, CA 94309, USA}
	
	\author{Qing-hai Wang}
	\email{qhwang@nus.edu.sg}
	\affiliation{Department of Physics, National University of Singapore, Singapore 117551, Singapore}
	
	\date{24 November 2026}

\begin{abstract}
	We investigate the Casimir effect of a complex scalar field on a two-dimensional sphere threaded by a fixed Wu--Yang magnetic monopole at the center. In this background the charged scalar field is a section of a nontrivial complex line bundle over \(S^2\), reflecting the monopole's nontrivial topology. We solve the Klein--Gordon equation analytically and compute the Casimir energy using a generalized Abel--Plana formula for multivalued functions. We find that the monopole can reverse the sign of the Casimir pressure, turning an attractive force into a repulsive one when the monopole strength is sufficiently large. This behavior persists across a broad range of curvature couplings and suggests that Wu--Yang monopoles may provide a source of repulsive vacuum stress, with possible implications for gravitational and cosmological settings.
\end{abstract}

\maketitle

\section{Introduction}
\label{sec:introduction}

Although magnetic monopoles have not been observed experimentally, they have remained a longstanding subject in theoretical physics~\cite{Dirac1931,Dirac1948,GoddardOlive1978,Preskill1983,BlagojevicSenjanovic1988,VilenkinShellard1994,Shnir2005}. A magnetic monopole is a pointlike particle carrying magnetic charge. Interest in such objects dates back to Dirac's seminal work~\cite{Dirac1931}, where he introduced the Dirac monopole and showed that its existence implies electric-charge quantization. The corresponding vector potential is singular along a semi-infinite line extending from the monopole, the so-called Dirac string. When the Dirac quantization condition is satisfied, however, this string can be shifted by a gauge transformation and is therefore unobservable.

A singularity-free formulation of the Abelian monopole was later developed by Wu and Yang~\cite{WuYang1975,WuYang1976}. Their construction is based on the observation that the direction of the Dirac string is gauge dependent. Instead of covering the space $\mathbb{R}^3$ surrounding the monopole with a single coordinate patch, one introduces two overlapping regions, $\mathcal{R}_a$ and $\mathcal{R}_b$, and defines the vector potential separately on each patch. The Wu--Yang construction thus provides a nonsingular description of the Abelian monopole at the price of a nontrivial bundle structure, and it highlights the close relation between topology and gauge physics.

After Dirac's original work, interest in monopoles persisted intermittently until the modern era began with the discovery of smooth solitonic monopole solutions by 't Hooft~\cite{tHooft1974} and Polyakov~\cite{Polyakov1974} in the Georgi--Glashow model~\cite{GeorgiGlashow1972}. Unlike the Dirac monopole, these non-Abelian monopoles are regular finite-energy solutions. Monopoles have since played important roles in particle physics, grand unified theories, and cosmology~\cite{Preskill1979,Preskill1983,VilenkinShellard1994}. Closely related monopolelike structures have also appeared in condensed-matter and synthetic systems, including rotating superfluids~\cite{Salomaa1987}, spin ice materials~\cite{Castelnovo2008,Gingras2009,Bramwell2009,Giblin2011}, chiral magnets~\cite{LinSaxena2016}, momentum-space Berry-curvature effects~\cite{Fang2003}, magnetic-needle analogs~\cite{Beche2014}, superconducting qubit platforms~\cite{Zhang2017},  magneto-optical systems~\cite{Marques2024}, quantum Hall effect~\cite{Haldane1983,Jain2007,Zhou2018,DolanHunterMcCabe2020,GattuJain2025}.

The Casimir effect is one of the clearest manifestations of quantum vacuum fluctuations. It may be understood as the difference between zero-point energies in the presence and absence of nontrivial boundary conditions or global geometric constraints. Casimir first predicted this effect in 1948 for two parallel perfectly conducting plates in vacuum~\cite{Casimir1948}. Since then, it has been studied extensively; see Refs.~\cite{PlunienMullerGreiner1986,MostepanenkTrunov1988,Milton2001,BordagKlimchitskayaMohi2009,BordagMohideenMostep2001,KlimchitskayaMohideenMostep2009} for reviews. Following the landmark experiment of Lamoreaux~\cite{Lamoreaux1997}, the Casimir effect has been confirmed in a variety of experimental settings \cite{Hertlein2008,Munday2009,Paladugu2016,Somers2018,ZhaoStable2019,Schmidt2022,ZhangMagField2024}. It has found applications across quantum field theory, condensed-matter physics, atomic and molecular physics, gravity, and cosmology~\cite{BellucciSaharian2009,ZhaoChiral2009,GrushinCortijo2011,BezerraKlimchMostepRomero2011a,BezerraMostepMotaRomero2011b,Wilson2011,TseMacDonald2012,Felicetti2014,RodriguezLopezGrushin2014,BezerraMotaRomero2014,SchechterKamenev2014,Quach2015,Wilson2015,Hartmann2017,Song2017,Macri2018,Ishikawa2020,Ishikawa2021,Chernodub2022,NakayamaSuzuki2023,JiangJing2025}. 

Casimir energies also arise in topologically nontrivial spaces~\cite{Milton2001,BordagKlimchitskayaMohi2009,BordagMohideenMostep2001,MamaevTrunov1979b,MamaevTrunov1979a,MamaevTrunov1980,Boyer1968,Davies1972,MiltonDeRaadSchwinger1978,Ford1975,MamaevTrunov1979a,HerdeiroSampaio2006,HerdeiroRibeiroSampaio2008,Valuyan2018,BenderMilton1994,BordagKirsten1996,BordagElizalde1997,ElizaldeBordagKirsten1998,Hagen2000,CognolaElizaldeKirsten2001,Saharian2001,PufuSachdev2013}, where both topology and curvature affect the vacuum structure. Typical examples include $S^2$ and its higher-dimensional generalizations~\cite{Boyer1968,Davies1972,MiltonDeRaadSchwinger1978,Ford1975,MamaevTrunov1979a,HerdeiroSampaio2006,HerdeiroRibeiroSampaio2008,Valuyan2018,BenderMilton1994,BordagKirsten1996,BordagElizalde1997,ElizaldeBordagKirsten1998,Hagen2000,CognolaElizaldeKirsten2001,Saharian2001,PufuSachdev2013}. The sphere is especially useful because the corresponding Casimir energy can often be computed analytically, allowing direct comparison among different regularization and renormalization schemes. Another notable feature is that the Casimir force need not be attractive; depending on the geometry, topology, and field content, it may also be repulsive~\cite{Boyer1968,Davies1972,BalianDuplantier1978,MiltonDeRaadSchwinger1978,Levin2010,JiangWilczek2019}.

The possibility of a repulsive Casimir effect is particularly interesting in gravitational and cosmological settings. Repulsive vacuum stresses are relevant to late-time cosmic acceleration~\cite{Riess1998,Perlmutter1999}, to inflationary scenarios~\cite{ZeldovichStarobinsky1984}, and to bouncing cosmologies that avoid an initial singularity~\cite{HerdeiroSampaio2006}. They are also important in Kaluza--Klein models, where Casimir forces induced by compact extra dimensions can destabilize the compactification radius~\cite{Kaluza1921,Klein1926,AppelquistChodos1983}. Since the sign of the Casimir force is highly model dependent, it is natural to ask how it changes as one varies the continuous or discrete parameters of a given system.

Despite the extensive literature on Casimir effects, the influence of magnetic monopoles on Casimir energy has received comparatively little attention. There are studies of Casimir energies in gauge theories with monopole-related configurations~\cite{Chernodub2022}, and Ref.~\cite{deMello2002} analyzed vacuum polarization in global monopole spacetime in the presence of a Wu--Yang magnetic monopole. Appendix A of Ref.~\cite{PufuSachdev2013} computes the ground-state energy in the massless, conformally coupled limit ($m_0=0$, $\xi=1/8$) of the present model. This system is also of independent interest because closely related spherical monopole backgrounds arise in quantum Hall physics~\cite{Haldane1983,Jain2007,GattuJain2025,Zhou2018,DolanHunterMcCabe2020}.

In this paper we study a charged complex scalar field with general mass and curvature coupling on $S^2$ in the background of a Wu--Yang magnetic monopole located at the center. We solve the corresponding Klein--Gordon equation analytically and compute the Casimir energy using contour-integration methods. We then determine how the monopole modifies the vacuum energy density, the total Casimir energy, and the Casimir pressure. In particular, we show that sufficiently large monopole charge can qualitatively change the sign of the pressure, turning an attractive force into a repulsive one. This behavior persists across a broad range of curvature couplings and suggests that Wu--Yang monopoles may provide a source of repulsive vacuum stresses. Such repulsive vacuum stresses may be of interest in gravitational and cosmological settings.

The remainder of this paper is organized as follows. In Sec.~\ref{sec:model} we introduce the model and solve the Klein--Gordon equation in the Wu--Yang monopole background. In Sec.~\ref{sec:regularization} we regularize the vacuum energy density, and in Sec.~\ref{sec:renormalization} we renormalize it. In Sec.~\ref{sec:plots} we illustrate the resulting Casimir pressure as a function of the relevant physical parameters. We conclude in the final section.

Throughout the paper we use the mostly-plus metric convention and units $G=\hbar=c=1$, where $G$ is Newton's constant and $c$ is the speed of light.

\section{Model}
\label{sec:model}

We consider a complex scalar field $\Phi$ in the presence of a fixed external electromagnetic field, described by the action
\begin{equation}
	S=\int \rd^d x\,\sqrt{-g}\,\mathcal{L},
\end{equation}
where
$
	\rd^d x := \rd x^0\wedge\cdots\wedge \rd x^{d-1},
$
$d$ is the spacetime dimension and $g=\det(g_{\mu\nu})$. The Lagrangian density is
\begin{equation}
	\mathcal{L}=-(D_\mu\Phi)^\dagger (D^\mu\Phi)-(\xi R+m_0^2)\Phi^\dagger\Phi,
\end{equation}
with
\begin{equation*}
	D_\mu:=\nabla_\mu - \ri ZeA_\mu,
\end{equation*}
where $\nabla_\mu$ is the covariant derivative and $A_\mu$ is the external vector potential. Here $x=(x^0=t,x^1,\dots,x^{d-1})$, $R$ is the scalar curvature of spacetime, $m_0$ and $Ze$ are the mass and charge associated with the complex field, and $\xi$ is the curvature-coupling constant. In particular, $\xi=0$ corresponds to minimal coupling, whereas $\xi=\frac{1}{8}$ corresponds to conformal coupling in three spacetime dimensions~\cite{BirrellDavies1982}. Unless otherwise stated, we assume $m_0>0$. The massless, conformally coupled limit is considered only for comparison with Ref.~\cite{PufuSachdev2013} at the end of Sec.~\ref{sec:renormalization}.

Varying the action with respect to $\Phi^\dag$ yields the massive Klein--Gordon equation
\begin{equation}
	\left(D_\mu D^\mu - \xi R - m_0^2\right)\Phi=0.
	\label{eq:complexKG}
\end{equation}
The conjugate field $\Phi^\dagger$ satisfies the corresponding equation with the opposite charge, $Ze\to -Ze$. For the monopole potential introduced below, the Lorenz gauge condition $\nabla_\mu A^\mu=0$ is satisfied.

The energy-momentum tensor is obtained by varying the action with respect to the metric,
\begin{equation}
	T_{\mu\nu}:= -\frac{2}{\sqrt{-g}}\frac{\delta S}{\delta g^{\mu\nu}},
	\label{eq:Tmunu}
\end{equation}
and its diagonal components are (no sum over \(\mu\))
\begin{equation}
	\begin{aligned}
		T_{\mu\mu}
		={}&
		2(D_\mu\Phi)^\dagger D_\mu\Phi
		+g_{\mu\mu}\mathcal{L}
		\\
		&+
		2\xi
		\left(
		R_{\mu\mu}
		-\nabla_\mu\nabla_\mu
		+g_{\mu\mu}\nabla^\alpha\nabla_\alpha
		\right)
		\Phi^\dagger\Phi.
	\end{aligned}
\label{eq:Tmumu}	
\end{equation}

Wu and Yang's singularity-free construction covers $\mathbb{R}^3$ by two overlapping regions, $\mathcal{R}_a$ and $\mathcal{R}_b$~\cite{WuYang1975}. In spherical coordinates, with the monopole located at the origin, these regions are distinguished by their ranges of the polar angle $\theta$,
\begin{equation}
	\begin{aligned}
		\mathcal{R}_a &: 0\leq \theta < \frac{\pi}{2}+\delta, \\
		\mathcal{R}_b &: \frac{\pi}{2}-\delta < \theta \leq \pi, \\
		\mathcal{R}_{ab} &: \frac{\pi}{2}-\delta < \theta < \frac{\pi}{2}+\delta,
	\end{aligned}
\end{equation}
where $0<\delta<\frac{\pi}{2}$ and $\mathcal{R}_{ab}$ denotes the overlap region; the remaining coordinates take their standard ranges $r>0$ and $0\leq \phi<2\pi$. In this formulation the vector potential is represented by local connection one-forms on the two patches rather than by a single globally defined potential. Its nonvanishing components are
\begin{equation}
	(A_\phi)_a=\mathrm{g}(1-\cos\theta),
	\qquad
	(A_\phi)_b=-\mathrm{g}(1+\cos\theta),
	\label{eq:vector_potential}
\end{equation}
where $\mathrm{g}$ is the monopole strength. In the overlap region these potentials are related by the gauge transformation
\begin{equation}
	(A_\phi)_a = (A_\phi)_b+\frac{\ri}{Ze}S_{ab}\partial_\phi S_{ab}^{-1},
\end{equation}
with
\begin{equation}
	S_{ab}=\e^{2\ri q\phi},
	\qquad
	q=Ze\mathrm{g}=\frac{n_0}{2},
\end{equation}
where $n_0$ is an integer in units $\hbar=c=1$. Here $Ze$ denotes the charge of the particle coupled to the monopole, and the same quantization condition applies to complex scalar and spinor fields. Accordingly, the charged scalar field is a \textit{section} of the associated complex line bundle and Eq.~\eqref{eq:complexKG} must be analyzed separately in the two regions $\mathcal{R}_a$ and $\mathcal{R}_b$, with the scalar field represented by $\Phi_a$ and $\Phi_b$ on the two patches. In the overlap region they are related by the gauge transformation
\begin{equation}
	\Phi_a=S_{ab}\Phi_b.
\end{equation}

Following Wu and Yang~\cite{WuYang1976}, we introduce the monopole angular momentum operator
\begin{equation}
	\widehat{\mathbf{L}}_q:=\mathbf{r}\times(\hat{\mathbf{p}}-Ze\mathbf{A})-q\,\frac{\mathbf{r}}{r}.
\end{equation}
Equation~\eqref{eq:complexKG} may then be rewritten as
\begin{equation}
	\left[
	\partial_t^2
	-\frac{\partial_r(r^2\partial_r)}{r^2}
	+\frac{\widehat{\mathbf{L}}_q^2-q^2}{r^2}
	+\xi R
	+m_0^2
	\right]\Phi=0.
	\label{eq:phi_equation}
\end{equation}

The eigenvalue equations for the angular momentum operator and its $z$ component are
\begin{equation}
	\begin{aligned}
		\widehat{\mathbf{L}}_q^2 Y_{lm}^q(\theta,\phi)={}&l(l+1)Y_{lm}^q(\theta,\phi), \\
		\widehat{L}_{q,z}Y_{lm}^q(\theta,\phi)={}&m\,Y_{lm}^q(\theta,\phi).
	\end{aligned}
\end{equation}
Here $Y_{lm}^q(\theta,\phi)$ are the monopole harmonics~\cite{WuYang1976}, with
\begin{equation}
	l=|q|,|q|+1,|q|+2,\dots,
	\qquad
	m=-l,-l+1,\dots,l.
\end{equation}
They satisfy the normalization condition
\begin{equation}
	\int_0^\pi \sin\theta\,\rd\theta \int_0^{2\pi}\rd\phi\,|Y_{lm}^q(\theta,\phi)|^2=1,
\end{equation}
and are mutually orthogonal for fixed $q$.

To study the Casimir effect, we now specialize to a spatial two-sphere $S^2$ of radius $a_0$. This is a curved manifold with scalar curvature
\begin{equation}
	R=\frac{2}{a_0^2}.
\end{equation}
The corresponding spacetime metric is
\begin{equation}
	\rd s^2=g_{\mu\nu}\rd x^\mu \rd x^\nu
	=-\rd t^2 + a_0^2(\rd\theta^2+\sin^2\theta\,\rd\phi^2).
	\label{eq:metric}
\end{equation}
It follows that
\begin{equation}
	\sqrt{-g}=a_0^2\sin\theta.
\end{equation}
In this geometry, Eq.~\eqref{eq:phi_equation} reduces to
\begin{equation}
	\left(
	\partial_t^2
	+\frac{\widehat{\mathbf{L}}_q^2-q^2}{a_0^2}
	+\frac{2\xi}{a_0^2}
	+m_0^2
	\right)\Phi=0.
	\label{eq:phia0}
\end{equation}

A complete set of positive- and negative-frequency solutions of Eq.~\eqref{eq:phia0} is given by $u_{q,lm}$ and $u_{-q,lm}^{*}$, respectively, where
\begin{equation}
u_{q,lm}(x)
=\frac{1}{a_0\sqrt{2\omega_{ql}}}\,
\e^{-\ri\omega_{ql}t}Y_{lm}^{q}(\theta,\phi),
	\label{eq:orthonormal}
\end{equation}
with
\begin{equation}
	\omega_{ql}
	:= \frac{1}{a_0}\left[\left(l+\frac{1}{2}\right)^2+\beta^2\right]^{1/2},
	\label{eq:omegaJ}
\end{equation}
where 
$x=(t,\theta,\phi)$, and
\begin{equation}
	\beta^2:= a_0^2m_0^2-q^2+2\xi-\quarter.
	\label{eq:beta2}
\end{equation}
Complex conjugation maps $q\to -q$. Since the spectrum is even in $q$, as required by charge-conjugation symmetry, the particle and antiparticle sectors have the same frequency $\omega_{ql}$. We retain the charge labels explicitly: by Eq.~\eqref{eq:orthonormal}, both $u_{q,lm}$ and $u_{-q,lm}^{*}$ transform with the charge-$q$ bundle transition function, whereas their complex conjugates transform in the charge-$-q$ bundle. In the special case $q=0$ and $\xi=\frac{1}{8}$, the result reduces to that of the conformally coupled scalar field without a Wu--Yang magnetic monopole~\cite{Milton2001,BordagMohideenMostep2001,BordagKlimchitskayaMohi2009,MamaevTrunov1979a,MamaevTrunov1979b}.

At fixed $q$, one readily verifies that
\begin{equation}
	\begin{aligned}
	\left(u_{q,lm},u_{q,l'm'}\right)
	&=\delta_{ll'}\delta_{mm'},\\
	\left(u_{-q,lm}^{*},u_{-q,l'm'}^{*}\right)
	&=-\delta_{ll'}\delta_{mm'},\\
	\left(u_{q,lm},u_{-q,l'm'}^{*}\right)&=0.
	\end{aligned}
\end{equation}
with the Klein--Gordon inner product
\begin{equation}
	\begin{aligned}
		(f,g)
		:={}&
		\ri\int_\Sigma \rd\Sigma_\mu
		\left[
		f^*D^\mu g
		-
		g(D^\mu f)^*
		\right]\\
		={}&
		\ri\int \rd V\,
		\left(
		f^*\partial_t g-g\,\partial_t f^*
		\right),
	\end{aligned}
	\label{eq:KGproduct}
\end{equation}
where $\Sigma$ is a constant-$t$ hypersurface, $\rd\Sigma_\mu=(-1,0,0)\rd V$ is future-directed, and $\rd V=a_0^2\sin\theta\,\rd\theta\,\rd\phi$. The quantized complex scalar field can therefore be expanded as
\begin{equation}
	\begin{aligned}
		\Phi(x)
		={}&\sum_{lm}\left[
		u_{q,lm}(x)\,\hat{a}_{lm}
		+u_{-q,lm}^{*}(x)\,\hat{b}_{lm}^\dagger
		\right], \\
		\Phi^\dagger(x)
		={}&\sum_{lm}\left[
		u_{q,lm}^{*}(x)\,\hat{a}_{lm}^\dagger
		+u_{-q,lm}(x)\,\hat{b}_{lm}
		\right],
	\end{aligned}
	\label{eq:phiphidag}
\end{equation}
where $\hat{a}_{lm}$, $\hat{b}_{lm}$, $\hat{a}_{lm}^\dagger$, and $\hat{b}_{lm}^\dagger$ are two independent sets of annihilation and creation operators satisfying
\begin{equation}
\begin{aligned}
	\bigl[\hat a_{lm},\hat a_{l'm'}^\dagger\bigr]
	&=\delta_{ll'}\delta_{mm'},\\
	\bigl[\hat b_{lm},\hat b_{l'm'}^\dagger\bigr]
	&=\delta_{ll'}\delta_{mm'},\\
	\bigl[\hat a_{lm},\hat a_{l'm'}\bigr]
	&=\bigl[\hat a_{lm}^\dagger,\hat a_{l'm'}^\dagger\bigr]=0,\\
	\bigl[\hat b_{lm},\hat b_{l'm'}\bigr]
	&=\bigl[\hat b_{lm}^\dagger,\hat b_{l'm'}^\dagger\bigr]=0.
\end{aligned}
\label{eq:abdagger}
\end{equation}
and the vacuum state
\begin{equation}
	\hat{a}_{lm}|0\rangle=\hat{b}_{lm}|0\rangle=0.
\end{equation}

Note that $\beta^2$ may be negative. However, for Eq.~\eqref{eq:omegaJ} to be well defined, one must have
\begin{equation}
	\left(l+\half\right)^2+\beta^2
	=
	\left(n+|q|+\half\right)^2+\beta^2
	\ge 0
\end{equation}
for arbitrary $q$ and nonnegative integer $n$. This implies
\begin{equation}
	\beta^2\ge -\left(|q|+\half\right)^2,
\end{equation}
or, equivalently,
\begin{equation}
	a_0^2m_0^2+|q|+2\xi\ge 0.
	\label{eq:a0limit}
\end{equation}
We therefore restrict attention to the regime
\begin{equation}
	\xi>-\half(a_0^2m_0^2 + |q|).
\end{equation}
We take the inequality to be strict in order to exclude the zero-frequency lowest mode, which requires separate treatment. If $\xi<-|q|/2$, this condition imposes a lower bound on $a_0$,
\begin{equation}
	a_0 > \frac{1}{m_0}\sqrt{2|\xi|-|q|}.
\end{equation}

Substituting Eq.~\eqref{eq:phiphidag} into the energy-momentum tensor in Eq.~\eqref{eq:Tmumu} and integrating over the sphere gives
\begin{equation}
	\hat H=\sum_{lm}\omega_{ql}
	\left(\hat a_{lm}^{\dagger}\hat a_{lm}
	+\hat b_{lm}^{\dagger}\hat b_{lm}+1\right).
\end{equation}
The equal particle and antiparticle frequencies therefore yield
\begin{equation}
	E_0=\langle0|\hat H|0\rangle
	=\sum_{l=|q|}^{\infty}(2l+1)\omega_{ql}.
\end{equation}
The vacuum energy density is spatially uniform, and hence
\begin{equation}
	\begin{aligned}
		\mathcal{E}_\mathrm{bare}(a_0,m_0)
		:={}& \frac{E_0(a_0,m_0)}{4\pi a_0^2} \\
		={}& \frac{1}{2\pi a_0^2}\sum_{l=|q|}^{\infty}
		\left(l+\frac{1}{2}\right)\omega_{ql}.
	\end{aligned}
	\label{eq:E0}
\end{equation}

\section{Regularization}
\label{sec:regularization}

The sum in Eq.~\eqref{eq:E0} is divergent and must therefore be regularized. A convenient tool for this purpose is the Abel--Plana formula (APF), which, like the Euler--Maclaurin formula, relates a sum to an integral that can be evaluated efficiently by complex-analytic methods~\cite{DLMF}. For a well-behaved analytic function satisfying
\begin{equation}
	\lim_{y\to\infty}\e^{-2\pi y}|f(x\pm \ri y)|=0,
\end{equation}
the two standard forms of the APF are
\begin{equation}
	\begin{aligned}
		\sum_{n=0}^{\infty}f(n)
		={}&
		\int_0^\infty f(x)\,\rd x
		+\frac{1}{2}f(0) \\
		& 
		+\ri\int_{0^+}^{\infty}\rd t\,
		\frac{f(\ri t)-f(-\ri t)}{\e^{2\pi t}-1},\\
		\sum_{n=0}^{\infty}f\!\left(n+\half\right)
		={}&
		\int_0^\infty f(x)\,\rd x\\
		& 
		-\ri\int_{0}^{\infty}\rd t\,
		\frac{f(\ri t)-f(-\ri t)}{\e^{2\pi t}+1}.
	\end{aligned}
	\label{eq:APF}
\end{equation}
The first formula applies to sums over integers and the second to sums over half-odd integers.

The APF has been generalized to meromorphic functions with poles, yielding the generalized Abel--Plana formula (GAPF) \cite{Saharian2007}. In the present problem, however, we use another extension to multivalued functions with branch cuts. We present the corresponding Abel--Plana formula for multivalued functions (GAPF-MV), together with its proof, in Appendix~\ref{app:gapf}.

To apply the APF, we define
\begin{equation}
	n := l-|q|,
\end{equation}
and rewrite Eq.~\eqref{eq:E0} as
\begin{equation}
	\mathcal{E}_\mathrm{bare}(a_0,m_0)
	=
	\frac{1}{2\pi a_0^3}\sum_{n=0}^{\infty} f(n+\delta),
	\label{eq:E0f}
\end{equation}
where
\begin{equation}
	\delta:=|q|+\half,
\end{equation}
which is either a positive integer or a positive half-odd integer, and
\begin{equation}
	f(z):=z\left(z^2+\beta^2\right)^{1/2}.
	\label{eq:fz}
\end{equation}
Here $f(z)$ is a multivalued function. Depending on whether $q$ is a half-odd integer or an integer, Eq.~\eqref{eq:E0f} becomes a sum over integers or over half-odd integers, respectively. Since the corresponding forms of the APF differ in these two cases, as shown in Eq.~\eqref{eq:APF}, we treat them separately below.

\subsection{Half-odd-integer monopole charge}

We begin with the case in which $q$ is a half-odd integer, so that $\delta$ is a positive integer. The case of integer $q$ will be discussed subsequently. For $\beta^2>0$, the branch points of the square-root factor in $\omega_{ql}$ lie on the imaginary axis, and the APF in Eq.~\eqref{eq:APF} applies. The sum on the right-hand side of Eq.~\eqref{eq:E0f} then becomes~\cite{HerdeiroRibeiroSampaio2008}
\begin{equation}
	\begin{aligned}
		\sum_{n=0}^{\infty} f(n+\delta)
		={}&
		\int_0^\infty \rd t\,f(t)
		-\int_0^{\delta} \rd t\,f(t)
		+\frac{f(\delta)}{2}
		+I_{\mathbb{Z}},
	\end{aligned}
	\label{eq:IdivIfin}
\end{equation}
where
\begin{equation}
	I_{\mathbb{Z}}
	:=
	\ri\int_{0^+}^{\infty}
	\rd t\,
	\frac{f(\ri t+\delta)-f(-\ri t+\delta)}{\e^{2\pi t}-1}.
	\label{eq:IZ}
\end{equation}

For $\beta^2<0$, the function develops a branch cut on the real axis, and we use the GAPF-MV derived in Appendix~\ref{app:gapf}. Since no other singularity of $I_{\mathbb{Z}}$ lies on the branch cut, only one additional integral appears:
\begin{equation}
	\begin{aligned}
		\sum_{n=0}^{\infty} f(n+\delta)
		={}&
		\int_{|\beta|}^\infty \rd t\,f(t)
		-\int_{|\beta|}^{\delta} \rd t\,f(t)
		+\frac{f(\delta)}{2}
		\\
		&
		+\frac{1}{2}\int_{-|\beta|}^{|\beta|}\rd t\,
		\left[f^+(t)+f^-(t)\right]
		+I_{\mathbb{Z}},
	\end{aligned}
	\label{eq:half_q_neg_beta}
\end{equation}
where $f^\pm$ denote the boundary values above and below the cut,
\begin{equation}
	f^+(x):=\lim_{\varepsilon\downarrow 0}f(x+\ri\varepsilon),
	\qquad
	f^-(x):=\lim_{\varepsilon\downarrow 0}f(x-\ri\varepsilon).
\end{equation}
In fact, the integrand in the additional cut contribution cancels identically because of the square-root discontinuity.

Denote the first integral in Eq.~\eqref{eq:IdivIfin} by $I_{\mathrm{div}}$. Since it is divergent, it must be regularized. A standard procedure is to introduce an analytic regulator in a convergent domain and then continue the result analytically to the physical value. We therefore define
\begin{equation}
	I_{\mathrm{div}}(s)
	:=
	\int_0^\infty \rd t\, t\left(t^2+\beta^2\right)^{\frac{1}{2}-s}.
\end{equation}
For $\beta^2>0$, this integral converges for $\Re s>3/2$, giving
\begin{equation}
	\begin{aligned}
		I_{\mathrm{div}}(s)
		={}&
		\int_0^\infty \rd t\, t\left(t^2+\beta^2\right)^{\frac{1}{2}-s}
		\\
		={}&
		-\frac{1}{3-2s}\,|\beta|^{3-2s},
	\end{aligned}
	\label{eq:Idiv}
\end{equation}
in the convergent domain. Since $\beta$ is defined only through $\beta^2$ in Eq.~\eqref{eq:beta2}, we take $\beta>0$ for $\beta^2>0$ without loss of generality. The regularized value is then obtained by analytic continuation to $s=0$:
\begin{equation}
	I_{\mathrm{div}}^{\mathrm{Reg}}
	=
	I_{\mathrm{div}}(0)
	=
	-\third \beta^3,
	\qquad
	\beta^2>0.
	\label{eq:IdivReg}
\end{equation}

For $\beta^2\leq 0$, the corresponding divergent integral is
\begin{equation}
	\begin{aligned}
		I_{\mathrm{div}}(s)
		={}&
		\int_{|\beta|}^\infty \rd t\, t\left(t^2-|\beta|^2\right)^{\frac{1}{2}-s}
		\\
		={}&
		\frac{1}{2}\int_0^\infty \rd u\,u^{\frac{1}{2}-s}, \qquad u:=t^2-|\beta|^2.
	\end{aligned}
\end{equation}
The resulting integral is scaleless. Within the analytic-continuation prescription used here, its regularized value vanishes. At $\beta=0$, this agrees with the $\beta\to0^+$ limit of Eq.~\eqref{eq:IdivReg}. Thus,
\begin{equation}
	I_{\mathrm{div}}^{\mathrm{Reg}}=0,
	\qquad
	\beta^2\leq 0.
\end{equation}

There are two ways to evaluate $I_{\mathbb{Z}}$. One is by direct integration, since $f(z)$ is sufficiently simple. The other is by contour integration, which is applicable more generally. We present both derivations in Appendices~\ref{app:IZ} and \ref{app:directIZ}. They yield the same result,
\begin{equation}
	\begin{aligned}
		I_{\mathbb{Z}}
		={}&
		-\frac{f(\delta)}{2}
		+\int_{t_{\min}}^{\delta}\rd t\,f(t)
		-\sum_{i=1}^{\mathcal{N}} f(\delta-i)
		\\
		&
		+
		\begin{cases}\dsp
			-2\beta^3 \int_0^1 \rd\eta\,
			\frac{\eta\sqrt{1-\eta^2}}{\e^{2\pi \beta\eta}-1},
			& \beta^2>0,
			\\[10pt]
			0,
			& \beta=0,
			\\[10pt]
			\dsp
			|\beta|^3 \fint_0^1 \rd\eta\,
			\eta\sqrt{1-\eta^2}\cot(\pi|\beta|\eta),
			& \beta^2<0,
		\end{cases}
	\end{aligned}
\end{equation}
where $\fint$ denotes Cauchy principal value of the integral, $t_{\min}=0$ for $\beta^2\geq 0$ and $t_{\min}=|\beta|$ for $\beta^2<0$, and $\mathcal{N}$ is the largest integer less than $\delta-t_{\min}$,
\begin{equation}
	\mathcal{N}:=
	\begin{cases}
		\delta-t_{\min}-1, & \delta-t_{\min}\in\mathbb{Z},\\
		\lfloor \delta-t_{\min}\rfloor, & \delta-t_{\min}\notin\mathbb{Z},
	\end{cases}
	\label{eq:N}
\end{equation}
with $\lfloor\cdot\rfloor$ the floor function. A sum over $i=1,\ldots,\mathcal{N}$ is understood to vanish when $\mathcal{N}\leq0$.

Putting everything together, we obtain the regularized energy density
\begin{equation}
	\begin{aligned}
		\mathcal{E}_{\mathrm{reg}}^{(1)}(a_0,m_0)
		={}&
		-\frac{\beta^3}{6\pi a_0^3}
		-\frac{\beta^3}{\pi a_0^3}
		\int_0^1 \rd\eta\,
		\frac{\eta\sqrt{1-\eta^2}}{\e^{2\pi \beta\eta}-1}
		\\
		&
		-\frac{1}{2\pi a_0^3}
		\sum_{i=1}^{|q|-\half} f\left(|q|+\half-i\right),
	\end{aligned}
	\label{eq:VarEhalfplus}
\end{equation}
for $\beta^2>0$,
\begin{equation}
	\mathcal{E}_{\mathrm{reg}}^{(\beta=0)}(a_0,m_0)
	=
	-\frac{1}{2\pi a_0^3}\sum_{i=1}^{|q|-\half} f\left(|q|+\half-i\right),
	\label{eq:VarEhalfzero}
\end{equation}
for $\beta=0$, and
\begin{equation}
	\begin{aligned}
		\mathcal{E}_{\mathrm{reg}}^{(2)}(a_0,m_0)
		={}&
		\frac{|\beta|^3}{2\pi a_0^3}
		\fint_0^1 \rd\eta\,
		\eta\sqrt{1-\eta^2}\,\cot(\pi |\beta|\eta)
		\\
		&
		-\frac{1}{2\pi a_0^3}
		\sum_{i=1}^{\mathcal{N}} f\left(|q|+\half-i\right),
	\end{aligned}
	\label{eq:VarEhalfminus}
\end{equation}
for $\beta^2<0$.

\subsection{Integer monopole charge}

We next consider the case in which $q$ is an integer. The sum in Eq.~\eqref{eq:E0} may then be rewritten as
\begin{equation}
	\sum_{n=0}^{\infty} f(n+\delta)
	=
	\sum_{k=|q|}^{\infty} f\left(k+\half\right).
\end{equation}
As in the half-odd-integer case, the analysis depends on the sign of $\beta^2$. For integer $q$, three regimes must be distinguished:
\begin{equation}
	\beta^2\geq 0,
	\qquad
	-q^2\leq \beta^2<0,
	\qquad
	-\delta^2 < \beta^2<-q^2.
\end{equation}
For $\beta^2<0$, we use the GAPF-MV derived in Appendix~\ref{app:gapf}.

For $\beta^2\geq 0$, there is no branch cut on the real axis, so the APF in Eq.~\eqref{eq:APF} applies. One then finds
\begin{equation}
	\sum_{k=|q|}^{\infty} f\left(k+\half\right)
	=
	\int_0^\infty \rd t\,f(t)
	-\int_0^{|q|}\rd t\,f(t)
	+I_{\mathbb{Z}+\half},
\end{equation}
where
\begin{equation}
	I_{\mathbb{Z}+\half}
	:=
	-\ri\int_0^\infty \rd t\,
	\frac{f(\ri t+|q|)-f(-\ri t+|q|)}{\e^{2\pi t}+1}.
	\label{eq:IZhalf}
\end{equation}

For $-q^2\leq \beta^2<0$, a branch cut appears on the real axis, and we use the GAPF-MV in Eq.~\eqref{eq:ap-mv-half}. As in the half-odd-integer case, however, the additional integral along the cut vanishes. The result is therefore
\begin{equation}
	\sum_{k=|q|}^{\infty} f\left(k+\half\right)
	=
	\int_{|\beta|}^\infty \rd t\,f(t)
	-\int_{|\beta|}^{|q|}\rd t\,f(t)
	+I_{\mathbb{Z}+\half}.
\end{equation}

For the narrow range
$
	-\delta^2 < \beta^2<-q^2,
$
there is not only a branch cut on the real axis, but also the possibility that an additional singularity of the integrand in $I_{\mathbb{Z}+\half}$ lies on the cut. In this regime, the branch cut has crossed $x=|q|$. Application of the GAPF-MV yields
\begin{equation}
	\begin{aligned}
		&\sum_{k=|q|+1}^{\infty} f\left(k+\half\right)
		+\frac12\left[f^+\left(|q|+\half\right)+f^-\left(|q|+\half\right)\right]
		\\
		={}&
		\int_{|\beta|}^\infty \rd t\,f(t)
		+I_{\mathbb{Z}+\half} - \fint_{|q|}^{|\beta|}\rd t\,
		t \sqrt{|\beta|^2-t^2}\tan(\pi t).
	\end{aligned}
\end{equation}
In contrast to the preceding regime, the branch-cut contribution is nonvanishing and gives the last term in the above equation.

The divergent integral is the same as in the half-odd-integer monopole case and hence gives the same regularized value,
\begin{equation}
	I_{\mathrm{div}}^{\mathrm{Reg}}
	=
	\begin{cases}
		-\third\beta^3, & \beta^2\geq 0,\\
		0, & \beta^2<0.
	\end{cases}
\end{equation}

By contrast, the finite contribution $I_{\mathbb{Z}+\half}$ depends on the range of $\beta^2$. For $\beta^2\geq-q^2$, one finds
\begin{equation}
	\begin{aligned}
		I_{\mathbb{Z}+\half}
		={}&
		\int_{t_{\min}}^{|q|}\rd t\,f(t)
		-\sum_{i=1}^{\mathcal{N}} f(\delta-i)
		\\
		&
		+
		\begin{cases}\dsp
			2\beta^3 \int_0^1 \rd\eta\,
			\frac{\eta\sqrt{1-\eta^2}}{\e^{2\pi \beta\eta}+1},
			& \beta^2\geq 0,
			\\[10pt]
			\dsp
			-|\beta|^3 \fint_0^1 \rd\eta\,
			\eta\sqrt{1-\eta^2}\tan(\pi|\beta|\eta),
			& \beta^2<0.
		\end{cases}
	\end{aligned}
\end{equation}
Here $t_{\min}$ and $\mathcal{N}$ are defined in the same way as in the previous case. For the final regime, $-\delta^2 < \beta^2<-q^2$, one obtains
\begin{equation}
	I_{\mathbb{Z}+\half}
	=
	-|\beta|^3 \fint_0^{|q/\beta|}\rd\eta\,
	\eta\sqrt{1-\eta^2}\tan(\pi|\beta|\eta).
\end{equation}
The details of the two derivations of $I_{\mathbb{Z}+\half}$ are given in Appendix~\ref{app:IZhalf} and \ref{app:directIZ}.

The corresponding regularized energy densities are as follows. For $\beta^2>0$,
\begin{equation}
	\begin{aligned}
		\mathcal{E}_{\mathrm{reg}}^{(3)}(a_0,m_0)
		={}&
		-\frac{\beta^3}{6\pi a_0^3}
		+\frac{\beta^3}{\pi a_0^3}
		\int_0^1 \rd\eta\,
		\frac{\eta\sqrt{1-\eta^2}}{\e^{2\pi \beta\eta}+1}
		\\
		&
		-\frac{1}{2\pi a_0^3}
		\sum_{i=1}^{|q|} f\left(|q|-i+\half\right).
	\end{aligned}
	\label{eq:VarE3}
\end{equation}
Taking the limit $\beta\to 0$, this reduces to
\begin{equation}
	\mathcal{E}_{\mathrm{reg}}^{(\beta=0)}(a_0,m_0)
	=
	-\frac{1}{2\pi a_0^3}
	\sum_{i=1}^{|q|} f\left(|q|-i+\half\right).
\end{equation}
For $-q^2\leq\beta^2<0$,
\begin{equation}
	\begin{aligned}
		\mathcal{E}_{\mathrm{reg}}^{(4)}(a_0,m_0)
		={}&
		-\frac{|\beta|^3}{2\pi a_0^3}
		\fint_0^1 \rd\eta\,
		\eta\sqrt{1-\eta^2}\,
		\tan(\pi |\beta|\eta)
		\\
		&
		-\frac{1}{2\pi a_0^3}
		\sum_{i=1}^{\mathcal{N}} f\left(|q|-i+\half\right).
	\end{aligned}
		\label{eq:VarE4}
\end{equation}
Finally, for $-\delta^2 < \beta^2<-q^2$,
\begin{equation}
	\mathcal{E}_{\mathrm{reg}}^{(5)}(a_0,m_0)
	=
	-\frac{|\beta|^3}{2\pi a_0^3}
	\fint_0^{1}\rd\eta\,
	\eta\sqrt{1-\eta^2}\,
	\tan(\pi |\beta|\eta).
		\label{eq:VarE5}
\end{equation}

\section{Renormalization}
\label{sec:renormalization}

In this section we renormalize the vacuum energy density and derive the corresponding Casimir energy and pressure. We then analyze the dependence on the sphere radius $a_0$, the curvature-coupling parameter $\xi$, and the monopole charge $q$ in the next section.

The Casimir energy is defined relative to the flat-space reference vacuum. In the present problem, the flat-space limit is obtained by taking $a_0\to\infty$. For $\beta^2>0$, this limit is straightforward, and we renormalize the vacuum energy by subtracting its large-$a_0$ contribution. Holding the remaining parameters fixed, we have
\begin{equation}
	\beta^2=a_0^2m_0^2-q^2+2\xi-\quarter \simeq a_0^2m_0^2 \gg 1
\end{equation}
in the large-radius limit.

Accordingly, the dominant contribution to Eq.~\eqref{eq:VarEhalfplus} in the large-$a_0$ limit is
\begin{equation}
	\mathcal{E}_{\mathrm{reg}}^{(1)}(a_0,m_0) 
	=-\frac{m_0^3}{6\pi}+O(a_0^{-2}).
\end{equation}
Thus the large-radius limit of the energy density is
\begin{equation}
	\mathcal{E}_{\mathrm{free}}(m_0):=-\frac{m_0^3}{6\pi}.
	\label{eq:VarEfree}
\end{equation}
As in Ref.~\cite{HerdeiroRibeiroSampaio2008}, we subtract this term to impose the flat-space vacuum normalization,
\begin{equation}
	\mathcal{E}_{\mathrm{Cas}}^{(1)}(a_0,m_0)
	:=
	\mathcal{E}_{\mathrm{reg}}^{(1)}(a_0,m_0)-\mathcal{E}_{\mathrm{free}}(m_0).
\end{equation}

The corresponding total energies are $E_{\mathrm{reg}}=4\pi a_0^2\mathcal{E}_{\mathrm{reg}}$ and $E_{\mathrm{Cas}}=4\pi a_0^2\mathcal{E}_{\mathrm{Cas}}$; their dependence on $q$ and $\xi$ is usually suppressed. This normalization is understood together with the analytic continuation adopted in Sec.~\ref{sec:regularization}. To relate the prescription to covariant renormalization~\cite{Vassilevich2003,DecaniniFolacci2008}, define
\begin{equation}
	\begin{aligned}
		H&:=-D_{S^2}^{\,2}+m_0^2+\xi R,\\
		\zeta_H(s)&:=\sum_{l=|q|}^{\infty}(2l+1)(\omega_{ql}^2)^{-s},
	\end{aligned}
\end{equation}
where $D_{S^2}^{\,2}$ is the gauge-covariant Laplacian and the sum is analytically continued from its domain of convergence. For the complex scalar, $E_{\mathrm{reg}}=\zeta_H(-1/2)$. This is also the energy obtained from the covariant Euclidean one-loop determinant: for $K=-\partial_\tau^2+H$ on a Euclidean time interval of length $T$ with periodic boundary conditions, the spectral zeta function per unit time obeys
\begin{equation}
	\lim_{T\to\infty}\frac{\zeta_K(s)}{T}
	=\frac{\Gamma(s-1/2)}{2\sqrt{\pi}\,\Gamma(s)} \zeta_H(s-1/2).
\end{equation}
Since $\zeta_H$ is regular at $-1/2$, the complex-scalar determinant gives $-\lim_{T\to\infty}\zeta_K'(0)/T=\zeta_H(-1/2)$. Our complete prescription is thus
\begin{equation}
	E_{\mathrm{Cas}}=\zeta_H(-1/2)+\frac23 m_0^3a_0^2,
\label{eq:zeta-normalization}
\end{equation}
with no additional finite local additions or subtractions.

This convention retains the finite curvature- and magnetic-field-dependent contributions generated by the scalar. These can be identified through the heat-kernel expansion.\footnote{We thank the anonymous referee for suggesting this heat-kernel analysis and providing the large-radius energy and pressure expansions used here.} Factoring out the mass term, the heat trace of the stated spectrum has the small-$t$ expansion
\begin{equation}
	\begin{aligned}
		K_0(t)&:=\tr\e^{-t(H-m_0^2)}\\
		&\sim\frac{a_0^2}{t}+\left(\frac13-2\xi\right)
		+\frac{t}{a_0^2}C(q,\xi)+O(t^2/a_0^4).
	\end{aligned}
\label{eq:heat-trace}
\end{equation}
Inserting this expansion, with $C(q,\xi)$ given in Eq.~\eqref{eq:C-asymptotic}, into the proper-time representation
\begin{equation}
	\zeta_H(s)=\frac{1}{\Gamma(s)}
	\int_0^\infty\rd t\,t^{s-1}\e^{-m_0^2t}K_0(t),
\end{equation}
initially defined for $\Re s>1$, analytically continuing to $s=-1/2$, and removing the flat contribution $-2a_0^2m_0^3/3$ gives, at fixed $m_0>0$, $q$, and $\xi$, 
\begin{equation}
	E_{\mathrm{Cas}}
	=m_0\left(\frac13-2\xi\right) - \frac{C(q,\xi)}{2m_0a_0^2}+O(a_0^{-4}),
\label{eq:energy-asymptotic}
\end{equation}
where
\begin{equation}
	C(q,\xi):=\frac1{15}-\frac{2\xi}{3}+2\xi^2-\frac{q^2}{6}.
\label{eq:C-asymptotic}
\end{equation}
The first term arises from the finite local contribution proportional to $m_0R$; it is independent of the radius and gives no radius pressure. The next term contains the finite induced curvature-squared and magnetic-field contributions, proportional locally to curvature-squared invariants divided by $m_0$ and to $(Ze)^2F_{\mu\nu}F^{\mu\nu}/m_0$, respectively, where $F_{\mu\nu}=\partial_\mu A_\nu-\partial_\nu A_\mu$. These are finite loop contributions, not additional ultraviolet subtractions required in three spacetime dimensions. They are retained even though their energy densities vanish as $a_0\to\infty$. Thus the flat-space limit alone does not fix the finite curvature normalization; the analytic continuation and the absence of further finite local terms complete our convention. The reported stress is the scalar vacuum contribution in this convention. Applications with dynamical gravity or gauge fields additionally require the renormalized local terms in the background action. The expansion~\eqref{eq:energy-asymptotic} assumes $m_0>0$.

For Case~(1), namely half-odd-integer $q$ with $\beta^2>0$, the resulting Casimir energy density is
\begin{equation}
	\begin{aligned}
		\mathcal{E}_{\mathrm{Cas}}^{(1)}(a_0,m_0)
		={}&
		-\frac{\beta^3-a_0^3m_0^3}{6\pi a_0^3}
		-\frac{\beta^3}{\pi a_0^3}
		\int_{0^+}^1 \rd\eta\,
		\frac{\eta\sqrt{1-\eta^2}}{\e^{2\pi \eta\beta}-1}
		\\
		&
		-\frac{1}{2\pi a_0^3}\sum_{i=1}^{\mathcal{N}} f(\delta-i).
	\end{aligned}
\end{equation}

In terms of the sphere area, the renormalized Casimir energy is
\begin{equation}
	E_{\mathrm{Cas}}(a_0,m_0):=S\,\mathcal{E}_{\mathrm{Cas}}(a_0,m_0),
\end{equation}
with
\begin{equation}
	S=4\pi a_0^2.
\end{equation}
For this case, we obtain
\begin{equation}
	\begin{aligned}
		E_{\mathrm{Cas}}^{(1)}
		={}&
		-\frac{2(\beta^3-a_0^3m_0^3)}{3a_0}
		-\frac{2}{a_0}\sum_{i=1}^{\mathcal{N}}f(\delta-i)
		\\
		&
		-\frac{4\beta^3}{a_0}\int_{0^+}^1 \rd\eta\,
		\frac{\eta\sqrt{1-\eta^2}}{\e^{2\pi \eta\beta}-1}.
	\end{aligned}
\end{equation}

The corresponding renormalized pressure is defined by
\begin{equation}
	P_{\mathrm{Cas}}(a_0,m_0):=-\frac{\partial E_{\mathrm{Cas}}(a_0,m_0)}{\partial S},
\end{equation}
where the partial derivative is taken at fixed $q$, $m_0$, and $\xi$. At fixed scalar charge $Ze$, fixing $q=Ze\mathrm{g}$ fixes the monopole flux, whereas the physical magnetic-field strength is $B=\mathrm{g}/a_0^2$ and varies with the radius. For the present case, it reads
\begin{equation}
	\begin{aligned}
		P_{\mathrm{Cas}}^{(1)}
		=&
		-\frac{1}{12\pi a_0^3}
		\left( \beta^3
		- 3 a_0^2m_0^2\beta
		+2 a_0^3m_0^3\right)
		\\
		&
		-\frac{\beta}{2\pi a_0^3}\left(
		\beta^2 - 3a_0^2m_0^2
		\right)
		\int_{0^+}^1 \rd\eta\,
		\frac{\eta\sqrt{1-\eta^2}}{\e^{2\pi \eta\beta}-1}
		\\
		&
		-\frac{m_0^2\beta^2}{a_0}
		\int_{0^+}^1 \rd\eta\,
		\frac{\e^{2\pi \eta\beta}\eta^2\sqrt{1-\eta^2}}
		{(\e^{2\pi \eta\beta}-1)^2}
		\\
		&
		-\frac{1}{4\pi a_0^3} \sum_{i=1}^{\mathcal{N}} \left[f(\delta-i)
		-a_0^2m_0^2 \frac{(\delta-i)^2}{f(\delta-i)}\right],
	\end{aligned}
\end{equation}
where we have used
\begin{equation}
	\frac{\partial \beta}{\partial a_0}
	=
	\frac{a_0m_0^2}{\beta},
	\qquad
	\frac{\partial f(\delta-i)}{\partial a_0}
	=
	a_0m_0^2\frac{(\delta-i)^2}{f(\delta-i)}.
\end{equation}

The large-radius limit is more subtle for $\beta^2<0$, since taking $a_0\to\infty$ with all other parameters fixed inevitably drives $\beta^2$ positive and hence takes the system outside the regime under consideration. We therefore use the same flat-space subtraction, as prescribed by Eq.~\eqref{eq:zeta-normalization}. This prescription is also consistent with the continuity of the regularized energy density at $\beta=0$; see Eqs.~\eqref{eq:VarEhalfplus}, \eqref{eq:VarEhalfzero}, and \eqref{eq:VarEhalfminus}. Thus, for $\beta^2<0$, the flat-space vacuum energy density is again given by Eq.~\eqref{eq:VarEfree}.

With this understanding, the renormalized energy density for Case~(2), namely half-odd-integer $q$ with
$
	-\left(|q|+\half\right)^2 < \beta^2<0,
$
follows from Eq.~\eqref{eq:VarEhalfminus}. The corresponding Casimir energy and pressure are
\begin{equation}
	\begin{aligned}
		E_{\mathrm{Cas}}^{(2)}
		={}&\frac{2}{3}a_0^2m_0^3 + \frac{2|\beta|^3}{a_0}
		\fint_0^1 \rd\eta\,
		\eta\sqrt{1-\eta^2}\cot(\pi |\beta|\eta)\\
		&
		-\frac{2}{a_0}\sum_{i=1}^{\mathcal{N}}f(\delta-i),
	\end{aligned}
\end{equation}
and
\begin{equation}
	\begin{aligned}
		P_{\mathrm{Cas}}^{(2)}
		={}&-\frac{m_0^3}{6\pi}
		+\frac{|\beta|^3}{4\pi a_0^3}
		\fint_0^1 \rd\eta\,
		\eta\sqrt{1-\eta^2}\cot(\pi |\beta|\eta)
		\\
		&
		+\frac{m_0^2|\beta|}{4\pi a_0} 
		\fint_0^1 \rd\eta\,
		\frac{\eta\cot(\pi |\beta|\eta)}{\sqrt{1-\eta^2}}\\
		&
		-\frac{1}{4\pi a_0^3}\sum_{i=1}^{\mathcal{N}}\left[f(\delta-i)
		- a_0^2 m_0^2\frac{(\delta-i)^2}{f(\delta-i)}\right],
	\end{aligned}
\end{equation}
where we have used
\begin{equation}
	\frac{\partial |\beta|}{\partial a_0}
	=
	-\frac{a_0m_0^2}{|\beta|},
\end{equation}
as well as
\begin{equation}
	\begin{aligned}
		& \frac{\partial}{\partial a_0}
		\left[
		|\beta|^3\fint_0^1 \rd\eta\,
		\eta\sqrt{1-\eta^2}\cot(\pi |\beta|\eta)
		\right]\\
		={}&
		\frac{\partial}{\partial a_0}
		\left[
		\fint_0^{|\beta|}\rd t\,t\sqrt{|\beta|^2-t^2}\cot(\pi t)
		\right]
		\\
		={}&
		-a_0m_0^2|\beta|
		\fint_0^1 \rd\eta\,
		\frac{\eta\cot(\pi |\beta|\eta)}{\sqrt{1-\eta^2}}.
	\end{aligned}
\end{equation}

The corresponding results for integer monopole charge are similar. For completeness, we list the corresponding Casimir energy and pressure below.

For $\beta^2\geq0$,
\begin{equation}
	\begin{aligned}
		E_{\mathrm{Cas}}^{(3)}
		={}&-\frac{2(\beta^3-a_0^3m_0^3)}{3a_0} 
		+ \frac{4\beta^3}{a_0}\int_0^1 \rd\eta\,
		\frac{\eta\sqrt{1-\eta^2}}{\e^{2\pi \eta\beta}+1} \\
		&
		-\frac{2}{a_0}\sum_{i=1}^{|q|}f(\delta-i),
	\end{aligned}
\end{equation}
and
\begin{equation}
	\begin{aligned}
		P_{\mathrm{Cas}}^{(3)}
		={}&
		-\frac{1}{12\pi a_0^3} \left(\beta^3 
		- 3 a_0^2 m_0^2 \beta + 2a_0^3 m_0^3\right)
		\\
		&
		+ \frac{\beta}{2\pi a_0^3} \left(
		\beta^2 - 3 a_0^2m_0^2
		\right)
		\int_0^1 \rd\eta\,
		\frac{\eta\sqrt{1-\eta^2}}{\e^{2\pi \eta\beta}+1}
		\\
		&
		+ \frac{m_0^2\beta^2}{a_0}
		\int_0^1 \rd\eta\,
		\frac{\e^{2\pi \eta\beta}\eta^2\sqrt{1-\eta^2}}
		{(\e^{2\pi \eta\beta}+1)^2}\\
		&
		-\frac{1}{4\pi a_0^3}\sum_{i=1}^{|q|}\left[f(\delta-i)
		- a_0^2m_0^2\frac{(\delta-i)^2}{f(\delta-i)}\right].
	\end{aligned}
\end{equation}

For $-q^2\leq \beta^2<0$, we obtain
\begin{equation}
	\begin{aligned}
		E_{\mathrm{Cas}}^{(4)}
		={}& \frac{2}{3}a_0^2m_0^3
		-\frac{2|\beta|^3}{a_0}
		\fint_0^1 \rd\eta\,
		\eta\sqrt{1-\eta^2}\tan(\pi |\beta|\eta)\\
		& 
		-\frac{2}{a_0}\sum_{i=1}^{\mathcal{N}}f(\delta-i),
	\end{aligned}
\end{equation}
and
\begin{equation}
	\begin{aligned}
		P_{\mathrm{Cas}}^{(4)}
		={}& -\frac{m_0^3}{6\pi}
		-\frac{|\beta|^3}{4\pi a_0^3}
		\fint_0^1 \rd\eta\,
		\eta\sqrt{1-\eta^2}\tan(\pi |\beta|\eta)
		\\
		&
		-\frac{m_0^2 |\beta|}{4\pi a_0} 
		\fint_0^1 \rd\eta\,
		\frac{\eta\tan(\pi |\beta|\eta)}{\sqrt{1-\eta^2}}\\
		&
		-\frac{1}{4\pi a_0^3}\sum_{i=1}^{\mathcal{N}}\left[f(\delta-i) -a_0^2m_0^2\frac{(\delta-i)^2}{f(\delta-i)} \right].
	\end{aligned}
\end{equation}

Finally, for $-\delta^2 < \beta^2<-|q|^2$, we have
\begin{align}
	E_{\mathrm{Cas}}^{(5)}
	={}& \frac{2}{3}a_0^2m_0^3
	-\frac{2|\beta|^3}{a_0}
	\fint_0^{1}\rd\eta\,
	\eta\sqrt{1-\eta^2}\tan(\pi |\beta|\eta),
\end{align}
and
\begin{equation}
	\begin{aligned}
		P_{\mathrm{Cas}}^{(5)}
		={}&-\frac{m_0^3}{6\pi}
		-
		\frac{|\beta|^3}{4\pi a_0^3}
		\fint_0^{1}\rd\eta\,
		\eta\sqrt{1-\eta^2}\tan(\pi |\beta|\eta)
		\\
		&
		-\frac{m_0^2|\beta|}{4\pi a_0}
		\fint_0^{1}\rd\eta\,
		\frac{\eta\tan(\pi |\beta|\eta)}{\sqrt{1-\eta^2}}.
	\end{aligned}	
\end{equation}

When a branch endpoint coincides with a kernel pole or the number of terms in the finite sum changes, the pressure is defined by the continuous limit of the complete expression, combining the integral and finite-sum contributions before taking the limit. These apparent singularities do not correspond to physical singularities within the strictly positive-frequency domain. Appendix~\ref{app:numerical-evaluation} provides a convergent spectral representation and its numerical implementation.

As an independent consistency check, we now consider the massless, conformally coupled limit, $m_0=0$ and $\xi=1/8$, for which $\beta^2=-q^2$. The common limit with the free-scalar calculation in Appendix A of Ref.~\cite{PufuSachdev2013} is
\begin{equation}
	a_0E_{\mathrm{Cas}}(a_0,0,\xi=1/8,q)
	=\frac{\Delta_Q}{N_b},\qquad Q=2|q|,
\label{eq:conformal-comparison}
\end{equation}
where $\Delta_Q$ is the monopole scaling dimension and $N_b$ is the number of complex scalar fields ($N_b=1$ here). Equivalently, this is $\mathcal F_Q^\infty$ evaluated with the free-scalar parameter $a_Q^2=-Q^2/4$ in that reference, rather than at its interacting saddle point. The flat-space subtraction vanishes in this limit. Numerical evaluation of the analytically continued spectral sum reproduces the free-scalar results of Ref.~\cite{PufuSachdev2013} to the precision quoted there.

Having obtained the analytic expressions for the Casimir pressure, we now turn to its dependence on $a_0$, $\xi$, and $q$.

\section{Parameter Dependence}
\label{sec:plots}

In this section we analyze the Casimir pressure as a function of $a_0$, $q$, and $\xi$. The large-radius expansion~\eqref{eq:energy-asymptotic} provides an analytic criterion for the asymptotic sign of the pressure. Differentiating at fixed $m_0>0$, $q$, and $\xi$, with $S=4\pi a_0^2$, gives
\begin{equation}
\begin{aligned}
	P_{\mathrm{Cas}}
	&=-\frac{1}{8\pi a_0}\frac{\partial E_{\mathrm{Cas}}}{\partial a_0}=-\frac{C(q,\xi)}{8\pi m_0a_0^4}+O(a_0^{-6}).
\end{aligned}
\label{eq:pressure-asymptotic}
\end{equation}
Thus $C>0$ gives an attractive large-radius tail approaching zero from below, whereas $C<0$ gives a repulsive tail approaching zero from above. At $C=0$, the next asymptotic term determines the sign.

To organize the coupling dependence, we use the smallest nonzero monopole charge, $|q|=1/2$. For this sector,
\begin{equation}
	C(1/2,\xi)=2\xi^2-\frac23\xi+\frac1{40} = 2(\xi-\xi_1)(\xi-\xi_2),
\end{equation}
whose two zeros are
\begin{equation}
	\begin{aligned}
		\xi_1&=\frac16-\frac{\sqrt{55}}{60}\simeq0.0430634,\\
		\xi_2&=\frac16+\frac{\sqrt{55}}{60}\simeq0.290270.
	\end{aligned}
\label{eq:xi-boundaries}
\end{equation}
We call $\xi<\xi_1$, $\xi_1<\xi<\xi_2$, and $\xi>\xi_2$ the low-, intermediate-, and high-$\xi$ regimes, respectively. These boundaries classify the large-radius pressure in the $|q|=1/2$ sector; they do not assert a uniform pressure sign or monotonicity over all radii. Since
\begin{equation}
	C(q,\xi)=C(1/2,\xi)-\frac{q^2-1/4}{6},
\end{equation}
all nonzero allowed charges have repulsive large-radius tails in the intermediate regime. Outside that regime, sufficiently large $|q|$ can still make the tail repulsive. The zero-flux sector always has an attractive large-radius tail, since $C(0,\xi)=2(\xi-1/6)^2+1/90>0$.

\begin{figure}[htbp]
\centering
\includegraphics[width=0.45\textwidth]{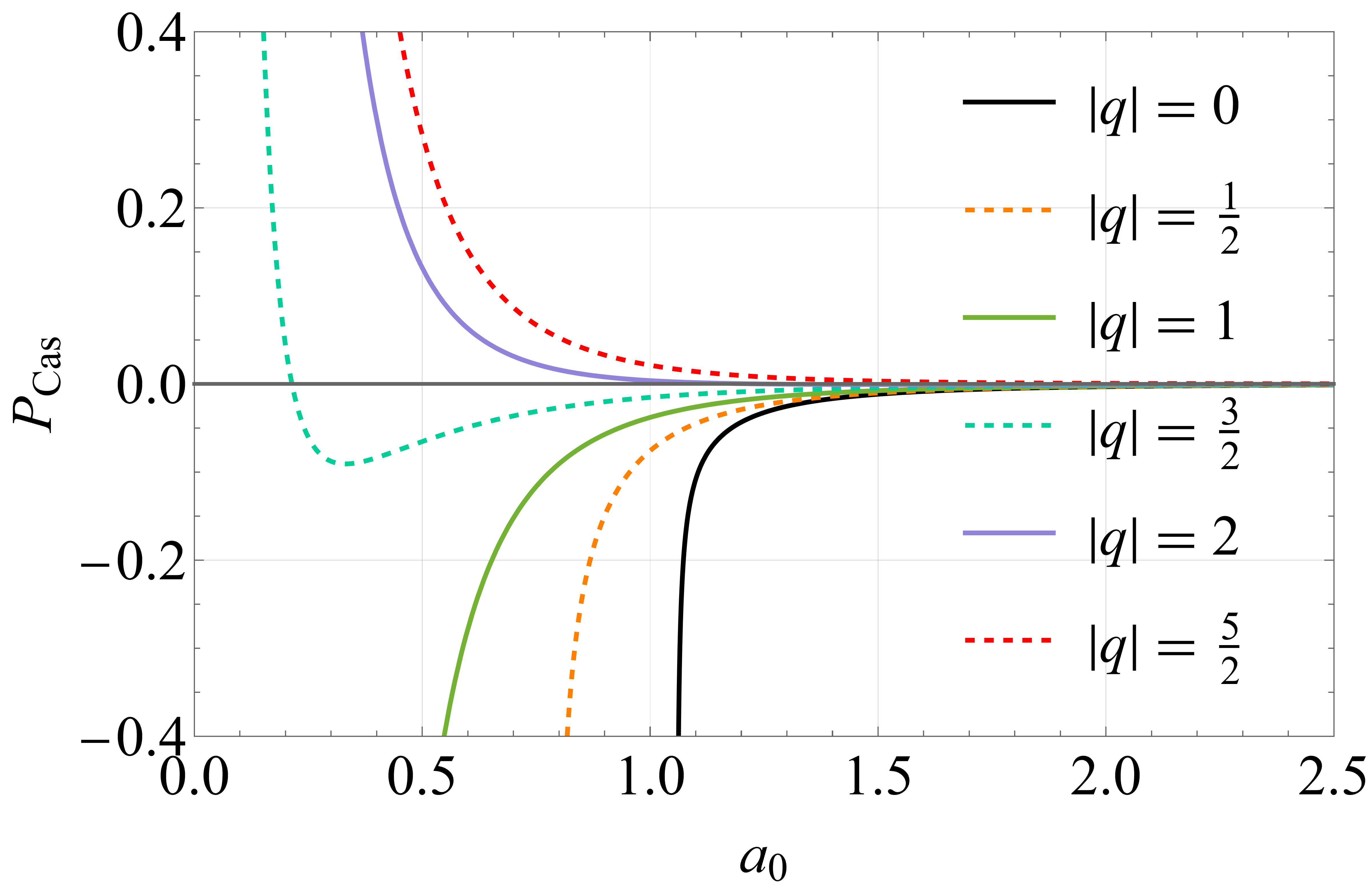}
	\caption{Casimir pressure $P_{\rm Cas}$ as a function of the sphere radius $a_0$ for $|q|=0,\frac12,1,\frac32,2,\frac52$, with $m_0=1$ and $\xi=-0.56$. Positive (negative) pressure corresponds to a repulsive (attractive) Casimir force.}
\label{fig:Pcas_xi}
\end{figure}

\begin{figure}[htbp]
\centering
\includegraphics[width=0.45\textwidth]{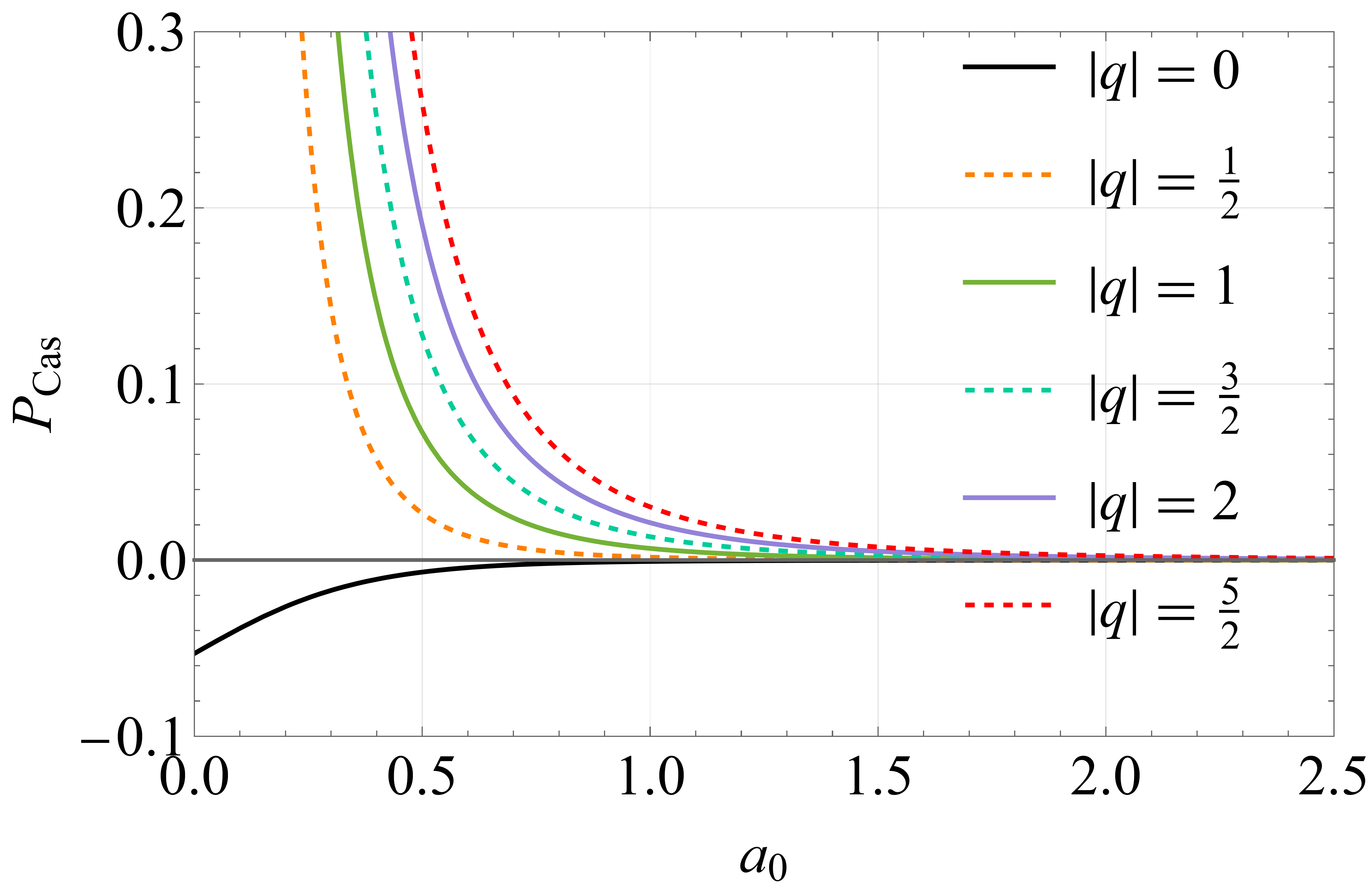}
	\caption{Same as Fig.~\ref{fig:Pcas_xi}, but for conformal coupling, $\xi=1/8$.}
\label{fig:Pcasxi0.125}
\end{figure}

\begin{figure}[htbp]
\centering
\includegraphics[width=0.45\textwidth]{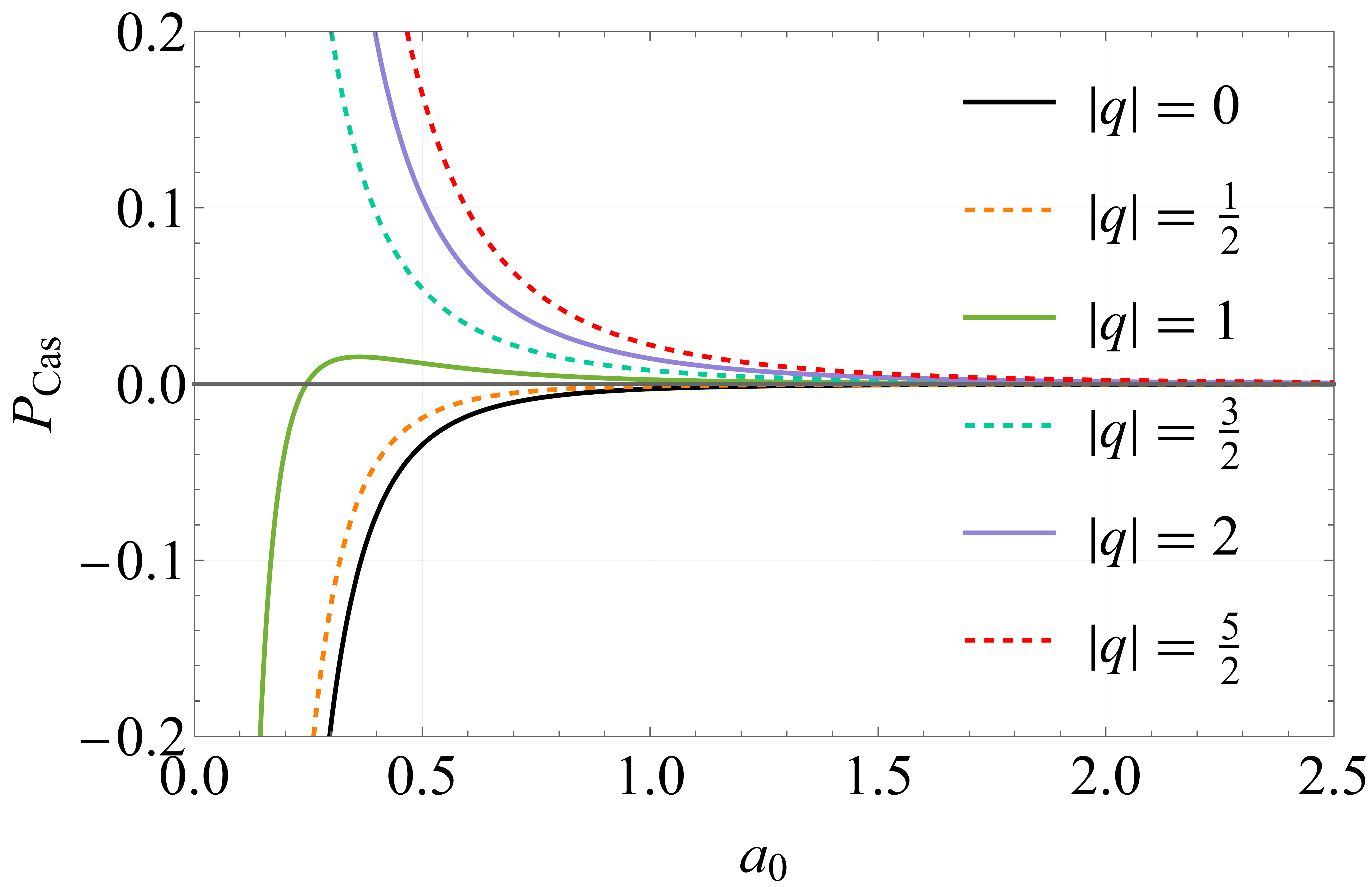}
	\caption{Same as Fig.~\ref{fig:Pcas_xi}, but for $\xi=0.35$.}
\label{fig:Pcasxi0.35}
\end{figure}

For illustration, Figs.~\ref{fig:Pcas_xi}--\ref{fig:Pcasxi0.35} show $P_{\mathrm{Cas}}(a_0,m_0)$ for $m_0=1$,
\begin{equation}
	|q|=0,\half,1,\txt\frac32,2,\txt\frac52,
\end{equation}
and representative couplings $\xi=-0.56,1/8,0.35$. All radii satisfy the strictly positive-frequency condition following Eq.~\eqref{eq:a0limit}. Positive and negative pressure correspond to repulsive and attractive forces, respectively. The figures illustrate the finite-radius behavior, while Eq.~\eqref{eq:pressure-asymptotic} determines the large-radius tail.

\paragraph*{Low-$\xi$ regime, $\xi<\xi_1$.}
For $\xi=-0.56$, Fig.~\ref{fig:Pcas_xi} shows attractive pressure for $|q|=0,1/2,1$ over the displayed ranges. For $|q|=3/2$, the pressure changes from positive to negative as $a_0$ increases. This crossing identifies a local minimum of the scalar vacuum energy under radius variations at fixed $q$, $m_0$, and $\xi$; it does not establish stability of the complete system, including background-field and gravitational contributions. The $|q|=2$ and $5/2$ curves also approach zero from below, although these tails are not resolved on the scale of Fig.~\ref{fig:Pcas_xi}.

Minimal coupling, $\xi=0$, also lies in this regime. For $m_0=1$ and $|q|=1/2$, the pressure changes from positive to negative with increasing radius, so a nonzero monopole does not ensure repulsive pressure at all radii.

\paragraph*{Intermediate-$\xi$ regime, $\xi_1<\xi<\xi_2$.}
The representative conformally coupled case, $\xi=1/8$, is shown in Fig.~\ref{fig:Pcasxi0.125}. Over the displayed radius ranges, the zero-flux curve is negative, whereas the nonzero-charge curves are positive and decrease with radius. Equation~\eqref{eq:pressure-asymptotic} also gives positive large-radius tails for all nonzero allowed charges. The asymptotic criterion alone does not establish positivity or monotonicity at every finite radius throughout this coupling interval.

\paragraph*{High-$\xi$ regime, $\xi>\xi_2$.}
For $\xi=0.35$, Fig.~\ref{fig:Pcasxi0.35} shows attractive pressure for $|q|=0,1/2$. The $|q|=1$ curve crosses from negative to positive as $a_0$ increases and subsequently decreases toward zero; its large-radius tail is positive since $C(1,0.35)=-0.088333\ldots$. For $|q|=3/2$, the pressure is positive and decreases over the displayed range, and its asymptotic tail is also positive. These examples demonstrate how increasing the monopole charge can turn an attractive scalar vacuum pressure into a repulsive one, while the sign at a given finite radius must be determined from the complete expression.

\section{Conclusion}
\label{sec:conclusion}

We have studied the Casimir effect of a complex scalar field on a two-dimensional sphere in the presence of a Wu--Yang magnetic monopole fixed at the center. In this background the scalar field is naturally described as a section rather than an ordinary function, reflecting the nontrivial bundle structure of the Wu--Yang construction. Solving the Klein--Gordon equation in terms of monopole harmonics, we obtained the mode spectrum and evaluated the vacuum energy using contour-integration techniques together with the Abel--Plana formula for multivalued functions.

We derived analytic expressions for the regularized and renormalized Casimir energy density, as well as for the corresponding Casimir energy and pressure. The results show that the monopole can substantially modify the vacuum stress. In particular, for negative, zero, and positive curvature coupling, a sufficiently strong Wu--Yang monopole can reverse the sign of the Casimir pressure, turning an attractive force into a repulsive one. In some parameter regimes the pressure changes sign from positive to negative as the radius increases, identifying a local minimum of the scalar vacuum energy at fixed $q$, $m_0$, and $\xi$. This statement concerns that energy contribution under radius variations and does not establish stability of the complete system, including background-field and gravitational contributions. These findings suggest that Wu--Yang magnetic monopoles may provide a source of repulsive vacuum stresses in curved compact geometries, with potential relevance to gravitational and cosmological settings.

\appendix

\section{Generalized Abel--Plana formula}
\label{app:gapf}

In this appendix we derive two generalizations of the Abel--Plana formula, first for meromorphic functions (GAPF) and then for multivalued functions possessing a branch cut on the real axis (GAPF-MV). The former extension can be found in the literature; see \cite{Saharian2007}. We include it here for completeness. To the best of our knowledge, the extension to the multivalued case used here has not been presented previously in this form. It is used in the main text for the cases with $\beta^2<0$, see Eqs.~\eqref{eq:VarEhalfminus}, \eqref{eq:VarE4}, and \eqref{eq:VarE5}. 

This appendix is self-contained and can be read independently of the main text; some notation is therefore reused with different meanings. We consider functions \(f(z)\) and \(g(z)\) of \(z=x+\ri y\) on the strip \(a\leq x\leq b\), with poles \(z_{f,i}\) and \(z_{g,i}\) in the open strip \(a<x<b\). Superscripts $+$, $-$, and $*$ label poles in the upper half-strip, lower half-strip, and on the real axis, respectively. Thus, \(z_{g,i}^{+}\) and \(z_{g,i}^{-}\) lie in the upper and lower half-strips, while \(z_{g,i}^{*}\) lies on the real axis. We define \(\sigma(z)\equiv\mathrm{sgn}(\mathrm{Im}\,z)\), with \(\mathrm{sgn}(x)=x/|x|\) for \(x\neq0\) and \(\mathrm{sgn}(0)=0\), and use \(\fint\) to denote a principal-value integral.

\subsection{Meromorphic functions}
\label{app:gapf-mero}

\paragraph*{Lemma (GAPF).}
\emph{Let $f(z)$ and $g(z)$ be meromorphic on the strip and satisfy}
\begin{equation}
	\lim_{H\rightarrow\infty}\int^{b\pm\ri H}_{a\pm\ri H}[g(z)\pm f(z)]\,\rd z=0.
	\label{eq:ap-cond1}
\end{equation}
\emph{Then}
\begin{equation}
	\begin{aligned}
		\int^b_{a}f(x)\,\rd x 
		={}& \mathbb{R}[f(z),g(z)] \\
		&-\frac{1}{2}\fint^{+\ri \infty}_{-\ri \infty}\rd z\,
		\bigl[g(u)+\sigma(z)f(u)\bigr]^{u=b+z}_{u=a+z},
	\end{aligned}
	\label{eq:ap-lemma1}
\end{equation}
\emph{where}
$$\bigl[p(u)\bigr]^{u=\beta}_{u=\alpha}:=p(\beta) - p(\alpha)$$
\emph{and}
\begin{equation}
	\begin{aligned}
		\mathbb{R}[f(z),g(z)]
		:={}& \pi\ri \sum_{i}{\mathrm{Res}}\,g(z_{g,i})
		+ \pi\ri \sum_{j}{\mathrm{Res}}\,f(z_{f,j}^+) \\
		& - \pi\ri \sum_{k}{\mathrm{Res}}\,f(z_{f,k}^-).
	\end{aligned}
	\label{eq:ap-R}
\end{equation}
\emph{Here ${\mathrm{Res}}\,g(z_{g,i}):=\underset{z=z_{g,i}}{\mathrm{Res}}\,g(z)$ denotes the residue of $g(z)$ at $z=z_{g,i}$.}

\begin{figure}[htbp]
	\centering
	\begin{tikzpicture}[scale=1.1, >=stealth, thick]
		\def\a{-3.0}
		\def\b{1.2}
		\def\H{2.0}
		\def\eps{0.15}
		\def\rpole{0.22}
		\def\poleone{-2.1}
		\def\poletwo{-1.1}
		
		\draw (\a, \H) -- (\a, \eps);
		\draw (\a, -\eps) -- (\a, -\H);
		
		\draw[->] (\a, \H) -- (\a, 1.4);
		\draw[->] (\a,-\eps ) -- (\a, -1.4);
		
		\draw (\a, -\H) -- (\b, -\H);
		
		\draw (\b, -\H) -- (\b, -\eps);
		\draw (\b, \eps) -- (\b, \H);
		
		\draw[->] (\b, -\H) -- (\b, -1.4);
		\draw[->] (\b, \eps) -- (\b, 1.4);
		
		\draw (\b, \H) -- (\a, \H);
		
		\node[above left] at (\a, \H) {$a+\mathrm{i}H$};
		\node[above right] at (\b, \H) {$b+\mathrm{i}H$};
		\node[below left] at (\a, -\H) {$a-\mathrm{i}H$};
		\node[below right] at (\b, -\H) {$b-\mathrm{i}H$};
		
		\draw[dotted, thick] (\a-0.7, 0) -- (\b+0.9, 0);
		\draw[->,very thick] (\b+0.8, 0) -- (\b+0.9, 0);
		\node[below left] at (\a, -0.2) {$a$};
		\node[below right] at (\b, -0.2) {$b$};
		
		\draw[dash dot] (\a, \eps) -- (\poleone - \rpole, \eps);
		\draw[->] (\poleone - \rpole-0.4, \eps) -- (\poleone - \rpole-0.25, \eps);
		\node[above] at ({(\a + \poleone - \rpole)/2}, \eps + 0.15) {$\gamma_1^+$};
		
		\draw (\poleone - \rpole, \eps) arc (180:0:\rpole);
		\node[above] at (\poleone, \rpole + \eps + 0.08) {$G_1^+$};
		
		\draw[dash dot] (\poleone + \rpole, \eps) -- (\poletwo - \rpole, \eps);
		\draw[->] (\poletwo - \rpole-0.3, \eps) -- (\poletwo - \rpole-0.15, \eps);
		\node[above] at ({(\poleone + \poletwo)/2}, \eps + 0.15) {$\gamma_2^+$};
		
		\draw (\poletwo - \rpole, \eps) arc (180:0:\rpole);
		\node[above] at (\poletwo, \rpole + \eps + 0.08) {$G_2^+$};
		
		\draw[ dash dot] (\poletwo + \rpole, \eps) -- (\b, \eps);
		\draw[ ->] (\b-0.9, \eps) -- (\b-0.7, \eps);
		
		\draw[->] (\b-0.7, -\eps) -- (\b-0.9, -\eps);
		\draw[ dash dot] (\b, -\eps) -- (\poletwo + \rpole, -\eps);
		\draw (\poletwo + \rpole, -\eps) arc (0:-180:\rpole);
		\node[below] at (\poletwo, -\rpole - \eps - 0.08) {$G_2^-$};
		
		\draw[ ->] (\poletwo - \rpole-0.15, -\eps) -- (\poletwo - \rpole-0.3, -\eps);
		\draw[ dash dot] (\poletwo - \rpole, -\eps) -- (\poleone + \rpole, -\eps);
		\node[below] at ({(\poleone + \poletwo)/2}, -\eps - 0.15) {$\gamma_2^-$};
		
		\draw (\poleone + \rpole, -\eps) arc (0:-180:\rpole);
		\node[below] at (\poleone, -\rpole - \eps - 0.08) {$G_1^-$};
		
		\draw [ dash dot](\poleone - \rpole, -\eps) -- (\a, -\eps);
		\draw [ ->](\poleone - \rpole-0.25, -\eps) -- (\poleone - \rpole-0.4, -\eps);
		\node[below] at ({(\a + \poleone - \rpole)/2}, -\eps - 0.15) {$\gamma_1^-$};
		
		\node at (\poleone, 0) {$\ast$};
		\node at (\poletwo, 0) {$\ast$};
		
		\node at (-0.85,1.6) {$C_H^{+}$};
		\node at (-0.85,-1.6) {$C_H^{-}$};
	\end{tikzpicture}
	\caption{Integration contours used in Eqs.~\eqref{eq:fCHplus} and \eqref{eq:fCHminus}. The horizontal edges lie at $\operatorname{Im}z=\pm H$, with $H\to\infty$ taken at the end. The symbol $*$ denotes a real-axis pole of $f(z)$. The segments $\gamma_i^\pm$ run just above and below the real axis between neighboring poles, while $G_i^\pm$ are the corresponding semicircular indentations.}
	\label{fig:GAPF-half}
\end{figure}
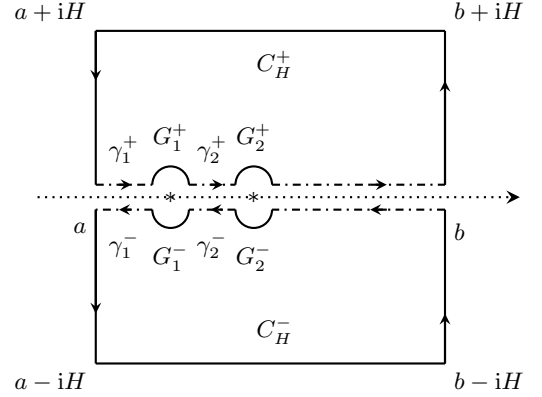

\paragraph*{Proof.}
Consider the rectangular contour $C_H$, traversed counterclockwise, with vertices at $a\pm\ri H$ and $b\pm\ri H$. The residue theorem gives
\begin{equation}
	\int_{C_H}g(z)\,\rd z=2\pi\ri\sum_{i}{\mathrm{Res}}\,g(z_{g,i}),
	\label{eq:ap-resg}
\end{equation}
where poles lying on the integration contour are excluded by infinitesimal indentations, and the sum runs over the poles enclosed by $C_H$. A pole lying exactly on the edges $\Re z=a$ or $b$ is treated by indenting the contour with a small semicircle, contributing half its residue. Denoting by $C_H^{+}$ and $C_H^{-}$ the upper and lower halves of the contour, we may write
\begin{equation}
	\begin{aligned}
		\fint_{C_H}g(z)\,\rd z
		={}&\sum_{\alpha=\pm}\int_{C^\alpha_H}\rd z\,[g(z)+\alpha f(z)]\\
		&-\sum_{\alpha=\pm}\alpha\int_{C^\alpha_H}\rd z\,f(z).
	\end{aligned}
	\label{eq:ap-split}
\end{equation}
The closed contours in the upper and lower half planes are
\begin{equation}
	C^{\pm}=C^{\pm}_H+\sum_i\bigl(\gamma^{\pm}_i+G^{\pm}_i\bigr),
	\label{eq:ap-mv-contour}
\end{equation}
where the segments $\gamma_i^{\pm}$ run along the upper/lower banks of the real axis between consecutive poles, and $G_i^{\pm}$ are the corresponding small semicircular deformations that indent each pole (Fig.~\ref{fig:GAPF-half}). Applying the residue theorem separately to $f(z)$ on the upper and lower halves gives
\begin{align}
	\int_{C^+}f(z)\,\rd z 
	={}&\int_{C^+_H}f(z)\,\rd z+\sum_i\!\left[\int_{G^+_i}\!f\,\rd z+\int_{\gamma^+_i}\!f\,\rd z\right] \notag\\
	={}&2\pi\ri\sum_{j}{\mathrm{Res}}\,f(z^+_{f,j}),\label{eq:fCHplus}\\
	\int_{C^-}f(z)\,\rd z 
	={}&\int_{C^-_H}f(z)\,\rd z+\sum_i\!\left[\int_{G^-_i}\!f\,\rd z+\int_{\gamma^-_i}\!f\,\rd z\right]\notag\\
	={}&2\pi\ri\sum_{k}{\mathrm{Res}}\,f(z^-_{f,k}), \label{eq:fCHminus}
\end{align}

Subtracting Eq.~\eqref{eq:fCHminus} from Eq.~\eqref{eq:fCHplus} yields
\begin{equation}
	\begin{aligned}
		&\int_{C^+_H}f(z)\,\rd z-\int_{C^-_H}f(z)\,\rd z \\
		={}&2\pi\ri\sum_{j}{\mathrm{Res}}\,f(z^+_{f,j}) - 2\pi\ri\sum_{k}{\mathrm{Res}}\,f(z^-_{f,k})
		- 2\fint^b_{a}f(x)\,\rd x,
	\end{aligned}
	\label{eq:ap-fdiff}
\end{equation}
where the contributions from the segments $\gamma_i^{\pm}$ combine into a principal-value integral in the limit of vanishing semicircle radius. Because $f(z)$ is single-valued, the two semicircles indent the pole with the same (clockwise) sense, so their half-residue contributions are equal and cancel in the difference above. This cancellation fails when $f$ is multivalued, as discussed in Sec.~\ref{app:gapf-mv}.

From the geometry of the contour,
\begin{align}
	\int_{C^+_H}[g+f]\,\rd z
	={}&\int^{\ri H}_{0^+}\rd z\,[g(u)+f(u)]^{u=b+z}_{u=a+z}\notag\\
	&-\int^{b+\ri H}_{a+\ri H}\rd z\,[g(z)+f(z)],
	\label{eq:ap-C+}\\
	\int_{C^-_H}[g-f]\,\rd z
	={}&-\int^{-\ri H}_{0^-}\rd z\,[g(u)-f(u)]^{u=b+z}_{u=a+z}\notag\\
	&+\int^{b-\ri H}_{a-\ri H}\rd z\,[g(z)-f(z)].
	\label{eq:ap-C-}
\end{align}
Combining Eqs.~\eqref{eq:ap-resg}, \eqref{eq:ap-split}, \eqref{eq:ap-fdiff}, \eqref{eq:ap-C+}, and \eqref{eq:ap-C-}, and taking $H\to\infty$, gives
\begin{align}
	\int^b_{a}f(x)\,\rd x
	={}&\mathbb{R}[f,g]
	-\frac{1}{2}\int^{+\ri \infty}_{0^+}\rd z\,[g(u)+f(u)]^{u=b+z}_{u=a+z}\notag\\
	&-\frac{1}{2}\int^{0^-}_{-\ri \infty}\rd z\,[g(u)-f(u)]^{u=b+z}_{u=a+z},
\end{align}
which is precisely Eq.~\eqref{eq:ap-lemma1}. \hfill$\square$

\paragraph*{Theorem (GAPF).}
\emph{If, in addition to Eq.~\eqref{eq:ap-cond1},}
\begin{equation}
	\lim_{b\rightarrow\infty}\int^{b\pm\ri \infty}_{b}[g(z)\pm f(z)]\,\rd z=0,
	\label{eq:ap-cond2}
\end{equation}
\emph{then}
\begin{equation}
	\lim_{b\rightarrow\infty}\Bigl\{\int^b_{a}f(x)\,\rd x-\mathbb{R}[f,g]\Bigr\}
	=\frac{1}{2}\fint^{a+\ri \infty}_{a-\ri \infty}\rd z\,[g(z)+\sigma(z)f(z)].
	\label{eq:ap-thm1}
\end{equation}
This follows immediately by taking $b\to\infty$ in Eq.~\eqref{eq:ap-lemma1}. Equation~\eqref{eq:ap-thm1} constitutes the generalized Abel--Plana formula (GAPF).

\subsection{Applications of GAPF}
\label{app:gapf-apps}

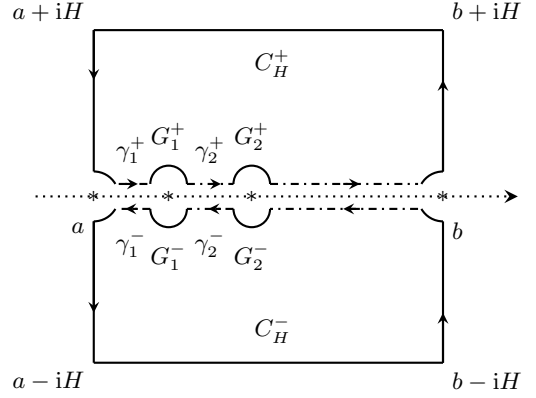
\begin{figure}[htbp]
	\centering
	\begin{tikzpicture}[scale=1.1, >=stealth, thick]
		\def\a{-3.0}
		\def\b{1.2}
		\def\H{2.0}
		\def\eps{0.15}
		\def\rc{0.22}
		\def\r0{0.3}
		\def\poleA{-2.1}
		\def\poleB{-1.1}
		
		\draw (\a, \H) -- (\a, \r0);
		\draw (\a, \r0) arc(90:30:\r0);
		\draw (\a, -\r0) arc(-90:-30:\r0);
		\draw (\a, -\r0) -- (\a, -\H);
		
		\draw[->] (\a, \H) -- (\a, 1.4);
		\draw[->] (\a,-\r0 ) -- (\a, -1.4);
		
		\draw (\a, -\H) -- (\b, -\H);
		
		\draw (\b, -\H) -- (\b, -\r0);
		\draw (\b, \r0) arc (90:150:\r0);
		\draw (\b, -\r0) arc (-90:-150:\r0);
		\draw (\b, \r0) -- (\b, \H);
		
		\draw[->] (\b, -\H) -- (\b, -1.4);
		\draw[->] (\b, \r0) -- (\b, 1.4);
		
		\draw (\b, \H) -- (\a, \H);
		
		\node[above left] at (\a, \H) {$a+\mathrm{i}H$};
		\node[above right] at (\b, \H) {$b+\mathrm{i}H$};
		\node[below left] at (\a, -\H) {$a-\mathrm{i}H$};
		\node[below right] at (\b, -\H) {$b-\mathrm{i}H$};
		
		\draw[dotted, thick] (\a-0.7, 0) -- (\b+0.9, 0);
		\draw[->,very thick] (\b+0.8, 0) -- (\b+0.9, 0);
		\node[below left] at (\a, -0.2) {$a$};
		\node[below right] at (\b, -0.2) {$b$};
		
		\draw[<-] (\poleA - \rc-0.15, \eps) -- (\poleA - \rc-0.3, \eps);
		\draw[ dash dot] (\a + \r0, \eps) -- (\poleA - \rc, \eps);
		\node[above] at ({(\a + \poleA)/2}, \eps + 0.14) {$\gamma_1^+$};
		
		\draw (\poleA - \rc, \eps) arc (180:0:\rc);
		\node[above] at (\poleA, \rc + \eps + 0.08) {$G_1^+$};
		
		\draw[->] (\poleA + \rc+0.3, \eps) -- (\poleB - \rc-0.15, \eps);
		\draw[ dash dot] (\poleA + \rc, \eps) -- (\poleB - \rc, \eps);
		\node[above] at ({(\poleA + \poleB)/2}, \eps + 0.14) {$\gamma_2^+$};
		
		\draw (\poleB - \rc, \eps) arc (180:0:\rc);
		\node[above] at (\poleB, \rc + \eps + 0.08) {$G_2^+$};
		
		\draw[->](\b - \r0 -0.9, \eps) -- (\b - \r0-0.7, \eps);
		\draw[ dash dot](\poleB + \rc, \eps) -- (\b - \r0, \eps);
		
		\draw[->] (\b - \r0-0.7, -\eps) -- (\b - \r0 -0.9, -\eps);
		\draw[ dash dot] (\b - \r0, -\eps) -- (\poleB + \rc, -\eps);
		
		\draw (\poleB + \rc, -\eps) arc (0:-180:\rc);
		\node[below] at (\poleB, -\rc - \eps - 0.08) {$G_2^-$};
		
		\draw[->] (\poleB - \rc-0.15, -\eps) -- (\poleB - \rc-0.3, -\eps);
		\draw[ dash dot] (\poleB - \rc, -\eps) -- (\poleA + \rc, -\eps);
		\node[below] at ({(\poleA + \poleB)/2}, -\eps - 0.14) {$\gamma_2^-$};
		
		\draw (\poleA + \rc, -\eps) arc (0:-180:\rc);
		\node[below] at (\poleA, -\rc - \eps - 0.08) {$G_1^-$};
		
		\draw [->](\poleA - \rc -0.15, -\eps) -- (\poleA - \rc -0.3, -\eps);
		\draw [ dash dot](\poleA - \rc , -\eps) -- (\a + \r0, -\eps);
		\node[below] at ({(\a + \poleA)/2}, -\eps - 0.14) {$\gamma_1^-$};
		
		\node at (\poleA, 0) {$\ast$};
		\node at (\poleB, 0) {$\ast$};
		\node at (\a, 0) {$\ast$};
		\node at (\b, 0) {$\ast$};
		
		\node at (-.85,1.6) {$C_H^{+}$};
		\node at (-.85,-1.6) {$C_H^{-}$};
		
	\end{tikzpicture}
	\caption{Integration contours for sums over integers. The symbol $*$ denotes the simple poles of the integer Abel--Plana kernel at $z=k\in\mathbb{Z}$. Here $a$ and $b$ are integers, and the endpoint poles are treated by quarter-circle indentations on each half-contour.}
	\label{fig:GAPF-int}
\end{figure}

Setting $a=m$, $b=n$ for integers $m,n$, and choosing
\begin{equation}
	g(z)=-\ri\cot(\pi z)\,f(z),
	\label{eq:ap-gint}
\end{equation}
with $f(z)$ analytic, endows $g(z)$ with simple poles at every integer, while the endpoints $z=m,n$ are indented by small semicircles (Fig.~\ref{fig:GAPF-int}).

From Eq.~\eqref{eq:ap-R},
\begin{equation}
	\mathbb{R}[f,g]=\sum^{n-1}_{k=m+1}f(k),
\end{equation}
while the semicircles at $z=m$ and $z=n$ contribute $\tfrac{1}{2}f(m)$ and $\tfrac{1}{2}f(n)$, respectively. Using
\begin{equation}
	\begin{aligned}
		g(z)+f(z)
		={}&
		\frac{2f(z)}{1-\e^{-2\pi \ri z}},
		\\
		g(z)-f(z)
		={}&
		\frac{2f(z)}{\e^{2\pi \ri z}-1},
	\end{aligned}
	\label{eq:cot}
\end{equation}
one finds
\begin{equation}
	\frac{1}{2}\fint^{m+\ri \infty}_{m-\ri \infty}\rd z\,[g+\sigma f]
	=-\ri\int^{\infty}_{0^+}\rd t\,
	\frac{f(m+\ri t)-f(m-\ri t)}{\e^{2\pi t}-1},
	\label{eq:ap-vertint}
\end{equation}
so that Eq.~\eqref{eq:ap-lemma1} becomes
\begin{equation}
	\begin{aligned}
		\sum^{n}_{k=m}f(k)
		={}&\frac{1}{2}f(m)+\frac{1}{2}f(n)+\int^{n}_{m}f(x)\,\rd x \\
		&-\ri\int^{\infty}_{0^+}\rd t \left[\frac{f(u+\ri t)-f(u-\ri t)}{\e^{2\pi t}-1}\right]_{u=m}^{u=n}.
	\end{aligned}
\end{equation}
Letting $n\rightarrow\infty$ and setting $m=0$ recovers the standard form,
\begin{equation}
	\sum^{\infty}_{k=0}f(k)
	=\frac{1}{2}f(0)+\int^{\infty}_{0}f(x)\,\rd x
	+\ri\int^{\infty}_{0^+}\rd t\,\frac{f(\ri t)-f(-\ri t)}{\e^{2\pi t}-1},
\end{equation}
provided $\lim_{y\to\infty}\e^{-2\pi|y|}|f(x+\ri y)|=0$ uniformly on any finite $x$ interval.

For sums over half-odd integers, we choose the same endpoints and set 
\begin{equation}
	g(z)=\ri\tan(\pi z)\,f(z),
\end{equation}
$g(z)$ acquires simple poles at $z=k+\tfrac{1}{2}$. In this case 
$$\mathbb{R}[f,g]=\sum^{n-1}_{k=m}f(k+\tfrac{1}{2}),$$ 
and using
\begin{equation}
	\begin{aligned}
		g(z)+f(z)
		={}&
		\frac{2f(z)}{1+\e^{-2\pi \ri z}},
		\\
		g(z)-f(z)
		={}&
		-\frac{2f(z)}{1+\e^{2\pi \ri z}},
	\end{aligned}
	\label{eq:tan}
\end{equation}
one obtains, for $m=0$ and $n\rightarrow\infty$, 
\begin{equation}
	\sum^{\infty}_{k=0}f\!\left(k+\tfrac{1}{2}\right)
	=\int^{\infty}_{0}f(x)\,\rd x
	-\ri\int^{\infty}_{0}\rd t\,\frac{f(\ri t)-f(-\ri t)}{\e^{2\pi t}+1}.
\end{equation}

\subsection{Extension to multivalued functions}
\label{app:gapf-mv}

\begin{figure}[htbp]
	\centering
	\begin{tikzpicture}[scale=1.1, >=stealth, thick]
		\def\a{-3.0}
		\def\b{1.2}
		\def\H{2.0}
		\def\eps{0.2}
		\def\epsa{0.1}
		\def\d{1.5}
		\def\rpole{0.22}
		\def\poleone{-2.1}
		\def\poletwo{0}
		
		\draw (\a, \H) -- (\a, \eps);
		\draw (\a, -\eps) -- (\a, -\H);
		
		\draw[->] (\a, \H) -- (\a, 1.4);
		\draw[->] (\a,-\eps ) -- (\a, -1.4);
		
		\draw (\a, -\H) -- (\b, -\H);
		
		\draw (\b, -\H) -- (\b, -\epsa);
		\draw (\b, \epsa) -- (\b, \H);
		
		\draw[->] (\b, -\H) -- (\b, -1.4);
		\draw[->] (\b, \epsa) -- (\b, 1.4);
		
		\draw (\b, \H) -- (\a, \H);
		
		\node[above left] at (\a, \H) {$a+\mathrm{i}H$};
		\node[above right] at (\b, \H) {$b+\mathrm{i}H$};
		\node[below left] at (\a, -\H) {$a-\mathrm{i}H$};
		\node[below right] at (\b, -\H) {$b-\mathrm{i}H$};
		
		\draw[dotted, thick] (\a-0.7, 0) -- (\b+0.9, 0);
		\draw[->,very thick] (\b+0.8, 0) -- (\b+0.9, 0);
		\node[below left] at (\a, -0.2) {$a$};
		\node[below right] at (\b, -0.2) {$b$};
		
		\draw[] (\a, \eps) -- (\poleone - \rpole, \eps);
		\draw[->] (\poleone - \rpole-0.4, \eps) -- (\poleone - \rpole-0.25, \eps);
		\node[above] at ({(\a + \poleone - \rpole)/2}, \eps + 0.15) {$\gamma_1^+$};
		
		\draw (\poleone - \rpole, \eps) arc (180:0:\rpole);
		\node[above] at (\poleone, \rpole + \eps + 0.08) {$G_1^+$};
		
		\draw[] (\poleone + \rpole, \eps) -- (\a+\d, \eps);
		\draw[->] (\a+\d-0.25, \eps) -- (\a+\d-0.1, \eps);
		\draw[ dash dot] (\a+\d, \epsa) -- (\poletwo - \rpole, \epsa);
		\node[above] at ({(\poleone + \poletwo)/2}, \epsa + 0.15) {$\gamma_2^+$};
		
		\draw (\a+\d, \eps) -- (\a+\d, \epsa);
		\draw (\poletwo - \rpole, \epsa) arc (180:0:\rpole);
		\node[above] at (\poletwo, \rpole + \epsa + 0.08) {$G_2^+$};
		
		\draw[ dash dot] (\poletwo + \rpole, \epsa) -- (\b, \epsa);
		\draw[ ->] (\b-0.8, \epsa) -- (\b-0.6, \epsa);
		
		\draw[dash dot] (\b, -\epsa) -- (\poletwo + \rpole, -\epsa);
		\draw[ <-] (\b-0.8, -\epsa) -- (\b-0.6, -\epsa);
		\draw (\poletwo + \rpole, -\epsa) arc (0:-180:\rpole);
		\node[below] at (\poletwo, -\rpole - \epsa - 0.08) {$G_2^-$};
		
		\draw[ dash dot] (\poletwo - \rpole, -\epsa) -- (\a+\d, -\epsa);
		\draw[<-] (\a+\d-0.25, -\eps) -- (\a+\d-0.1, -\eps);
		\draw[] (\a+\d, -\eps) -- (\poleone + \rpole, -\eps);
		\node[below] at ({(\poleone + \poletwo)/2}, -\epsa - 0.15) {$\gamma_2^-$};
		
		\draw (\a+\d, -\eps) -- (\a+\d, -\epsa);
		\draw (\poleone + \rpole, -\eps) arc (0:-180:\rpole);
		\node[below] at (\poleone, -\rpole - \eps - 0.08) {$G_1^-$};
		
		\draw (\poleone - \rpole, -\eps) -- (\a, -\eps);
		\draw[<-] (\poleone - \rpole-0.4, -\eps) -- (\poleone - \rpole-0.25, -\eps);
		\node[below] at ({(\a + \poleone - \rpole)/2}, -\eps - 0.15) {$\gamma_1^-$};
		
		\draw[decorate,decoration={zigzag, segment length=4pt, amplitude=2pt}] (\a-0.8,0) -- (\a+\d,0);
		
		\node at (\a+\d,-\eps-0.8) {$a\!+\!d $};
		\draw[->](\a+\d,-\eps-0.66)--(\a+\d,-\eps);
		
		\node at (\poleone, 0) {$\ast$};
		\node at (\poletwo, 0) {$\ast$};
		
		\node at (-.85,1.6) {$C_H^{+}$};
		\node at (-.85,-1.6) {$C_H^{-}$};
	\end{tikzpicture}
	\caption{Integration contour for the multivalued case of Sec.~\ref{app:gapf-mv}. The branch cut (zigzag line) lies on $[a,a+d]$. The segments $\gamma_i^\pm$ run just above and below the real axis. Their contributions do not cancel across the branch cut, whereas they do cancel appropriately outside the cut. The semicircles $G_i^\pm$ indent the real-axis singularities of $g(z)$.}
	\label{fig:contour-mv}
\end{figure}
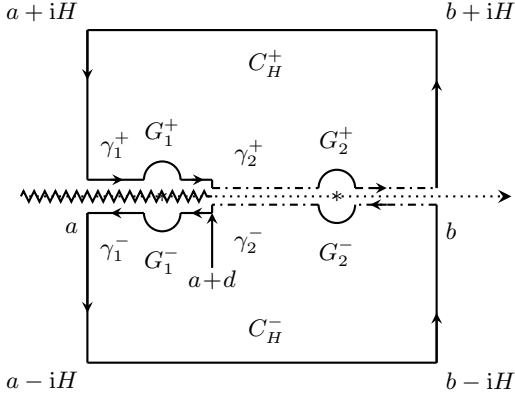

We now generalize the analysis to multivalued \(f(z)\) (and, correspondingly, \(g(z)\)), with a branch cut along the real interval \([a,a+d]\subset[a,b]\) (Fig.~\ref{fig:contour-mv}). This is precisely the situation encountered in Case~(5) of Sec.~\ref{sec:regularization}. For simplicity, we assume that \(f(z)\) has no other singularities in the strip \(a<x<b\) and that all singularities of \(g(z)\) in this strip lie on the real axis. The extension to functions with poles off the real axis is straightforward but cumbersome and is not considered here.

Let \(f^\pm(z)\) and \(g^\pm(z)\) denote the corresponding analytic branches in the upper and lower half-planes, respectively. Since neither $f(z)$ nor $g(z)$ has singularities off the real axis, the closed contour integrals of $f^\pm(z)$ and $g^\pm(z)$ along $C^\pm$ each vanish individually, and hence so does any linear combination of them:
\begin{align}
	0
	={}&\sum_{\alpha=\pm}\int_{C^\alpha}\rd z\,[g^\alpha(z)+\alpha f^\alpha(z)]  -\sum_{\alpha=\pm}\alpha\int_{C^\alpha}\rd z\,f^\alpha(z).
	\label{eq:ap-mv-sum}
\end{align}
Separating each segment of these integrals, we obtain
\begin{align}
	0
	={}&\int_{C^+_H}\frac{[g^+(z)+f^+(z)]}{2}\,\rd z
	+\int_{C^-_H}\frac{[g^-(z)-f^-(z)]}{2}\,\rd z\notag\\
	&+\sum_i\int_{G^+_i+\gamma^+_i}\frac{[g^+(z)+f^+(z)]}{2}\,\rd z\notag\\
	&+\sum_i\int_{G^-_i+\gamma^-_i}\frac{[g^-(z)-f^-(z)]}{2}\,\rd z.
	\label{eq:ap-mv-R}
\end{align}
The first two terms in Eq.~\eqref{eq:ap-mv-R} are evaluated exactly as in the meromorphic case, with $f,g\to f^\pm,g^\pm$:
\begin{align}
	\int_{C^+_H}[g^++f^+]\,\rd z
	={}&\int^{\ri H}_{0^+}\rd z\,[g^+(u)+f^+(u)]^{u=b+z}_{u=a+z}\notag\\
	&-\int^{b+\ri H}_{a+\ri H}\rd z\,[g^+(z)+f^+(z)],
	\label{eq:ap-mv-C+}\\
	\int_{C^-_H}[g^--f^-]\,\rd z
	={}&-\int^{-\ri H}_{0^-}\rd z\,[g^-(u)-f^-(u)]^{u=b+z}_{u=a+z}\notag\\
	&+\int^{b-\ri H}_{a-\ri H}\rd z\,[g^-(z)-f^-(z)].
	\label{eq:ap-mv-C-}
\end{align}
The remaining $\gamma^\pm_i$ and $G^\pm_i$ contributions in Eq.~\eqref{eq:ap-mv-R} depend on whether the corresponding segment lies on the branch cut (BC); four cases arise.

\emph{(a) $\gamma^\alpha_i$ near the cut.} The upper and lower segments do not cancel:
\begin{equation}
	\begin{aligned}
		&\sum_{\gamma_i\in\mathrm{BC}}\!\left[\int_{\gamma^+_i}\frac{g^++f^+}{2}\,\rd z
		+\int_{\gamma^-_i}\frac{g^--f^-}{2}\,\rd z\right] \\
		={}&\fint^{a+d}_{a}\rd x\,\frac{g^+(x)-g^-(x)+f^+(x)+f^-(x)}{2}.
	\end{aligned}
	\label{eq:ap-mv-caseA}
\end{equation}

\emph{(b) $\gamma^\alpha_i$ away from the cut.} Here $\gamma^+_i$ and $\gamma^-_i$ overlap with opposite orientation, $g^+=g^-$ and $f^+=f^-$, so
\begin{equation}
	\sum_{\gamma_i\notin\mathrm{BC}}[\cdots]=\int^{b}_{a+d}f(x)\,\rd x.
	\label{eq:ap-mv-caseB}
\end{equation}

\emph{(c) $G^\alpha_i$ near the cut ($z_i\in\mathrm{BC}$).} Since $g^+\neq g^-$ and $f(z)$ has no singularities other than the cut, each semicircle contributes minus half the residue of $g^\alpha$:
\begin{equation}
	\sum_{z_i\in\mathrm{BC}}[\cdots]
	=-\pi\ri\sum_{z_i\in\mathrm{BC}}
	\bigl[\mathrm{Res}\,g^+(z^+_i)+\mathrm{Res}\,g^-(z^-_i)\bigr],
	\label{eq:ap-mv-caseC}
\end{equation}
with $z^\pm_i=z_i\pm\ri \varepsilon$.

\emph{(d) $G^\alpha_i$ away from the cut ($z_i\notin\mathrm{BC}$).} Here $g^+=g^-=g$, and the two semicircles combine into a full clockwise contour:
\begin{equation}
	\sum_{z_i\notin\mathrm{BC}}[\cdots]
	=-2\pi\ri\sum_{z_i\notin\mathrm{BC}}\mathrm{Res}\,g(z_i).
	\label{eq:ap-mv-caseD}
\end{equation}

Collecting Eqs.~\eqref{eq:ap-mv-R}--\eqref{eq:ap-mv-caseD} and discarding the horizontal segments as $H\to\infty$ by virtue of Eq.~\eqref{eq:ap-cond1}, we obtain the following result.

\paragraph*{Lemma (GAPF-MV).}
\emph{If}
\begin{equation}
	\lim_{H\rightarrow\infty}\int^{b\pm\ri H}_{a\pm\ri H}
	[g^\pm(z)\pm f^\pm(z)]\,\rd z=0,
	\label{eq:ap-mv-cond1}
\end{equation}
\emph{then}
\begin{align}
	&\fint^{a+d}_{a}\rd x\,\frac{g^+(x)-g^-(x)+f^+(x)+f^-(x)}{2}\notag\\
	&
	+\int^{b}_{a+d}f(x)\,\rd x\notag\\
	={}& \frac{\pi\ri}{2}\sum_{z_i\in\mathrm{BC}}
	\left[\mathrm{Res}\,g^+(z^+_i)+\mathrm{Res}\,g^-(z^-_i) \right] + \pi\ri\sum_{z_i\notin\mathrm{BC}}\mathrm{Res}\,g(z_i) \notag\\
	&-\frac{1}{2}\fint^{+\ri \infty}_{-\ri \infty}\rd z\,
	[g(u)+\sigma(z)f(u)]^{u=b+z}_{u=a+z}.
	\label{eq:ap-mv-lemma}
\end{align}

\paragraph*{Theorem (GAPF-MV).}
\emph{If, in addition, Eq.~\eqref{eq:ap-cond2} holds, then}
\begin{align}
	&\fint^{a+d}_{a}\rd x\,\frac{g^+(x)-g^-(x)+f^+(x)+f^-(x)}{2}
	+\int^{\infty}_{a+d}f(x)\,\rd x\notag\\
	={}&\pi\ri\sum_{z_i\in\mathrm{BC}}
	\frac{\mathrm{Res}\,g^+(z^+_i)+\mathrm{Res}\,g^-(z^-_i)}{2} \notag\\
	& +\pi\ri\sum_{z_i\notin\mathrm{BC}}\mathrm{Res}\,g(z_i)
	+\frac{1}{2}\fint^{a+\ri \infty}_{a-\ri \infty}\rd z\,[g(z)+\sigma(z)f(z)].
	\label{eq:ap-mv-thm}
\end{align}
Equations~\eqref{eq:ap-mv-lemma} and \eqref{eq:ap-mv-thm} give the GAPF-MV for a cut on $[a,a+d]$; branch cuts placed elsewhere in $[a,b]$ are treated analogously.

\subsection{Applications of the GAPF-MV}
\label{app:gapf-mv-int}

We now apply the GAPF-MV to sums over integers and over half-odd integers. Taking $a=m$, $b=n$ with $m,n$ integers, and
\begin{equation}
	g(z)=-\ri \cot(\pi z)\,f(z),
	\label{eq:ap-mv-gint}
\end{equation}
which has simple poles at every integer, we note that the pole at $z=a$ lies on the cut while $z=b$ does not. In the following sums, ``$z_i\in\mathrm{BC}$'' denotes singularities in the interior of the branch cut and excludes the endpoint $z=a$, whose contribution is written separately. Accounting for these two corner contributions, Eq.~\eqref{eq:ap-mv-thm} becomes
\begin{align}
	&\fint^{a+d}_{a}\rd x\,\frac{g^+(x)-g^-(x)+f^+(x)+f^-(x)}{2}
	+\fint^{\infty}_{a+d}f(x)\,\rd x\notag\\
	={}&\frac{\pi\ri}{2} \sum_{z_i\in\mathrm{BC}} 
	\left[\mathrm{Res}\,g^+(z^+_i)+\mathrm{Res}\,g^-(z^-_i)\right] +\pi\ri \sum_{z_i\notin\mathrm{BC}}\mathrm{Res}\,g(z_i) \notag\\
	& +\frac{\pi\ri}{4}\left[\mathrm{Res}\,g^+(a^+)+\mathrm{Res}\,g^-(a^-)\right]
	+ \frac{\pi\ri}{2} \,\mathrm{Res}\,g(b)\notag\\
	&+\frac{1}{2}\fint^{a+\ri \infty}_{a-\ri \infty}\rd z\,[g(z)+\sigma(z)f(z)].
	\label{eq:ap-mv-int-master}
\end{align}
Since \(f(z)\) has no poles in the strip, the residue contributions arise solely from the Abel--Plana kernel and reduce to
\begin{align}
	\pi\ri\sum_{z_i\notin\mathrm{BC}}\!\mathrm{Res}\,g(z_i)
	+\frac{\pi\ri}{2} \,\mathrm{Res}\,g(b)
	={}&\sum_{k\notin\mathrm{BC}}f(k)+\frac{f(b)}{2},\notag\\
	\frac{\pi\ri}{2}\sum_{z_i\in\mathrm{BC}} \left[\mathrm{Res}\,g^+(z^+_i)+\mathrm{Res}\,g^-(z^-_i)\right]
	={}&\sum_{k\in\mathrm{BC}}\frac{f^+(k)+f^-(k)}{2},\notag\\
	\frac{\pi\ri}{4} \left[\mathrm{Res}\,g^+(a^+)+\mathrm{Res}\,g^-(a^-)\right]
	={}&\frac{f^+(a)+f^-(a)}{4}.
	\label{eq:ap-mv-int-res}
\end{align}
Applying Eq.~\eqref{eq:cot}, the vertical integral evaluates to
\begin{align}
	&\frac{1}{2}\int^{a+\ri \infty}_{a-\ri \infty}\rd z\,[g+\sigma f]\notag\\	
	={}&\int^{m+\ri \infty}_{m+\ri 0^+}\rd z\,\frac{f^+(z)}{1-\e^{-2\pi\ri z}}
	+\int^{m+\ri 0^-}_{m-\ri \infty}\rd z\,\frac{f^-(z)}{\e^{2\pi\ri z}-1}\notag\\
	={}&-\ri\int^{\infty}_{0^+}\rd t\,
	\frac{f^+(m+\ri t)-f^-(m-\ri t)}{\e^{2\pi t}-1}.
	\label{eq:ap-mv-int-vert}
\end{align}
Substituting Eqs.~\eqref{eq:ap-mv-int-res}--\eqref{eq:ap-mv-int-vert} into Eq.~\eqref{eq:ap-mv-int-master} gives
\begin{align}
	&\fint^{a+d}_{a}\rd x\!\left[\frac{f^+(x)}{1-\e^{-2\pi\ri x}}
	-\frac{f^-(x)}{\e^{2\pi\ri x}-1}\right]+\int^{\infty}_{a+d}f(x)\,\rd x\notag\\
	={}&\sum_{k\notin\mathrm{BC}}f(k)+\sum_{k\in\mathrm{BC}}\frac{f^+(k)+f^-(k)}{2}
	+\frac{f^+(a)+f^-(a)}{4}\notag\\
	&-\ri\int^{\infty}_{0^+}\rd t\,
	\frac{f^+(m+\ri t)-f^-(m-\ri t)}{\e^{2\pi t}-1}.
	\label{eq:ap-mv-int}
\end{align}
This is the GAPF-MV for a sum over integers; For $d=0$, so that $f^+=f^-$ and the branch cut is absent, Eq.~\eqref{eq:ap-mv-int} reduces to the ordinary Abel--Plana formula, as required.

For a sum over half-odd integers, we keep the same endpoints and choose
\begin{equation}
	g(z)=\ri \tan(\pi z)\,f(z),
	\label{eq:ap-mv-ghalf}
\end{equation}
so that $g(z)$ has simple poles at $z=k+\tfrac{1}{2}$. We focus on Case~(5) of Sec.~\ref{sec:regularization}, for which $0<d<\tfrac{1}{2}$; other values of $d$ are treated identically. Here
\begin{align}
	\pi\ri\sum_{z_i\notin\mathrm{BC}}\mathrm{Res}\,g(z_i)
	=\sum_{k+\frac{1}{2}\notin\mathrm{BC}}f\!\left(k+\tfrac{1}{2}\right),
\end{align}
and
\begin{equation}
	\begin{aligned}
		& \quad  \frac{\pi\ri}{2}\sum_{z_i\in\mathrm{BC}} \left[\mathrm{Res}\,g^+(z^+_i)+\mathrm{Res}\,g^-(z^-_i)\right]\\
		={}& \frac12 \sum_{k+\frac{1}{2}\in\mathrm{BC}}
		\left[f^+(k+\tfrac{1}{2})+f^-(k+\tfrac{1}{2})\right].
	\end{aligned}
	\label{eq:ap-mv-half-res}
\end{equation}
Applying Eq.~\eqref{eq:tan}, the vertical integral becomes
\begin{equation}
	\begin{aligned}
		& \frac{1}{2}\int^{a+\ri \infty}_{a-\ri \infty}\rd z\,[g+\sigma f] \\
		={}&\int^{m+\ri \infty}_{m}\rd z\,\frac{f^+(z)}{1+\e^{-2\pi\ri z}} 
		-\int^{m}_{m-\ri \infty}\rd z\,\frac{f^-(z)}{\e^{2\pi\ri z}+1} \\
		={}&\ri\int^{\infty}_{0}\rd t\,
		\frac{f^+(m+\ri t)-f^-(m-\ri t)}{\e^{2\pi t}+1}.
	\end{aligned}
	\label{eq:ap-mv-half-vert}
\end{equation}
Substituting Eqs.~\eqref{eq:ap-mv-half-res}--\eqref{eq:ap-mv-half-vert} into Eq.~\eqref{eq:ap-mv-thm},
\begin{align}
	&\fint^{a+d}_{a}\rd x\!\left[\frac{f^+(x)}{1+\e^{-2\pi\ri x}}
	+\frac{f^-(x)}{\e^{2\pi\ri x}+1}\right]+\int^{\infty}_{a+d}f(x)\,\rd x\notag\\
	={}&\sum_{k+\frac{1}{2}\notin\mathrm{BC}}f\!\left(k+\tfrac{1}{2}\right)
	+\sum_{k+\frac{1}{2}\in\mathrm{BC}}
	\frac{f^+(k+\tfrac{1}{2})+f^-(k+\tfrac{1}{2})}{2}\notag\\
	&+\ri\int^{\infty}_{0}\rd t\,
	\frac{f^+(a+\ri t)-f^-(a-\ri t)}{\e^{2\pi t}+1}.
	\label{eq:ap-mv-half}
\end{align}
This is the GAPF-MV for a sum over half-odd integers; for $d=0$ it reduces to the ordinary Abel--Plana formula for half-odd integers, as expected.

Finally, we apply Eq.~\eqref{eq:ap-mv-half} to Case~(5) of the main text, in which the branch cut has crossed $x=a$, with $d=|\beta|-|q|<\tfrac{1}{2}$. Here $f(z)=z\sqrt{z^2-|\beta|^2}$, so that for $x<|\beta|$,
\begin{equation}
	f^\pm(x)=\pm\ri x\sqrt{|\beta|^2-x^2}.
	\label{eq:ap-mv-fpm}
\end{equation}
Substituting Eq.~\eqref{eq:ap-mv-fpm} into Eq.~\eqref{eq:ap-mv-half} with $a=|q|$ yields
\begin{equation}
	\begin{aligned}
		& \sum_{k\geq|q|}f\!\left(k+\tfrac{1}{2}\right) \\
		={}&\int^{\infty}_{|\beta|}f(x)\,\rd x
		-\fint^{|\beta|}_{|q|}\rd x\,x\sqrt{|\beta|^2-x^2}\,\tan(\pi x) +I_{\mathbb{Z}+\frac{1}{2}},
	\end{aligned}
	\label{eq:ap-mv-final}
\end{equation}
with
\begin{equation}
	I_{\mathbb{Z}+\frac{1}{2}}
	=-\ri\int^{\infty}_{0}\rd t\,
	\frac{f(|q|+\ri t)-f(|q|-\ri t)}{\e^{2\pi t}+1},
\end{equation}
as defined in Eq.~\eqref{eq:IZhalf}.

\section{Contour integrals for $I_{\mathbb{Z}}$}
\label{app:IZ}

We follow the contour-integration method described in Appendix B of Ref.~\cite{HerdeiroRibeiroSampaio2008}. Setting $z=\ri t$, we rewrite $I_{\mathbb{Z}}$ as
\begin{equation}
	I_{\mathbb{Z}}=
	-\int_{C_1} \rd z\,\frac{f(z+\delta)}{1-\e^{-2\pi \ri z}}
	+
	\int_{C_2} \rd z\,\frac{f(z+\delta)}{\e^{2\pi \ri z}-1}.
	\label{eq:I0C1C2}
\end{equation}
The contours $C_1$ and $C_2$ run along $z=\pm \ri t$ with $t\in(0,\infty)$ and are then deformed into closed contours enclosing regions in which the integrands are analytic. For brevity, we suppress the analytic regulator (or damping factor) used to justify the contour closure. It is understood that the regulator is retained during the contour deformation so that the regulated integrands vanish on the arcs at infinity, and is removed only after the finite contour contributions have been combined.

The denominators in Eq.~\eqref{eq:I0C1C2} vanish at $z\in\mathbb{Z}$. Since the branch structure of $f(z)$ depends on the sign of $\beta^2$, the cases $\beta^2\geq 0$ and $\beta^2<0$ must be treated separately.

\subsection{Case (1): $q\in\mathbb{Z}+\half$ and $\beta^2\geq 0$ }

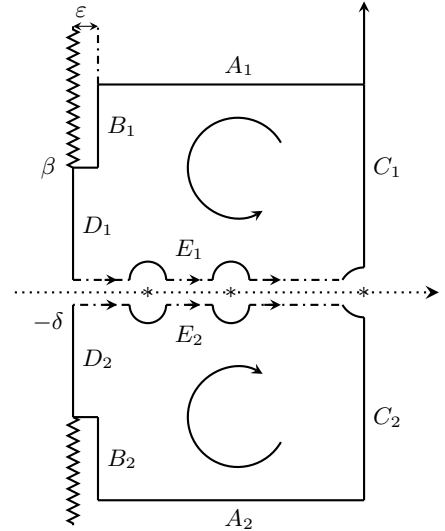
\begin{figure}[htbp]
	\centering
	\begin{tikzpicture}[scale=1.1, >=stealth, thick]
		\def\a{-3.0}
		\def\b{0.5}
		\def\H{2.5}
		\def\eps{0.15}
		\def\rc{0.22}
		\def\r0{0.3}
		\def\poleA{-2.1}
		\def\poleB{-1.1}
		
		\draw (\a+\r0, \H) -- (\a+\r0, 1.5);
		\draw(\a+\r0, 1.5) -- (\a, 1.5);
		\draw(\a, 1.5) -- (\a, \eps);
		\draw[dash dot](\a+\r0, \H) -- (\a+\r0, \H+0.7);
		\draw[<->,thin](\a,\H+0.7) -- (\a+\r0,\H+0.7);
		\node[above] at (\a+0.1,\H+0.7) {$\varepsilon$};
		\node[right] at (\a-\r0-0.2,1.5) {$\beta$};
		\node[right] at (\a,0.8) {$D_1$};
		\node[right] at (\a+\r0,2) {$B_1$};
		\node[left] at (\a,-.35) {$-\delta$};
		
		\draw (\a+\r0, -1.5) -- (\a+\r0, -\H);
		\draw (\a, -1.5) -- (\a+\r0, -1.5);
		\draw(\a, -1.5) -- (\a, -\eps);
		\node[right] at (\a,-0.8) {$D_2$};
		\node[right] at (\a+\r0,-2) {$B_2$};
		\draw (\a+\r0, -\H) -- (\b, -\H);
		\node[below] at (-1,-\H) {$A_2$};
		
		\draw[decorate,decoration={zigzag, segment length=4pt, amplitude=2pt}] (\a, 1.5) -- (\a, \H+0.7);
		\draw[decorate,decoration={zigzag, segment length=4pt, amplitude=2pt}] (\a, -1.5) -- (\a, -\H-0.3);
		
		\draw (\b, -\H) -- (\b, -\r0);
		\draw (\b, \r0) arc (90:150:\r0);
		\draw (\b, -\r0) arc (-90:-150:\r0);
		\draw (\b, \r0) -- (\b, \H);
		\node[right] at (\b,1.5) {$C_1$};
		\node[right] at (\b,-1.5) {$C_2$};
		
		\draw[->] (\b, \H) -- (\b, \H+1);
		
		\draw (\b, \H) -- (\a+\r0, \H);
		\node[above] at (-1,\H) {$A_1$};
		\draw[dotted, thick] (\a-0.7, 0) -- (\b+0.9, 0);
		\draw[->,very thick] (\b+0.8, 0) -- (\b+0.9, 0);
		
		\draw[<-] (\poleA - \rc-0.15, \eps) -- (\poleA - \rc-0.3, \eps);
		\draw[ dash dot] (\a , \eps) -- (\poleA - \rc, \eps);
		
		\draw (\poleA - \rc, \eps) arc (180:0:\rc);
		
		\draw[->] (\poleA + \rc+0.3, \eps) -- (\poleB - \rc-0.15, \eps);
		\draw[ dash dot] (\poleA + \rc, \eps) -- (\poleB - \rc, \eps);
		\node[above] at ({(\poleA + \poleB)/2}, \eps + 0.14) {$E_1$};
		
		\draw (\poleB - \rc, \eps) arc (180:0:\rc);
		
		\draw[->](\b - \r0 -0.9, \eps) -- (\b - \r0-0.7, \eps);
		\draw[ dash dot](\poleB + \rc, \eps) -- (\b - \r0, \eps);
		
		\draw[<-] (\b - \r0-0.7, -\eps) -- (\b - \r0 -0.9, -\eps);
		\draw[ dash dot] (\b - \r0, -\eps) -- (\poleB + \rc, -\eps);
		
		\draw (\poleB + \rc, -\eps) arc (0:-180:\rc);
		
		\draw[<-] (\poleB - \rc-0.15, -\eps) -- (\poleB - \rc-0.3, -\eps);
		\draw[ dash dot] (\poleB - \rc, -\eps) -- (\poleA + \rc, -\eps);
		\node[below] at ({(\poleA + \poleB)/2}, -\eps - 0.14) {$E_2$};
		
		\draw (\poleA + \rc, -\eps) arc (0:-180:\rc);
		
		\draw [<-](\poleA - \rc -0.15, -\eps) -- (\poleA - \rc -0.3, -\eps);
		\draw [ dash dot](\poleA - \rc , -\eps) -- (\a , -\eps);
		
		\node at (\poleA, 0) {$\ast$};
		\node at (\poleB, 0) {$\ast$};
		
		\node at (\b, 0) {$\ast$};
		\draw[->] (-0.5, 1.8) arc (30:300:0.6);
		\draw[->] (-0.5, -1.8) arc (330:60:0.6);
	\end{tikzpicture}
	\caption{Integration contour for Case~(1), with
		$q\in\mathbb{Z}+\frac12$ and $\beta^2>0$. The symbol $*$ denotes the simple poles of the integrand in $I_Z$ at $z=0,-1,-2,\ldots,-\delta+1$. The paths $E_1$ and $E_2$ coincide apart from semicircular indentations around these poles. The branch cuts extend from $z=-\delta\pm \ri\beta$ to $\pm \ri\infty$. Adapted from Fig.~15 of Ref.~\cite{HerdeiroRibeiroSampaio2008}.}
	\label{fig:beta_positive_contour}
\end{figure}

We first consider $\beta^2>0$; the case $\beta=0$ follows by taking the limit $\beta\to 0$. The contour is shown in Fig.~\ref{fig:beta_positive_contour}. It is similar to the left panel of Fig.~15 in Ref.~\cite{HerdeiroRibeiroSampaio2008}, except for the paths $E_1$ and $E_2$, since zeros of the denominator need not be canceled by zeros of the numerator. In particular, there are additional simple poles, including one at $z=0$, which contribute to the contour integral. This contribution was not taken into account in Ref.~\cite{HerdeiroRibeiroSampaio2008}.

We now evaluate the contributions from the individual contour segments shown in Fig.~\ref{fig:beta_positive_contour}.

\paragraph*{(1) Paths $A_1$ and $A_2$.}
These run backward along $z=\pm \ri H+x$, with $x\in[-\delta+\varepsilon,0]$ and $\varepsilon$ infinitesimal. The exponential factor in the denominator suppresses these contributions as $H\to\infty$, and hence
\begin{equation}
	\lim_{\varepsilon\to 0}(I_{A_1}+I_{A_2})=0.
\end{equation}

\paragraph*{(2) Paths $B_1$ and $B_2$.}
These run backward along $z=\pm \ri \tau-\delta+\varepsilon$, with $\tau\in[\beta,\infty)$. Note that
\begin{equation}
	\sqrt{\beta^2-\tau^2}=\ri\sqrt{\tau^2-\beta^2}
\end{equation}
along $B_1$, whereas
\begin{equation}
	\sqrt{\beta^2-\tau^2}=-\ri\sqrt{\tau^2-\beta^2}
\end{equation}
along $B_2$. Therefore,
\begin{align}
	\lim_{\varepsilon\to 0}(I_{B_1}+I_{B_2})
	={}&
	-\ri\int_{\beta}^{\infty}\rd\tau\,
	\frac{\tau\sqrt{\tau^2-\beta^2}}{\e^{2\pi \tau}-1}
	\nonumber\\
	&
	+\ri\int_{\beta}^{\infty}\rd\tau\,
	\frac{\tau\sqrt{\tau^2-\beta^2}}{\e^{2\pi \tau}-1}
	=0.
\end{align}

\paragraph*{(3) Paths $D_1$ and $D_2$.}
These run backward along $z=-\delta\pm \ri \tau$, with $\tau\in[\epsilon,\beta]$. One finds
\begin{align}
	\lim_{\epsilon\to 0}(I_{D_1}+I_{D_2})
	={}&
	\int_{0^+}^{\beta}\rd\tau\,
	\frac{\tau\sqrt{\beta^2-\tau^2}}{\e^{2\pi \tau}-1}
	+
	\int_{0^+}^{\beta}\rd\tau\,
	\frac{\tau\sqrt{\beta^2-\tau^2}}{\e^{2\pi \tau}-1}
	\nonumber\\
	={}&
	2\beta^3\int_{0^+}^1\rd\eta\,
	\frac{\eta\sqrt{1-\eta^2}}{\e^{2\pi \beta\eta}-1}.
\end{align}

\paragraph*{(4) Paths $E_1$ and $E_2$.}
These paths run along $z=-\delta+t\pm\ri\epsilon$, with $t\in[0,\delta)$. Since $z=-1,-2,\dots,-\delta+1$ are poles of the integrand, $E_1$ and $E_2$ must be deformed into small semicircles of radius $\iota$ around the poles as they pass along $(-\delta,0)$. Their combined contribution is
\begin{equation}
	\lim_{\iota\to 0}(I_{E_1}+I_{E_2})
	=
	-\int_0^\delta \rd x\,f(x)
	+\sum_{i=1}^{\delta-1} f(\delta-i).
\end{equation}

\paragraph*{(5) Pole contribution at the origin.}
Integrating along $z=\iota \e^{\pm \ri\theta}$, with $\theta\in[\pi,\pi/2]$ and $\iota$ infinitesimal, gives
\begin{equation}
	\lim_{\iota\to 0}(I_{P_1}+I_{P_2})=\frac{f(\delta)}{2}.
\end{equation}

Applying Cauchy's theorem and summing all contributions, we obtain
\begin{equation}
	\begin{aligned}
		I_{\mathbb{Z}}
		={}&
		-\frac{f(\delta)}{2}
		+\int_0^\delta \rd x\,f(x)
		-\sum_{i=1}^{\delta-1}f(\delta-i)
		\\
		&
		-2\beta^3\int_{0^+}^1 \rd\eta\,
		\frac{\eta\sqrt{1-\eta^2}}{\e^{2\pi \beta\eta}-1}.
	\end{aligned}
\end{equation}

\subsection{Case (2): $q\in\mathbb{Z}+\half$ and $-\delta^2 < \beta^2<0$}

\begin{figure}[htbp]
	\centering
	\begin{tikzpicture}[scale=1.1, >=stealth, thick]
		\def\a{-3.0}
		\def\b{1.3}  
		\def\H{2.5}  
		\def\eps{0.18}
		\def\rc{0.22}
		\def\r0{0.3}
		\def\poleA{-1.5}
		
		\def\zg{2.2}
		\draw (\a+\r0, \H) -- (\a+\r0, \eps);
		
		\draw[dash dot](\a, 0) -- (\a, \H+0.7);
		\draw[dash dot](\a+\r0, \H) -- (\a+\r0, \H+0.7);
		\draw[<->,thin](\a,\H+0.7) -- (\a+\r0,\H+0.7);
		
		\node[above] at (\a+0.1,\H+0.7) {$\varepsilon$};
		
		\node[right] at (\a+\r0+0.2,0.5) {$D_1$};
		\node[right] at (\a+\r0,1.5) {$B_1$};
		\node[left] at (\a,-0.35) {$-\delta$};
		
		\draw (\a+\r0, -\eps) -- (\a+\r0, -\H);
		
		\draw[dash dot](\a, 0) -- (\a, -\H-0.3);
		\node[right] at (\a+\r0+0.2,-0.5) {$D_2$};
		\node[right] at (\a+\r0,-1.5) {$B_2$};
		\draw (\a+\r0, -\H) -- (\b, -\H);
		\node[below] at (-1,-\H) {$A_2$};
		
		\draw (\b, -\H) -- (\b, -\r0);
		
		\draw (\b, -\r0) arc (-90:-270:\r0);
		\draw (\b, \r0) -- (\b, \H);
		\node[right] at (\b,1.5) {$C_1$};
		\node[right] at (\b,-1.5) {$C_2$};
		
		\draw[->] (\b, \H) -- (\b, \H+1);
		
		\draw (\b, \H) -- (\a+\r0, \H);
		\node[above] at (-1,\H) {$A_1$};
		
		\draw[dotted, thick] (\a+\zg, 0) -- (\b+0.9, 0);
		\draw[dotted, thick] (\a-\zg, 0) -- (\a-\zg-0.5, 0);
		\draw[->,very thick] (\b+0.8, 0) -- (\b+0.9, 0);
		\draw (\poleA+1.7,0) circle (\rc);
		
		\node [below]at (\a-\zg,-0.1) {$-\delta \!-\!\left\lvert \beta \right\rvert $};
		\node at (\poleA+1.7, 0) {$\ast$};
		
		\node at (\a+\zg+0.2,-\eps-0.55) {$-\delta \!+\! \left\lvert \beta \right\rvert $};
		\draw[->](\a+\zg+0.2,-\eps-0.40)--(\a+\zg,-\eps);
		
		\draw[->,ultra thick](\a+\zg+0.25, 0) -- (\a+\zg+0.45,0);
		
		\draw[decorate,decoration={zigzag, segment length=4pt, amplitude=2pt}] (\a-\zg, 0) -- (\a+\zg, 0);
		
		\draw[<-] (\poleA - \rc-0.15, \eps) -- (\poleA - \rc-0.3, \eps);
		\draw[ dash dot] (\a +\r0, \eps) -- (\poleA - \rc, \eps);
		\draw[ dash dot] (\poleA + \rc, \eps) -- (\a+\zg, \eps);
		\node at (\poleA+1.1, 0.43) {$E_1$};
		\node at (\poleA+1.7, 0.43) {$E_3$};
		\draw (\poleA - \rc, \eps) arc (180:0:\rc);
		
		\draw (\a+\zg, \eps) -- (\a+\zg, -\eps);
		
		\node at (\poleA+1.1, -0.43) {$E_2$};
		
		\node at (\poleA+1.7, -0.43) {$E_4$};
		\draw (\poleA + \rc, -\eps) arc (0:-180:\rc);
		\draw[ dash dot] (\poleA + \rc, -\eps) -- (\a+\zg, -\eps);
		
		\draw [<-](\poleA - \rc -0.15, -\eps) -- (\poleA - \rc -0.45, -\eps);
		\draw [ dash dot](\poleA - \rc , -\eps) -- (\a+\r0 , -\eps);
		
		\node at (\poleA, 0) {$\ast$};
		
		\node at (\b, 0) {$\ast$};
		\draw[->] (-0.2, 1.8) arc (30:300:0.6);
		\draw[->] (-0.2, -1.8) arc (330:60:0.6);
	\end{tikzpicture}
	\caption{Integration contour for Case~(2), with $q\in\mathbb{Z}+\frac12$ and $-\delta^2<\beta^2<0$. The branch cut on the real axis extends from $-\delta-|\beta|$ to $-\delta+|\beta|$. The symbol $*$ denotes simple poles of the integrand; in particular,		$z=0,-1,\ldots,-\mathcal{N}$ lie between the right branch point and the origin, where $\mathcal{N}$ is the largest integer less than $\delta-|\beta|$. The paths are indented around any poles lying on the branch cut or real axis.}
	\label{fig:beta_negative_contour}
\end{figure}
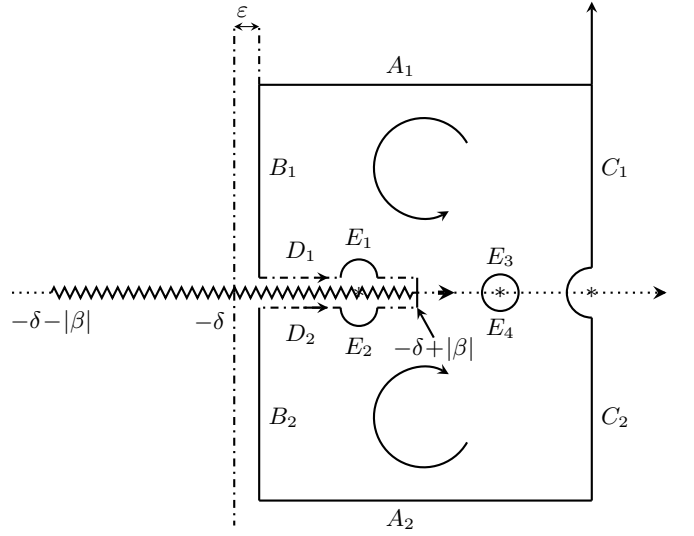

We next consider
$
	-\delta^2 < \beta^2=-|\beta|^2<0.
$
In this regime,
\begin{equation}
	f(z+\delta)=(z+\delta)\left[(z+\delta)^2-|\beta|^2\right]^{1/2},
\end{equation}
and the appropriate contour is shown in Fig.~\ref{fig:beta_negative_contour}. This contour is similar to that used in Ref.~\cite{HerdeiroRibeiroSampaio2008}, except that the paths $E_1$ and $E_2$ must be deformed around these poles of the integrand along the interval $(-\delta+|\beta|,0)$.

We again evaluate the contributions from the individual contour segments shown in Fig.~\ref{fig:beta_negative_contour}.
\paragraph*{(1) Paths $A_1$ and $A_2$.}
These contributions are the same as in the case $\beta^2\geq 0$ and therefore vanish.

\paragraph*{(2) Paths $B_1$ and $B_2$.}
These run backward along $z=\pm \ri \tau-\delta+\varepsilon$, with $\tau\in[0,\infty)$. Since
\begin{equation}
	\sqrt{-\tau^2-|\beta|^2}=\ri\sqrt{\tau^2+|\beta|^2}
\end{equation}
along $B_1$, whereas
\begin{equation}
	\sqrt{-\tau^2-|\beta|^2}=-\ri\sqrt{\tau^2+|\beta|^2}
\end{equation}
along $B_2$, one obtains
\begin{equation}
	\begin{aligned}
		\lim_{\varepsilon\to 0}(I_{B_1}+I_{B_2})
		={}&0.
	\end{aligned}
\end{equation}

\paragraph*{(3) Paths $D_1$ and $D_2$.}
These run along $z=-\delta+t\pm \ri\epsilon$, with $t\in(0,|\beta|]$, and must be deformed around the poles at $t\in\mathbb{Z}$. One finds
\begin{equation}
	\begin{aligned}
		& \lim_{\epsilon\to 0}(I_{D_1}+I_{D_2})\\
		={}&
		\ri\fint_0^{|\beta|}\rd t\,
		\frac{t\sqrt{|\beta|^2-t^2}}{\e^{-2\pi \ri t}-1}
		-\ri\fint_0^{|\beta|}\rd t\,
		\frac{t\sqrt{|\beta|^2-t^2}}{\e^{2\pi \ri t}-1}
		\\
		&
		+\lim_{\epsilon\to 0}(I_{\mathrm{sc}^+}+I_{\mathrm{sc}^-})
		\\
		={}&
		-|\beta|^3\fint_0^1 \rd\eta\,
		\eta\sqrt{1-\eta^2}\,
		\cot(\pi |\beta|\eta),
	\end{aligned}
\end{equation}
where we used the fact that the upper-semicircle contribution $I_{\mathrm{sc}^+}$ cancels the lower one $I_{\mathrm{sc}^-}$. If the branch-cut endpoint $-\delta+|\beta|$ coincides with a kernel pole, quarter-circle indentations replace the semicircular ones. The energy-contour integrand then behaves locally as $(z-z_*)^{-1/2}$, so the indentation contributions vanish as the square root of their radius. The same endpoint argument applies to Case~(4). This does not separately regularize the differentiated pressure integrands: their exceptional values are defined by the combined limit, or equivalently by the smooth spectral representation in Appendix~\ref{app:numerical-evaluation}, Eq.~\eqref{eq:smooth-spectral-pressure}.

\paragraph*{(4) Paths $E_1$ and $E_2$.}
These run along $z=-\delta+t$, with $t\in[|\beta|,\delta)$. Since $z=-1,-2,\cdots,-\mathcal{N}$ are poles of the integrand, the paths $E_1$ and $E_2$ must be deformed into small semicircles around these poles along $(-\delta+|\beta|,0)$, as shown in Fig.~\ref{fig:beta_negative_contour}. Their contribution is
\begin{equation}
	\lim_{\iota\to 0}(I_{E_1}+I_{E_2})
	=
	-\int_{|\beta|}^{\delta}\rd x\,f(x)
	+\sum_{i=1}^{\mathcal{N}}f(\delta-i).
\end{equation}

\paragraph*{(5) Pole contribution at the origin.}
This is the same as in the case $\beta^2\geq 0$:
\begin{equation}
	\lim_{\iota\to 0}(I_{P_1}+I_{P_2})=\frac{f(\delta)}{2}.
\end{equation}

Applying Cauchy's theorem and collecting all contributions, we obtain
\begin{equation}
	\begin{aligned}
		I_{\mathbb{Z}}
		={}&
		-\frac{f(\delta)}{2}
		+\int_{|\beta|}^{\delta}\rd x\,f(x)
		-\sum_{i=1}^{\mathcal{N}}f(\delta-i)
		\\
		&
		+|\beta|^3\fint_0^1 \rd\eta\,
		\eta\sqrt{1-\eta^2}\,
		\cot(\pi |\beta|\eta).
	\end{aligned}
\end{equation}

\section{Contour integrals for $I_{\mathbb{Z}+\half}$}
\label{app:IZhalf}

Introducing again the variable $z=\ri t$, the integral $I_{\mathbb{Z}+\half}$ can be written in contour form as
\begin{equation}
	I_{\mathbb{Z}+\half}=
	-\int_{C_1} \rd z\,\frac{f(z+|q|)}{1+\e^{-2\pi \ri z}}
	-\int_{C_2} \rd z\,\frac{f(z+|q|)}{\e^{2\pi \ri z}+1}.
	\label{eq:I_integer}
\end{equation}
The contours $C_1$ and $C_2$ run along $z=\pm \ri t$ with $t\in[0,\infty)$. Equation~\eqref{eq:I_integer} is analogous to Eq.~\eqref{eq:I0C1C2}, except that $\delta$ is replaced by $|q|$, the poles on the real axis now occur at
\begin{equation}
	z=n+\half,
	\qquad n\in\mathbb{Z},
\end{equation}
and there is no pole at the origin.

\subsection{Case (3): $q\in\mathbb{Z}$ and $\beta^2\geq 0$}

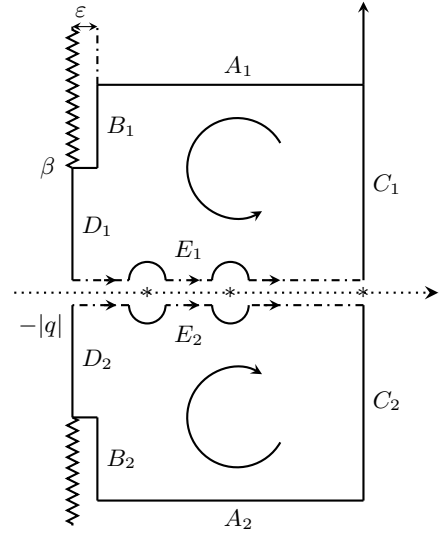
\begin{figure}[htbp]
	\centering
	\begin{tikzpicture}[scale=1.1, >=stealth, thick]
		\def\a{-3.0}
		\def\b{0.5}
		\def\H{2.5}   
		\def\eps{0.15} 
		\def\rc{0.22}  
		\def\r0{0.3} 
		\def\poleA{-2.1}
		\def\poleB{-1.1}
		
		\draw (\a+\r0, \H) -- (\a+\r0, 1.5);
		\draw(\a+\r0, 1.5) -- (\a, 1.5);
		\draw(\a, 1.5) -- (\a, \eps);
		\draw[dash dot](\a+\r0, \H) -- (\a+\r0, \H+0.7);
		\draw[<->,thin](\a,\H+0.7) -- (\a+\r0,\H+0.7);
		\node[above] at (\a+0.1,\H+0.7) {$\varepsilon$};
		\node[right] at (\a-0.5,1.5) {$\beta$};
		\node[right] at (\a,0.8) {$D_1$};
		\node[right] at (\a+\r0,2) {$B_1$};
		\node[left] at (\a,-0.4) {$-\!\left\lvert q\right\rvert $};
		
		\draw (\a+\r0, -1.5) -- (\a+\r0, -\H);
		\draw (\a, -1.5) -- (\a+\r0, -1.5);
		\draw(\a, -1.5) -- (\a, -\eps);
		\node[right] at (\a,-0.8) {$D_2$};
		\node[right] at (\a+\r0,-2) {$B_2$};
		\draw (\a+\r0, -\H) -- (\b, -\H);
		\node[below] at (-1,-\H) {$A_2$};
		
		\draw (\b, -\H) -- (\b, -\eps);
		
		\draw (\b, \eps) -- (\b, \H);
		\node[right] at (\b,1.3) {$C_1$};
		\node[right] at (\b,-1.3) {$C_2$};
		
		\draw[->] (\b, \H) -- (\b, \H+1);
		
		\draw (\b, \H) -- (\a+\r0, \H);
		\node[above] at (-1,\H) {$A_1$};
		\draw[dotted, thick] (\a-0.7, 0) -- (\b+0.9, 0);
		\draw[->,very thick] (\b+0.8, 0) -- (\b+0.9, 0);
		
		\draw[decorate,decoration={zigzag, segment length=4pt, amplitude=2pt}] (\a, 1.5) -- (\a, \H+0.7);
		\draw[decorate,decoration={zigzag, segment length=4pt, amplitude=2pt}] (\a, -1.5) -- (\a, -\H-0.3);
		
		\draw[<-] (\poleA - \rc-0.15, \eps) -- (\poleA - \rc-0.3, \eps);
		\draw[ dash dot] (\a , \eps) -- (\poleA - \rc, \eps);

		\draw (\poleA - \rc, \eps) arc (180:0:\rc);
		
		\draw[->] (\poleA + \rc+0.3, \eps) -- (\poleB - \rc-0.15, \eps);
		\draw[ dash dot] (\poleA + \rc, \eps) -- (\poleB - \rc, \eps);
		\node[above] at ({(\poleA + \poleB)/2}, \eps + 0.14) {$E_1$};
		
		\draw (\poleB - \rc, \eps) arc (180:0:\rc);
		
		\draw[->](\b - \r0 -0.9, \eps) -- (\b - \r0-0.7, \eps);
		\draw[ dash dot](\poleB + \rc, \eps) -- (\b , \eps);
		
		\draw[<-] (\b - \r0-0.7, -\eps) -- (\b - \r0 -0.9, -\eps);
		\draw[ dash dot] (\b , -\eps) -- (\poleB + \rc, -\eps);
		
		\draw (\poleB + \rc, -\eps) arc (0:-180:\rc);
		
		\draw[<-] (\poleB - \rc-0.15, -\eps) -- (\poleB - \rc-0.3, -\eps);
		\draw[ dash dot] (\poleB - \rc, -\eps) -- (\poleA + \rc, -\eps);
		\node[below] at ({(\poleA + \poleB)/2}, -\eps - 0.14) {$E_2$};
		
		\draw (\poleA + \rc, -\eps) arc (0:-180:\rc);
		
		\draw [<-](\poleA - \rc -0.15, -\eps) -- (\poleA - \rc -0.3, -\eps);
		\draw [ dash dot](\poleA - \rc , -\eps) -- (\a , -\eps);
		
		\node at (\poleA, 0) {$\ast$};
		\node at (\poleB, 0) {$\ast$};
		
		\node at (\b, 0) {$\ast$};
		\draw[->] (-0.5, 1.8) arc (30:300:0.6);
		\draw[->] (-0.5, -1.8) arc (330:60:0.6);
	\end{tikzpicture}
	\caption{Integration contour for Case~(3), with $q\in\mathbb{Z}$ and $\beta^2>0$. The symbol $*$ denotes the simple poles at $z=-\frac12,-\frac32,\ldots,-|q|+\frac12$. The branch cuts extend from $z=-|q|\pm \ri\beta$ to $\pm \ri\infty$.}
	\label{fig:integer_q_beta_positive_contour}
\end{figure}

We choose the contour shown in Fig.~\ref{fig:integer_q_beta_positive_contour}. It is similar to Fig.~\ref{fig:beta_positive_contour} except that the poles on the real axis are now located at $z=n+\frac{1}{2}$ and $\delta$ is replaced by $|q|$. We evaluate the contributions from the different contour segments separately.

\paragraph*{(1) Paths $A_1$ and $A_2$.}
These run backward along $z=\pm \ri H+x$, with $x\in[-|q|+\varepsilon,0]$ and $\varepsilon$ infinitesimal. The exponential factor in the denominator suppresses these contributions as $H\to\infty$, and thus
\begin{equation}
	\lim_{\varepsilon\to 0}(I_{A_1}+I_{A_2})=0.
\end{equation}

\paragraph*{(2) Paths $B_1$ and $B_2$.}
These run backward along $z=\pm \ri \tau-|q|+\varepsilon$, with $\tau\in[\beta,\infty)$. Similar to Case (1), we have 
\begin{equation}
	\begin{aligned}
		\lim_{\varepsilon\to 0}(I_{B_1}+I_{B_2})
		={}&
		0.
	\end{aligned}
\end{equation}

\paragraph*{(3) Paths $D_1$ and $D_2$.}
These run backward along $z=-|q|\pm \ri t$, with $t\in[\epsilon,\beta]$. One obtains
\begin{equation}
	\begin{aligned}
		&\lim_{\epsilon\to 0}(I_{D_1}+I_{D_2})\\
		={}&
		-\int_0^{\beta}\rd\tau\,
		\frac{\tau\sqrt{\beta^2-\tau^2}}{\e^{2\pi \tau}+1}
		-\int_0^{\beta}\rd\tau\,
		\frac{\tau\sqrt{\beta^2-\tau^2}}{\e^{2\pi \tau}+1}
		\\
		={}&
		-2\beta^3\int_0^1\rd\eta\,
		\frac{\eta\sqrt{1-\eta^2}}{\e^{2\pi \beta\eta}+1}.
	\end{aligned}
\end{equation}

\paragraph*{(4) Paths $E_1$ and $E_2$.}
These run along $z=-|q|+t$, with $t\in[0,|q|]$. The points
\begin{equation}
	z=-|q|+\frac12,\,-|q|+1+\frac12,\dots,-\frac12
\end{equation}
are poles of the integrand, and the paths $E_1$ and $E_2$ must therefore be deformed into small semicircles around them as they pass along $[-|q|,0]$. Their contribution is
\begin{equation}
	\lim_{\iota\to 0}(I_{E_1}+I_{E_2})
	=
	-\int_0^{|q|}\rd x\,f(x)
	+\sum_{i=1}^{|q|}f\left(|q|-i+\half\right).
\end{equation}

Applying Cauchy's theorem and summing all contributions, we obtain
\begin{equation}
	\begin{aligned}
		I_{\mathbb{Z}+\half}
		={}&
		\int_0^{|q|}\rd x\,f(x)
		-\sum_{i=1}^{|q|}f\left(|q|-i+\frac12\right)
		\\
		&
		+2\beta^3\int_0^1\rd\eta\,
		\frac{\eta\sqrt{1-\eta^2}}{\e^{2\pi \beta\eta}+1}.
	\end{aligned}
\end{equation}

\subsection{Case (4): $q\in\mathbb{Z}$ and $-|q|^2\leq\beta^2<0$}

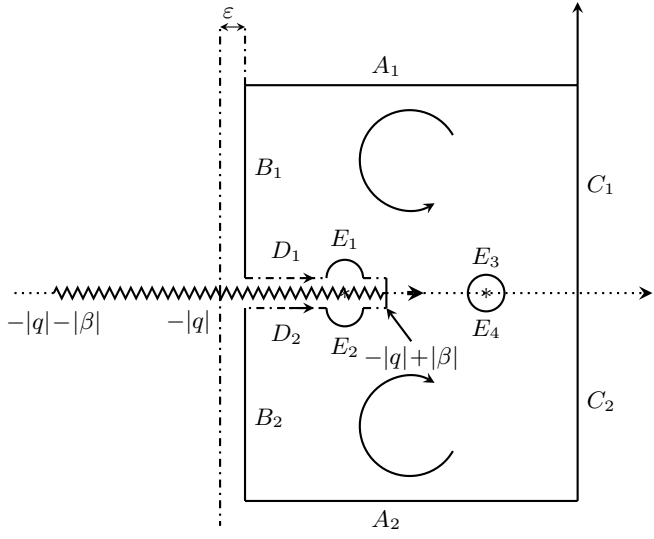
\begin{figure}[htbp]
	\centering
	\begin{tikzpicture}[scale=1.1, >=stealth, thick]
		\def\a{-3.0} 
		\def\b{1.3}  
		\def\H{2.5} 
		\def\eps{0.18} 
		\def\rc{0.22}  
		\def\r0{0.3}  
		\def\poleA{-1.5}
		
		\def\zg{2.}
		\draw (\a+\r0, \H) -- (\a+\r0, \eps);
		
		\draw[dash dot](\a, 0) -- (\a, \H+0.7);
		\draw[dash dot](\a+\r0, \H) -- (\a+\r0, \H+0.7);
		\draw[<->,thin](\a,\H+0.7) -- (\a+\r0,\H+0.7);
		
		\node[above] at (\a+0.1,\H+0.7) {$\varepsilon$};
		
		\node[right] at (\a+\r0+0.2,0.5) {$D_1$};
		\node[right] at (\a+\r0,1.5) {$B_1$};
		\node[left] at (\a,-0.35) {$-\!\left\lvert q\right\rvert $};
		
		\draw (\a+\r0, -\eps) -- (\a+\r0, -\H);
		
		\draw[dash dot](\a, 0) -- (\a, -\H-0.3);
		\node[right] at (\a+\r0+0.2,-0.5) {$D_2$};
		\node[right] at (\a+\r0,-1.5) {$B_2$};
		\draw (\a+\r0, -\H) -- (\b, -\H);
		\node[below] at (-1,-\H) {$A_2$};
		
		\draw (\b, -\H) -- (\b, 0);
		
		\draw (\b, 0) -- (\b, \H);
		\node[right] at (\b,1.3) {$C_1$};
		\node[right] at (\b,-1.3) {$C_2$};
		
		\draw[->] (\b, \H) -- (\b, \H+1);
		
		\draw (\b, \H) -- (\a+\r0, \H);
		\node[above] at (-1,\H) {$A_1$};
		\draw[dotted, thick] (\a+\zg, 0) -- (\b+0.9, 0);
		\draw[dotted, thick] (\a-\zg, 0) -- (\a-\zg-0.5, 0);
		\draw[->,very thick] (\b+0.8, 0) -- (\b+0.9, 0);
		\draw (\poleA+1.7,0) circle (\rc);
		
		\node[below] at (\a-\zg,-0.1) {$-\!\left\lvert q\right\rvert \!-\! \left\lvert \beta \right\rvert $};
		\node at (\poleA+1.7, 0) {$\ast$};
		
		\node at (\a+\zg+0.3,-\eps-0.6) {$-\!\left\lvert q\right\rvert \!+\! \left\lvert \beta \right\rvert $};
		\draw[->](\a+\zg+0.3,-\eps-0.4)--(\a+\zg,-\eps);
		\draw[->,ultra thick](\a+\zg+0.25, 0) -- (\a+\zg+0.45,0);
		
		\draw[decorate,decoration={zigzag, segment length=4pt, amplitude=2pt}] (\a-\zg, 0) -- (\a+\zg, 0);
		
		\draw[<-] (\poleA - \rc-0.15, \eps) -- (\poleA - \rc-0.3, \eps);
		\draw[ dash dot] (\a +\r0, \eps) -- (\poleA - \rc, \eps);
		\draw[ dash dot] (\poleA + \rc, \eps) -- (\a+\zg, \eps);
		\node at (\poleA+1, 0.43) {$E_1$};
		\node at (\poleA+1.7, 0.43) {$E_3$};
		\draw (\poleA - \rc, \eps) arc (180:0:\rc);
		
		\draw (\a+\zg, \eps) -- (\a+\zg, -\eps);
		
		\node at (\poleA+1, -0.43) {$E_2$};
		
		\node at (\poleA+1.7, -0.43) {$E_4$};
		\draw (\poleA + \rc, -\eps) arc (0:-180:\rc);
		\draw[ dash dot] (\poleA + \rc, -\eps) -- (\a+\zg, -\eps);
		
		\draw [<-](\poleA - \rc -0.15, -\eps) -- (\poleA - \rc -0.45, -\eps);
		\draw [ dash dot](\poleA - \rc , -\eps) -- (\a+\r0 , -\eps);
		
		\node at (\poleA, 0) {$\ast$};
		
		\draw[->] (-0.2, 1.9) arc (30:300:0.6);
		\draw[->] (-0.2, -1.9) arc (330:60:0.6);
	\end{tikzpicture}
	\caption{Integration contour for Case~(4), with $q\in\mathbb{Z}$ and $-|q|^2<\beta^2<0$, so that $|\beta|<|q|$. The real-axis branch cut extends from $-|q|-|\beta|$ to $-|q|+|\beta|$. The symbol $*$ denotes the simple poles $z = -\mathcal{N} + \frac12, - \mathcal{N} + \frac32, \ldots, - \frac12$, with $\mathcal{N}$ defined in Eq.~\eqref{eq:N}.}
	\label{fig:integer_q_beta_negative_contour}
\end{figure}

In this case we use the contour shown in Fig.~\ref{fig:integer_q_beta_negative_contour}, which is the analog of Fig.~\ref{fig:beta_negative_contour} with poles at $z=n+\frac12$ and with $\delta$ replaced by $|q|$. We assume $-|q|^2\leq \beta^2<0$, i.e.\ $|\beta|\leq |q|$, and again evaluate the contour contributions separately.

\paragraph*{(1) Paths $A_1$ and $A_2$.}
These contributions are the same as in the case $\beta^2\geq 0$ and therefore vanish.

\paragraph*{(2) Paths $B_1$ and $B_2$.}
These run backward along $z=\pm \ri \tau-|q|+\varepsilon$, with $\tau\in[0,\infty)$. The same as in Case (2), one finds
\begin{equation}
	\begin{aligned}
		\lim_{\varepsilon\to 0}(I_{B_1}+I_{B_2})
		={}&
		0.
	\end{aligned}
\end{equation}

\paragraph*{(3) Paths $D_1$ and $D_2$.}
These run along $z=-|q|+t\pm \ri\epsilon$, with $t\in[\epsilon,|\beta|]$. One obtains
\begin{equation}
	\begin{aligned}
		&\lim_{\epsilon\to 0}(I_{D_1}+I_{D_2})
		\\
		={}&
		-\ri\fint_0^{|\beta|}\rd t\,
		\frac{t\sqrt{|\beta|^2-t^2}}{1+\e^{-2\pi \ri t}}
		+\ri\fint_0^{|\beta|}\rd t\,
		\frac{t\sqrt{|\beta|^2-t^2}}{\e^{2\pi \ri t}+1}
		\\
		={}&
		|\beta|^3\fint_0^1\rd\eta\,
		\eta\sqrt{1-\eta^2}\,
		\tan(\pi |\beta|\eta).
	\end{aligned}
\end{equation}

\paragraph*{(4) Paths $E_1$ and $E_2$.}
These run along $z=-|q|+t$, with $t\in[|\beta|,|q|)$. The poles are located at
\begin{equation}
	\txt z=-\mathcal{N} +\frac12,
	-\mathcal{N}+\frac32,
	\cdots,
	-\frac12,
\end{equation}
where $\mathcal{N}$ is defined in Eq.~\eqref{eq:N}. The paths $E_1$ and $E_2$ must therefore be deformed into small semicircles around these poles along $(-|q|+|\beta|,0)$, as shown in Fig.~\ref{fig:integer_q_beta_negative_contour}. Their contribution is
\begin{equation}
	\lim_{\iota\to 0}(I_{E_1}+I_{E_2})
	=
	-\int_{|\beta|}^{|q|}\rd x\,f(x)
	+\sum_{i=1}^{\mathcal{N}}f(\delta-i).
\end{equation}

Applying Cauchy's theorem and collecting all contributions, we obtain
\begin{align}
	I_{\mathbb{Z}+\half}
	={}&
	\int_{|\beta|}^{|q|}\rd x\,f(x)
	-\sum_{i=1}^{\mathcal{N}}f(\delta-i)
	\nonumber\\
	&
	-|\beta|^3\fint_0^1\rd\eta\,
	\eta\sqrt{1-\eta^2}\tan(\pi |\beta|\eta).
\end{align}

\subsection{Case (5): $q\in\mathbb{Z}$ and $-\delta^2 < \beta^2<-|q|^2$}

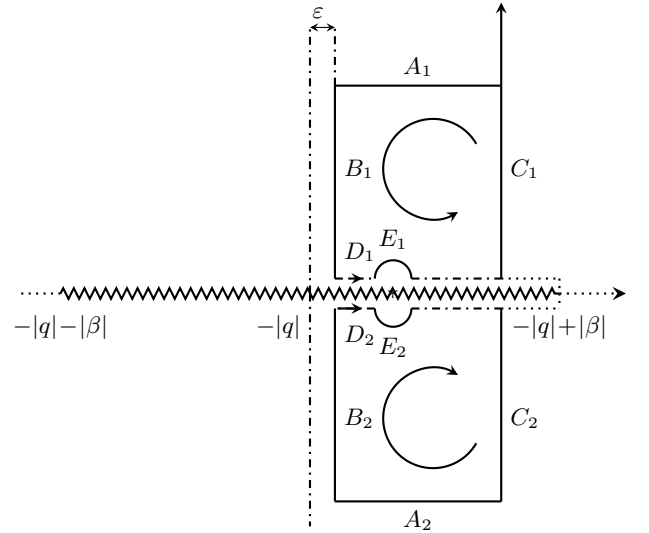
\begin{figure}[htbp]
	\centering
	\begin{tikzpicture}[scale=1.1, >=stealth, thick]
		\def\a{-2.0}  
		\def\b{0.3}  
		\def\H{2.5} 
		\def\eps{0.18}  
		\def\rc{0.22}
		\def\r0{0.3}   
		\def\poleA{-1}
		
		\def\zg{3}
		\draw (\a+\r0, \H) -- (\a+\r0, \eps);
		
		\draw[dash dot](\a, 0) -- (\a, \H+0.7);
		\draw[dash dot](\a+\r0, \H) -- (\a+\r0, \H+0.7);
		\draw[<->,thin](\a,\H+0.7) -- (\a+\r0,\H+0.7);
		
		\node[above] at (\a+0.1,\H+0.7) {$\varepsilon$};
		
		\node[right] at (\a+\r0,0.5) {$D_1$};
		\node[right] at (\a+\r0,1.5) {$B_1$};
		\node[left] at (\a,-0.4) {$-\!\left\lvert q\right\rvert $};
		
		\draw (\a+\r0, -\eps) -- (\a+\r0, -\H);
		
		\draw[dash dot](\a, 0) -- (\a, -\H-0.3);
		\node[right] at (\a+\r0,-0.5) {$D_2$};
		\node[right] at (\a+\r0,-1.5) {$B_2$};
		\draw (\a+\r0, -\H) -- (\b, -\H);
		\node[below] at (-0.7,-\H) {$A_2$};
		
		\draw (\b, -\H) -- (\b, -\eps);
		
		\draw (\b, \eps) -- (\b, \H);
		\node[right] at (\b,1.5) {$C_1$};
		\node[right] at (\b,-1.5) {$C_2$};
		
		\draw[->] (\b, \H) -- (\b, \H+1);
		
		\draw (\b, \H) -- (\a+\r0, \H);
		\node[above] at (-0.7,\H) {$A_1$};
		\draw[dotted, thick] (\a+\zg, 0) -- (\b+1.4, 0);
		\draw[dotted, thick] (\a-\zg, 0) -- (\a-\zg-0.5, 0);
		\draw[->,very thick] (\b+1.4, 0) -- (\b+1.5, 0);
		
		\node[below] at (\a-\zg,-0.15) {$-\!\left\lvert q\right\rvert \!-\!\left\lvert \beta \right\rvert $};
		
		\node[below] at (\a+\zg,-0.15) {$-\!\left\lvert q\right\rvert \!+\! \left\lvert \beta \right\rvert $};
		
		\draw[decorate,decoration={zigzag, segment length=4pt, amplitude=2pt}] (\a-\zg, 0) -- (\a+\zg, 0);
		
		\draw[<-] (\poleA - \rc-0.15, \eps) -- (\poleA - \rc-0.3, \eps);
		\draw[ dash dot] (\a +\r0, \eps) -- (\poleA - \rc, \eps);
		\draw[ dash dot] (\poleA + \rc, \eps) -- (\b, \eps);
		\draw[ dotted] (\b, \eps) -- (\a+\zg, \eps);
		\node at (\poleA+0.7, 0.43) {$E_1$};
		
		\draw (\poleA - \rc, \eps) arc (180:0:\rc);
		
		\draw [dotted](\a+\zg, \eps) -- (\a+\zg, -\eps);
		
		\node at (\poleA+0.7, -0.43) {$E_2$};
		
		\draw (\poleA + \rc, -\eps) arc (0:-180:\rc);
		\draw[ dash dot] (\poleA + \rc, -\eps) -- (\b, -\eps);
		\draw[ dotted] (\b, -\eps) -- (\a+\zg, -\eps);
		
		\draw [<-](\poleA - \rc -0.15, -\eps) -- (\poleA - \rc -0.45, -\eps);
		\draw [ dash dot](\poleA - \rc , -\eps) -- (\a+\r0 , -\eps);
		
		\node at (\poleA, 0) {$\ast$};
		
		\draw[->] (-0, 1.8) arc (30:300:0.6);
		\draw[->] (-0, -1.8) arc (330:60:0.6);
	\end{tikzpicture}
	\caption{Integration contour for Case~(5), with $q\in\mathbb{Z}$ and $-\delta^2<\beta^2<-|q|^2$, so that		$|q|<|\beta|<\delta$. The real-axis branch cut extends from $-|q|-|\beta|$ to $-|q|+|\beta|$ and therefore crosses the origin.		The symbol $*$ denotes simple poles of the integrand on $(-|q|,0)$.}
	\label{fig:integer_q_beta_crossing_contour}
\end{figure}

Finally, consider
$
	-\delta^2 < \beta^2<-|q|^2,
$
for which $|\beta|>|q|$. The relevant contour is shown in Fig.~\ref{fig:integer_q_beta_crossing_contour}, where the right endpoint of the branch cut on the real axis lies at
$
	z=-|q|+|\beta|\in\left[0,\half\right).
$
In this regime $\delta\geq |\beta|>|q|$. The branch cut
$
	[-|q|-|\beta|,\,-|q|+|\beta|]
$
extends across the origin, so the contours $C_1$ and $C_2$ must be deformed around the segment from $-|q|+|\beta|$ to $0$, as shown in Fig.~\ref{fig:integer_q_beta_crossing_contour}. Equation~\eqref{eq:ap-mv-final} is then replaced by
\begin{equation}
	\begin{aligned}
		\sum_{k=|q|}^{\infty} f\left(k+\frac12\right)
		={}&
		\int_{|\beta|}^{\infty} f(x)\,\rd x + I_{\mathbb{Z}+\half}
		\\
		&
		- \fint_{|q|}^{|\beta|}\rd t\, t \sqrt{|\beta|^2-t^2}\tan(\pi t).
	\end{aligned}
\end{equation}

The contribution from the paths $D_1$ and $D_2$ is obtained by integrating along $z=-|q|+t\pm \ri\epsilon$, with $t\in[0,|q|]$:
\begin{equation}
	\begin{aligned}
		&\lim_{\epsilon\to 0}(I_{D_1}+I_{D_2})\\
		={}&
		\ri
		\fint_0^{|q|}\rd t\,
		\frac{t\sqrt{|\beta|^2-t^2}}{\e^{2\pi \ri t}+1}
		-\ri
		\fint_0^{|q|}\rd t\,
		\frac{t\sqrt{|\beta|^2-t^2}}{1+\e^{-2\pi \ri t}}
		\\
		={}&
		|\beta|^3\fint_0^{|q/\beta|}\rd\eta\,
		\eta\sqrt{1-\eta^2}\tan(\pi |\beta|\eta).
	\end{aligned}
\end{equation}
It is clear that
\begin{equation}
	\lim_{\epsilon\to 0}(I_{D_1}+I_{D_2})=0
\end{equation}
when $q=0$.

Applying Cauchy's theorem and summing all contributions, we obtain
\begin{equation}
	I_{\mathbb{Z}+\half}
	=
	-|\beta|^3\fint_0^{|q/\beta|}\rd\eta\,
	\eta\sqrt{1-\eta^2}\,
	\tan(\pi |\beta|\eta).
\end{equation}

\section{Direct Evaluation of $I_{\mathbb{Z}}$ and $I_{\mathbb{Z}+\half}$}
\label{app:directIZ}

In this appendix we present an alternative direct method for evaluating the integrals $I_{\mathbb{Z}}$ in Eq.~\eqref{eq:IZ} and $I_{\mathbb{Z}+\half}$ in Eq.~\eqref{eq:IZhalf}. This method provides a useful check on the contour-integral calculation. Although closed-form expressions are difficult to obtain in most cases, numerical evaluation confirms agreement between the two approaches.

From Eq.~\eqref{eq:fz}, one has
\begin{align}
	f(z+\delta)
	={}&
	(z+\delta)\left[(z+\delta)^2+\beta^2\right]^{1/2}
	\nonumber\\
	={}&
	(z+\delta)\sqrt{(z-z^+)(z-z^-)},
\end{align}
where the branch points are
\begin{equation}
	z^\pm=-\delta\pm\sqrt{-\beta^2}.
\end{equation}
Here
\begin{equation}
	\sqrt{-\beta^2}
	=
	\begin{cases}
		\ri\beta, & \beta^2\geq 0,\\
		|\beta|, & -\delta^2\leq \beta^2<0.
	\end{cases}
\end{equation}
Moreover,
\begin{equation}
	\sqrt{(z+\delta)^2+\beta^2}
	=
	|(z+\delta)^2+\beta^2|^{1/2}\e^{\ri \vartheta}
\end{equation}
with
$$\vartheta := 
\half\bigl[\operatorname{Arg}(z-z^+)+\operatorname{Arg}(z-z^-)\bigr].$$
For $t>0$, this gives
\begin{equation}
	\begin{aligned}
		f(\delta+\ri t)
		={}&
		(\delta+\ri t)
		\sqrt{\delta^2+\beta^2-t^2+2\ri t\delta}
		\\
		={}&
		(\delta+\ri t)\left[a_1(\delta,t)+\ri a_2(\delta,t)\right]
		\\
		={}&
		\delta a_1-t a_2
		+\ri\left(t a_1+\delta a_2\right),
		\label{eq:fdelta+it}
	\end{aligned}
\end{equation}
and
\begin{align}
	f(\delta-\ri t)
	={}&
	\delta a_1-t a_2
	-\ri\left(t a_1+\delta a_2\right).
	\label{eq:fdelta-it}
\end{align}
Here $a_1$ and $a_2$ are real nonnegative functions satisfying
\begin{equation}
	a_1^2-a_2^2=\delta^2+\beta^2-t^2,
	\qquad
	a_1a_2=t\delta.
	\label{eq:a12}
\end{equation}
Thus
\begin{equation}
	\begin{aligned}
		& a_{1,2}(\delta,t)\\
		={}&
		\left[
		\frac{
			\sqrt{(\delta^2+\beta^2-t^2)^2+4t^2\delta^2}
			\pm(\delta^2+\beta^2-t^2)}
		{2}
		\right]^{1/2}.
	\end{aligned}
\end{equation}

For half-odd-integer $q$, substituting Eqs.~\eqref{eq:fdelta+it} and \eqref{eq:fdelta-it} into Eq.~\eqref{eq:IZ} yields
\begin{equation}
	I_{\mathbb{Z}}
	=
	-2\int_{0^+}^{\infty}\rd t\,
	\frac{t\,a_1(\delta,t)+\delta\,a_2(\delta,t)}
	{\e^{2\pi t}-1}.
	\label{eq:IZdirect}
\end{equation}
Similarly, for integer $q$ one obtains
\begin{equation}
	I_{\mathbb{Z}+\half}
	=
	2\int_0^{\infty}\rd t\,
	\frac{t\,a_1(|q|,t)+|q|\,a_2(|q|,t)}
	{\e^{2\pi t}+1}.
	\label{eq:IZhalfdirect}
\end{equation}

As a simple check, consider the case without a Wu--Yang magnetic monopole, $q=0$, for which
\begin{equation}
	\beta^2=a_0^2m_0^2+2\xi-\quarter.
\end{equation}
Equation~\eqref{eq:a12} gives
\begin{equation}
	a_{1,2}(0,t)
	=
	\left[
	\frac{|\beta^2-t^2|\pm(\beta^2-t^2)}{2}
	\right]^{1/2}.
	\label{eq:a120t}
\end{equation}
For $\beta^2>0$, one has
\begin{equation}
	a_1(0,t)=\sqrt{\beta^2-t^2},\qquad a_2(0,t)=0,
	\qquad t^2<\beta^2,
\end{equation}
whereas
\begin{equation}
	a_1(0,t)=0,\qquad a_2(0,t)=\sqrt{t^2-\beta^2},
	\qquad t^2>\beta^2.
\end{equation}
Substitution into Eq.~\eqref{eq:IZhalfdirect} gives
\begin{equation}
	I_{\mathbb{Z}+\half}
	=
	2\int_0^\beta\rd t\,
	\frac{t\sqrt{\beta^2-t^2}}{\e^{2\pi t}+1}
	=
	2\beta^3\int_0^1\rd\eta\,
	\frac{\eta\sqrt{1-\eta^2}}{\e^{2\pi\beta\eta}+1},
\end{equation}
in agreement with the contour-integral result. On the other hand, if
\begin{equation}
	-\quarter < \beta^2=-|\beta|^2<0,
\end{equation}
then $a_1(0,t)=0$ and $a_2(0,t)=\sqrt{t^2-\beta^2}$, so Eq.~\eqref{eq:IZhalfdirect} gives
$
	I_{\mathbb{Z}+\half}=0.
$
This is again consistent with the result obtained in Appendix~\ref{app:IZhalf}. The remaining cases can be checked numerically and agree with the contour-integral method.

\section{Numerical evaluation near apparent singularities}
\label{app:numerical-evaluation}
When a branch endpoint reaches a pole of the cotangent or tangent kernel, or when $\mathcal N$ changes, the pressure is defined by the continuous limit of the complete expression, with the appropriate finite sum on each side. For $m_0\ne0$, an individual pressure integrand can behave as $(1-\eta)^{-3/2}$ at a coincident endpoint. It is then neither an ordinary convergent integral nor a Cauchy principal value at that endpoint. The integral and finite-sum contributions must be combined before taking the parameter limit.

An equivalent convergent spectral representation provides an unambiguous evaluation at and near these points. Let $b=\beta^2$, $x_n=n+\delta$, and let $\zeta(s,\delta)$ denote the Hurwitz zeta function. Define
\begin{equation}
	\begin{aligned}
		G_\delta(b)={}&2\sum_{n=0}^\infty \left[x_n\sqrt{x_n^2+b}-x_n^2-\frac b2\right]\\
		&+2\zeta(-2,\delta)+b\zeta(0,\delta).
	\end{aligned}
\label{eq:smooth-spectral-energy}
\end{equation}
The first two large-$n$ terms have been subtracted and restored by analytic continuation, so this is the same spectral prescription as Eq.~\eqref{eq:zeta-normalization}. Its derivative is
\begin{equation}
	G_\delta'(b)=\sum_{n=0}^\infty \left[\frac{x_n}{\sqrt{x_n^2+b}}-1\right] +\zeta(0,\delta).
\label{eq:smooth-spectral-derivative}
\end{equation}
The energy and pressure are therefore
\begin{equation}
	\begin{aligned}
		E_{\mathrm{Cas}}&=\frac{G_\delta(b)}{a_0} + \frac23 a_0^2m_0^3,\\
		P_{\mathrm{Cas}}&=\frac{G_\delta(b)-2a_0^2m_0^2G_\delta'(b)}{8\pi a_0^3} -\frac{m_0^3}{6\pi}.
	\end{aligned}
\label{eq:smooth-spectral-pressure}
\end{equation}
Both sums converge uniformly on compact subsets of $b>-\delta^2$, the strictly positive-frequency domain. Their smoothness makes explicit that an auxiliary endpoint-pole crossing is not a physical singularity. For numerical stability the summands can be written as
\begin{equation}
	\begin{aligned}
		x_n\sqrt{x_n^2+b}-x_n^2-\frac b2 
		&=-\frac{b^2}{2(\sqrt{x_n^2+b}+x_n)^2},\\
		\frac{x_n}{\sqrt{x_n^2+b}}-1 
		&=-\frac{b}{\sqrt{x_n^2+b}(\sqrt{x_n^2+b}+x_n)}.
	\end{aligned}
\end{equation}
After summing the first $N$ terms, the remaining sum in $G_\delta$ is evaluated as
\begin{equation}
	2\sum_{k=2}^\infty \binom{1/2}{k}b^k \zeta(2k-2,N+\delta),
\end{equation}
with $N$ chosen so that $|b|/(N+\delta)^2<1$; differentiating this tail gives the corresponding tail in $G_\delta'$.

Numerical checks at representative endpoint-pole crossings confirm that the pressure remains finite and varies smoothly within the strictly positive-frequency domain. The results are stable under increasing the number of directly summed modes and retained tail terms.


\begin{thebibliography}{150}

\bibitem{Dirac1931}
	P.~A.~M.~Dirac,
	Quantised singularities in the electromagnetic field,
	Proc.\ R.\ Soc.\ London A \textbf{133}, 60 (1931).

\bibitem{Dirac1948}
	P.~A.~M.~Dirac,
	The Theory of Magnetic Poles,
	Phys.\ Rev.\ \textbf{74}, 817 (1948).

\bibitem{GoddardOlive1978}
	P.~Goddard and D.~I.~Olive,
	Magnetic monopoles in gauge field theories,
	Rep.\ Prog.\ Phys.\ \textbf{41}, 1357 (1978).

\bibitem{Preskill1983}
	J.~Preskill,
	Monopoles in the very early universe,
	in \textit{The Very Early Universe}, edited by G.~W.~Gibbons, S.~W.~Hawking, and S.~T.~C.~Siklos,
	(Cambridge University Press, Cambridge, 1983), pp.~119-146.

\bibitem{BlagojevicSenjanovic1988}
	M.~Blagojevi\'{c} and P.~Senjanovi\'{c},
	The quantum field theory of electric and magnetic charge,
	Phys.\ Rep.\ \textbf{157}, 233 (1988).

\bibitem{VilenkinShellard1994}
	A.~Vilenkin and E.~P.~S.~Shellard,
	\textit{Cosmic Strings and Other Topological Defects}
	(Cambridge University Press, Cambridge, 1994).

\bibitem{Shnir2005}
	Y.~M.~Shnir,
	\textit{Magnetic Monopoles}
	(Springer, Berlin, 2005).

\bibitem{WuYang1975}
	T.~T.~Wu and C.~N.~Yang,
	Concept of nonintegrable phase factors and global formulation of gauge fields,
	Phys.\ Rev.\ D \textbf{12}, 3845 (1975).

\bibitem{WuYang1976}
	T.~T.~Wu and C.~N.~Yang,
	Dirac monopole without strings: Monopole harmonics,
	Nucl.\ Phys.\ B \textbf{107}, 365 (1976).

\bibitem{tHooft1974}
	G.~'t~Hooft,
	Magnetic monopoles in unified gauge theories,
	Nucl.\ Phys.\ B \textbf{79}, 276 (1974).

\bibitem{Polyakov1974}
	A.~M.~Polyakov,
	Particle spectrum in the quantum field theory,
	Pis'ma Zh.\ Eksp.\ Teor.\ Fiz.\ \textbf{20}, 430 (1974).

\bibitem{GeorgiGlashow1972}
	H.~Georgi and S.~L.~Glashow,
	Unity of all elementary-particle forces,
	Phys.\ Rev.\ D \textbf{6}, 2977 (1972).

\bibitem{Preskill1979}
	J.~Preskill,
	Cosmological production of superheavy monopoles,
	Phys.\ Rev.\ Lett.\ \textbf{43}, 1365 (1979).

\bibitem{Salomaa1987}
	M.~M.~Salomaa,
	Monopoles in the rotating superfluid helium-3 A-B interface,
	Nature \textbf{326}, 367 (1987).

\bibitem{Castelnovo2008}
	C.~Castelnovo, R.~Moessner, and S.~L.~Sondhi,
	Magnetic monopoles in spin ice,
	Nature (London) \textbf{451}, 42 (2008).

\bibitem{Gingras2009}
	M.~J.~P.~Gingras,
	Observing monopoles in a magnetic analog of ice,
	Science \textbf{326}, 375 (2009).

\bibitem{Bramwell2009}
	S.~T.~Bramwell, S.~R.~Giblin, S.~Calder, R.~Aldus, D.~Prabhakaran, and T.~Fennell,
	Measurement of the charge and current of magnetic monopoles in spin ice,
	Nature (London) \textbf{461}, 956 (2009).

\bibitem{Giblin2011}
	S.~R.~Giblin, S.~T.~Bramwell, P.~C.~W.~Holdsworth, D.~Prabhakaran, and I.~Terry,
	Creation and measurement of long-lived magnetic monopole currents in spin ice,
	Nat.\ Phys.\ \textbf{7}, 252 (2011).

\bibitem{LinSaxena2016}
	S.-Z.~Lin and A.~Saxena,
	Dynamics of Dirac strings and monopolelike excitations in chiral magnets under a current drive,
	Phys.\ Rev.\ B \textbf{93}, 060401(R) (2016).

\bibitem{Fang2003}
	Z.~Fang, N.~Nagaosa, K.~S.~Takahashi, A.~Asamitsu, R.~Mathieu,
	T.~Ogasawara, H.~Yamada, M.~Kawasaki, Y.~Tokura, and K.~Terakura,
	The anomalous Hall effect and magnetic monopoles in momentum space,
	Science \textbf{302}, 92 (2003).

\bibitem{Beche2014}
	A.~B\'{e}ch\'{e}, R.~Van~Boxem, G.~Van~Tendeloo, and J.~Verbeeck,
	Magnetic monopole field exposed by electrons,
	Nat.\ Phys.\ \textbf{10}, 26 (2014).

\bibitem{Zhang2017}
	Z.-L.~Zhang, M.-F.~Chen, H.-Z.~Wu, and Z.-B.~Yang,
	Quantum simulation of Abelian Wu--Yang monopoles in spin-1/2 systems,
	Laser Phys.\ Lett.\ \textbf{14}, 045205 (2017).

\bibitem{Marques2024}
	M.~I.~Marqu\'{e}s, S.~Edelstein, P.~A.~Serena, B.~Castillo L\'{o}pez de Larrinzar, and A.~Garcia-Mart\'{i}n,
	Magneto-optical particles in isotropic spinning fields mimic magnetic monopoles,
	Phys.\ Rev.\ Lett.\ \textbf{133}, 046901 (2024).

\bibitem{Haldane1983}
	F.~D.~M.~Haldane,
	Fractional quantization of the Hall effect: A hierarchy of incompressible quantum fluid states,
	Phys.\ Rev.\ Lett.\ \textbf{51}, 605 (1983).

\bibitem{Jain2007}
	J.~K.~Jain,
	\textit{Composite Fermions}
	(Cambridge University Press, Cambridge, 2007).

\bibitem{Zhou2018}
	X.-F.~Zhou, C.~Wu, G.-C.~Guo, R.~Wang, H.~Pu, and Z.-W.~Zhou,
	Synthetic Landau levels and spinor vortex matter on a Haldane spherical surface with a magnetic monopole,
	Phys.\ Rev.\ Lett.\ \textbf{120}, 130402 (2018).

\bibitem{DolanHunterMcCabe2020}
	B.~P.~Dolan and A.~Hunter-McCabe,
	Ground state wave functions for the quantum Hall effect on a sphere and the Atiyah--Singer index theorem,
	J.\ Phys.\ A: Math.\ Theor.\ \textbf{53}, 215306 (2020).

\bibitem{GattuJain2025}
	M.~Gattu and J.~K.~Jain,
	Unlocking new regimes in fractional quantum Hall effect with quaternions,
	Phys.\ Rev.\ Lett.\ \textbf{134}, 156501 (2025).

\bibitem{Casimir1948}
	H.~B.~G.~Casimir,
	On the attraction between two perfectly conducting plates,
	Proc.\ K.\ Ned.\ Akad.\ Wet.\ \textbf{51}, 793 (1948).

\bibitem{PlunienMullerGreiner1986}
	G.~Plunien, B.~M\"{u}ller, and W.~Greiner,
	The Casimir effect,
	Phys.\ Rep.\ \textbf{134}, 87 (1986).

\bibitem{MostepanenkTrunov1988}
	V.~M.~Mostepanenko and N.~N.~Trunov,
	The Casimir effect and its applications,
	Sov.\ Phys.\ Usp.\ \textbf{31}, 965 (1988).

\bibitem{Milton2001}
	K.~A.~Milton,
	\textit{The Casimir Effect: Physical Manifestations of Zero-Point Energy}
	(World Scientific, Singapore, 2001).

\bibitem{BordagKlimchitskayaMohi2009}
	M.~Bordag, G.~L.~Klimchitskaya, U.~Mohideen, and V.~M.~Mostepanenko,
	\textit{Advances in the Casimir Effect}
	(Oxford University Press, New York, 2009).

\bibitem{BordagMohideenMostep2001}
	M.~Bordag, U.~Mohideen, and V.~M.~Mostepanenko,
	New developments in the Casimir effect,
	Phys.\ Rep.\ \textbf{353}, 1 (2001).

\bibitem{KlimchitskayaMohideenMostep2009}
	G.~L.~Klimchitskaya, U.~Mohideen, and V.~M.~Mostepanenko,
	The Casimir force between real materials: Experiment and theory,
	Rev.\ Mod.\ Phys.\ \textbf{81}, 1827 (2009).

\bibitem{Lamoreaux1997}
	S.~K.~Lamoreaux,
	Demonstration of the Casimir force in the 0.6 to 6~$\mu$m range,
	Phys.\ Rev.\ Lett.\ \textbf{78}, 5 (1997);
	Erratum: Phys.\ Rev.\ Lett.\ \textbf{81}, 5475 (1998).

\bibitem{Hertlein2008}
	C.~Hertlein, L.~Helden, A.~Gambassi, S.~Dietrich, and C.~Bechinger,
	Direct measurement of critical Casimir forces,
	Nature \textbf{451}, 172 (2008).

\bibitem{Munday2009}
	J.~N.~Munday, F.~Capasso, and V.~A.~Parsegian,
	Measured long-range repulsive Casimir--Lifshitz forces,
	Nature \textbf{457}, 170 (2009).

\bibitem{Paladugu2016}
	S.~Paladugu, A.~Callegari, Y.~Tuna, L.~Barth, S.~Dietrich, A.~Gambassi, and G.~Volpe,
	Nonadditivity of critical Casimir forces,
	Nat.\ Commun.\ \textbf{7}, 11403 (2016).

\bibitem{Somers2018}
	D.~A.~T.~Somers, J.~L.~Garrett, K.~J.~Palm, and J.~N.~Munday,
	Measurement of the Casimir torque,
	Nature \textbf{564}, 386 (2018).

\bibitem{ZhaoStable2019}
	R.~Zhao, L.~Li, S.~Yang, W.~Bao, Y.~Xia, P.~D.~Ashby,
	Y.~Wang, and X.~Zhang,
	Stable Casimir equilibria and quantum trapping,
	Science \textbf{364}, 984 (2019).

\bibitem{Schmidt2022}
	F.~Schmidt, A.~Callegari, A.~Daddi-Moussa-Ider, B.~Munkhbat, R.~Verre,
	T.~Shegai, M.~K\"{a}ll, H.~L\"{o}wen, A.~Gambassi, and G.~Volpe,
	Tunable critical Casimir forces counteract Casimir--Lifshitz attraction,
	Nat.\ Phys.\ \textbf{19}, 271 (2022).

\bibitem{ZhangMagField2024}
	Y.~Zhang, H.~Zhang, X.~Wang, Y.~Wang, Y.~Liu, S.~Li,
	T.~Zhang, C.~Fan, and C.~Zeng,
	Magnetic-field tuning of the Casimir force,
	Nat.\ Phys.\ \textbf{20}, 1282 (2024).

\bibitem{BellucciSaharian2009}
	S.~Bellucci and A.~A.~Saharian,
	Fermionic Casimir effect for parallel plates in the presence of compact dimensions with applications to nanotubes,
	Phys.~Rev.~D \textbf{80}, 105003 (2009).

\bibitem{ZhaoChiral2009}
	R.~Zhao, J.~Zhou, T.~Koschny, E.~N.~Economou, and C.~M.~Soukoulis,
	Repulsive Casimir force in chiral metamaterials,
	Phys.~Rev.~Lett.~\textbf{103}, 103602 (2009).

\bibitem{GrushinCortijo2011}
	A.~G.~Grushin and A.~Cortijo,
	Tunable Casimir repulsion with three-dimensional topological insulators,
	Phys.~Rev.~Lett.~\textbf{106}, 020403 (2011).

\bibitem{BezerraKlimchMostepRomero2011a}
	V.~B.~Bezerra, G.~L.~Klimchitskaya, V.~M.~Mostepanenko, and C.~Romero,
	Thermal Casimir effect in closed Friedmann universe revisited,
	Phys.~Rev.~D \textbf{83}, 104042 (2011).

\bibitem{BezerraMostepMotaRomero2011b}
	V.~B.~Bezerra, V.~M.~Mostepanenko, H.~F.~Mota, and C.~Romero,
	Thermal Casimir effect for neutrino and electromagnetic fields in the closed Friedmann cosmological model,
	Phys.~Rev.~D \textbf{84}, 104025 (2011).

\bibitem{Wilson2011}
	C.~M.~Wilson, G.~Johansson, A.~Pourkabirian, M.~Simoen,
	J.~R.~Johansson, T.~Duty, F.~Nori, and P.~Delsing,
	Observation of the dynamical Casimir effect in a superconducting circuit,
	Nature (London) \textbf{479}, 376 (2011).

\bibitem{TseMacDonald2012}
	W.-K.~Tse and A.~H.~MacDonald,
	Quantized Casimir force,
	Phys.~Rev.~Lett.~\textbf{109}, 236806 (2012).

\bibitem{Felicetti2014}
	S.~Felicetti, M.~Sanz, L.~Lamata, G.~Romero, G.~Johansson, P.~Delsing, and E.~Solano,
	Dynamical Casimir effect entangles artificial atoms,
	Phys.~Rev.~Lett.~\textbf{113}, 093602 (2014).

\bibitem{RodriguezLopezGrushin2014}
	P.~Rodriguez-Lopez and A.~G.~Grushin,
	Repulsive Casimir effect with Chern insulators,
	Phys.~Rev.~Lett.~\textbf{112}, 056804 (2014).

\bibitem{BezerraMotaRomero2014}
	V.~B.~Bezerra, H.~F.~Mota, and C.~Romero,
	Thermal Casimir effect in closed cosmological models with a cosmic string,
	Phys.~Rev.~D \textbf{89}, 024015 (2014).

\bibitem{SchechterKamenev2014}
	M.~Schecter and A.~Kamenev,
	Phonon-mediated Casimir interaction between mobile impurities in one-dimensional quantum liquids,
	Phys.~Rev.~Lett.~\textbf{112}, 155301 (2014).

\bibitem{Quach2015}
	J.~Q.~Quach,
	Gravitational Casimir effect,
	Phys.~Rev.~Lett.~\textbf{114}, 081104 (2015).

\bibitem{Wilson2015}
	J.~H.~Wilson, A.~A.~Allocca, and V.~Galitski,
	Repulsive Casimir force between Weyl semimetals,
	Phys.~Rev.~B \textbf{91}, 235115 (2015).

\bibitem{Hartmann2017}
	M.~Hartmann, G.-L.~Ingold, and P.~A.~Maia~Neto,
	Plasma versus Drude modeling of the Casimir force: Beyond the proximity force approximation,
	Phys.~Rev.~Lett.~\textbf{119}, 043901 (2017).

\bibitem{Song2017}
	G.~Song, J.~Xu, C.~Zhu, P.~He, Y.~Yang, and S.-Y.~Zhu,
	Casimir force between hyperbolic metamaterials,
	Phys.~Rev.~A \textbf{95}, 023814 (2017).

\bibitem{Macri2018}
	V.~Macr\`{i}, A.~Ridolfo, O.~Di~Stefano, A.~F.~Kockum, F.~Nori, and S.~Savasta,
	Nonperturbative dynamical Casimir effect in optomechanical systems: Vacuum Casimir--Rabi splittings,
	Phys.~Rev.~X \textbf{8}, 011031 (2018).

\bibitem{Ishikawa2020}
	T.~Ishikawa, K.~Nakayama, and K.~Suzuki,
	Casimir effect for lattice fermions,
	Phys.~Lett.~B \textbf{809}, 135713 (2020).

\bibitem{Ishikawa2021}
	T.~Ishikawa, K.~Nakayama, and K.~Suzuki,
	Lattice-fermionic Casimir effect and topological insulators,
	Phys.~Rev.~Res.~\textbf{3}, 023201 (2021).

\bibitem{Chernodub2022}
	M.~N.~Chernodub, V.~A.~Goy, A.~V.~Molochkov, and A.~S.~Tanashkin,
	Casimir boundaries, monopoles, and deconfinement transition in $(3+1)$-dimensional compact electrodynamics,
	Phys.~Rev.~D \textbf{105}, 114506 (2022).

\bibitem{NakayamaSuzuki2023}
	K.~Nakayama and K.~Suzuki,
	Dirac/Weyl-node-induced oscillating Casimir effect,
	Phys.~Lett.~B \textbf{843}, 138017 (2023).

\bibitem{JiangJing2025}
	T.-H.~Jiang and J.~Jing,
	Realizing a mechanical dynamical Casimir effect with a low-frequency oscillator,
	Phys.~Rev.~A \textbf{111}, 022811 (2025).

\bibitem{MamaevTrunov1979b}
	S.~G.~Mamaev and N.~N.~Trunov,
	Vacuum averages of the energy-momentum tensor of quantized fields on manifolds of various topology and geometry.~II,
	Sov.~Phys.~J.~\textbf{22}, 966 (1979).

\bibitem{MamaevTrunov1979a}
	S.~G.~Mamaev and N.~N.~Trunov,
	Vacuum expectation values of the energy-momentum tensor of quantized fields on manifolds with different topologies and geometries.~I,
	Sov.~Phys.~J.~\textbf{22}, 766 (1979).

\bibitem{MamaevTrunov1980}
	S.~G.~Mamaev and N.~N.~Trunov,
	Vacuum expectation values of the energy-momentum tensor of quantized fields on manifolds with different topologies and geometries.~III,
	Sov.~Phys.~J.~\textbf{23}, 551 (1980).

\bibitem{Boyer1968}	
	T.~H.~Boyer,
	Quantum Electromagnetic Zero-Point Energy of a Conducting Spherical Shell and the Casimir Model for a Charged Particle,
	Phys.~Rev.~\textbf{174}, 1764 (1968).

\bibitem{Davies1972}
	B.~Davies,
	Quantum Electromagnetic Zero‐Point Energy of a Conducting Spherical Shell,
	J.~Math.~Phys.~\textbf{13}, 1324 (1972).

\bibitem{MiltonDeRaadSchwinger1978}
	K.~A.~Milton, L.~L.~DeRaad Jr., and J.~Schwinger,
	Casimir self-stress on a perfectly conducting spherical shell,
	Ann.~Phys.~\textbf{115}, 388 (1978).

\bibitem{Ford1975}
	L.~H.~Ford,
	Quantum vacuum energy in general relativity,
	Phys.~Rev.~D \textbf{11}, 3370 (1975).

\bibitem{HerdeiroSampaio2006}
	C.~A.~R.~Herdeiro and M.~Sampaio,
	Casimir energy and a cosmological bounce,
	Class.~Quantum Grav.~\textbf{23}, 473 (2006).

\bibitem{HerdeiroRibeiroSampaio2008}
	C.~A.~R.~Herdeiro, R.~H.~Ribeiro, and M.~Sampaio,
	Scalar Casimir effect on a D-dimensional Einstein static universe,
	Class.~Quantum Grav.~\textbf{25}, 165010 (2008).

\bibitem{Valuyan2018}
	M.~A.~Valuyan,
	Casimir energy calculation for massive scalar field on spherical surfaces: an alternative approach,
	Can.~J.~Phys.~\textbf{96}, 1004 (2018).

\bibitem{BenderMilton1994}
	C.~M.~Bender and K.~A.~Milton,
	Scalar Casimir effect for a $D$-dimensional sphere,
	Phys.~Rev.~D \textbf{50}, 6547 (1994).

\bibitem{BordagKirsten1996}
	M.~Bordag and K.~Kirsten,
	Vacuum energy  in a spherical symmetric background field, 
	Phys.~Rev.~D \textbf{53}, 5753 (1996).

\bibitem{BordagElizalde1997}
	M.~Bordag, E.~Elizalde, K.~Kirsten, and S.~Leseduarte,
	Casimir energies for massive scalar fields in a spherical geometry,
	Phys.~Rev.~D \textbf{56}, 4896 (1997).

\bibitem{ElizaldeBordagKirsten1998}
	E.~Elizalde, M.~Bordag, and K.~Kirsten,
	Casimir energy for a massive fermionic quantum field with a spherical boundary,
	J.~Phys.~A: Math.~Gen.~\textbf{31}, 1743 (1998).

\bibitem{Hagen2000}
	C.~R.~Hagen,
	Casimir energy for spherical boundaries,
	Phys.~Rev.~D \textbf{61}, 065005 (2000).

\bibitem{CognolaElizaldeKirsten2001}
	G.~Cognola, E.~Elizalde, and K.~Kirsten,
	Casimir energies for spherically symmetric cavities,
	J.~Phys.~A: Math.~Gen.~\textbf{34}, 7311 (2001).

\bibitem{Saharian2001}
	A.~A.~Saharian,
	Scalar Casimir effect for $D$-dimensional spherically symmetric Robin boundaries,
	Phys.~Rev.~D \textbf{63}, 125007 (2001).

\bibitem{PufuSachdev2013}
	S.~S.~Pufu and S.~Sachdev,
	Monopoles in $2+1$-dimensional conformal field theories with global U(1) symmetry,
	J.~High Energ.~Phys.~\textbf{2013}, 127 (2013).

\bibitem{BalianDuplantier1978}
	R.~Balian and B.~Duplantier,
	Electromagnetic waves near perfect conductors.~II.~Casimir effect,
	Ann.~Phys.~\textbf{112}, 165 (1978).

\bibitem{Levin2010}
	M.~Levin, A.~P.~McCauley, A.~W.~Rodriguez, M.~T.~H.~Reid, and S.~G.~Johnson,
	Casimir repulsion between metallic objects in vacuum,
	Phys.~Rev.~Lett.~\textbf{105}, 090403 (2010).

\bibitem{JiangWilczek2019}
	Q.-D.~Jiang and F.~Wilczek,
	Chiral Casimir forces: Repulsive, enhanced, tunable,
	Phys.~Rev.~B \textbf{99}, 125403 (2019).

\bibitem{Riess1998}
	A.~G.~Riess \textit{et al.} (Supernova Search Team Collaboration),
	Observational evidence from supernovae for an accelerating universe and a cosmological constant,
	Astron.\ J.\ \textbf{116}, 1009 (1998).

\bibitem{Perlmutter1999}
	S.~Perlmutter \textit{et al.} (Supernova Cosmology Project Collaboration),
	Measurements of $\Omega$ and $\Lambda$ from 42 high-redshift supernovae,
	Astrophys.\ J.\ \textbf{517}, 565 (1999).

\bibitem{ZeldovichStarobinsky1984}
	Ya.~B.~Zel'dovich and A.~A.~Starobinsky,
	Quantum creation of a universe in a nontrivial topology,
	Sov.\ Astron.\ Lett.\ \textbf{10}, 135 (1984).

\bibitem{Kaluza1921}
	T.~Kaluza,
	On the Unification Problem in Physics,
	Sitz.\ Preuss.\ Akad.\ Wiss.\ Berlin (Math.\ Phys.) 966-972 (1921).

\bibitem{Klein1926}
	O.~Klein,
	Quantentheorie und f\"unfdimensionale Relativit\"atstheorie, 
	Z.\ Phys.\ \textbf{37}, 895 (1926).

\bibitem{AppelquistChodos1983}
	T.~Appelquist and A.~Chodos,
	The quantum dynamics of Kaluza--Klein theories,
	Phys.\ Rev.\ D \textbf{28}, 772 (1983).

\bibitem{deMello2002}
	E.~R.~Bezerra~de~Mello,
	Vacuum polarization effects in the global monopole spacetime in the presence of the Wu--Yang magnetic monopole,
	Classical Quantum Gravity \textbf{19}, 5141 (2002).

\bibitem{BirrellDavies1982}
	N.~D.~Birrell and P.~C.~W.~Davies,
	\textit{Quantum Fields in Curved Space}
	(Cambridge University Press, Cambridge, 1982).

\bibitem{DLMF}
	\textit{NIST Digital Library of Mathematical Functions}, edited by F.~W.~J.~Olver, A.~B.~Olde Daalhuis, D.~W.~Lozier, B.~I.~Schneider, R.~F.~Boisvert, C.~W.~Clark, B.~R.~Miller, and B.~V.~Saunders, http://dlmf.nist.gov/.

\bibitem{Saharian2007}
	A.~A.~Saharian,
	The generalized Abel--Plana formula with applications to Bessel functions and Casimir effect,
	arXiv:0708.1187 [hep-th].

\bibitem{Vassilevich2003}
	D.~V.~Vassilevich,
	Heat kernel expansion: user's manual,
	Phys.~Rep.~\textbf{388}, 279 (2003).

\bibitem{DecaniniFolacci2008}
	Y.~D\'ecanini and A.~Folacci,
	Hadamard renormalization of the stress-energy tensor for a quantized scalar field in a general spacetime of arbitrary dimension,
	Phys.~Rev.~D~\textbf{78}, 044025 (2008).

\end{thebibliography}
\end{document}